\documentclass[11pt,a4paper,english]{article}
\pdfoutput=1
\usepackage[latin1]{inputenc}
\usepackage{babel}
\usepackage{fancyhdr}
\usepackage[T1]{fontenc}
\usepackage{lmodern}
\usepackage{amsmath,amssymb}
\usepackage{siunitx}
\usepackage[version=3]{mhchem}
\usepackage{soul}
\usepackage{array,booktabs}
\usepackage{geometry}
\usepackage{graphicx}
\usepackage{icomma}
\usepackage{varioref}
\usepackage{mathtools}
\usepackage{xspace} 
\usepackage{rotating}
\usepackage{multirow}
\usepackage{caption}
\usepackage{nomencl} 
\usepackage{calc} 
\usepackage{subcaption} 
  \DeclareCaptionLabelSeparator{tilpasning}{:\quad}
\usepackage[nottoc,notlot,notlof]{tocbibind}
\usepackage[latin1]{inputenc}

\renewcommand{\si}[1]{\SI{#1}}

\newcommand{\fref}[1]{{\bf Figure \ref{#1}}}
\newcommand{\tref}[1]{{\bf Table \ref{#1}}}

\renewcommand{\deg}{\,^{\circ}}
\renewcommand{\d}{\,\text{d}}
\newcommand{\CA}{${}^{13}C\alpha{}$\xspace}
\newcommand{\CB}{${}^{13}C\beta$\xspace}
\newcommand{\C}{${}^{13}C'$\xspace}
\newcommand{\N}{${}^{15}N^{H}$\xspace}
\newcommand{\HN}{${}^{1}H^{N}$\xspace}
\newcommand{\HA}{${}^{1}H\alpha$\xspace}

\fancyheadoffset[LE,RO]{1pt} 
\fancypagestyle{concon}{
\fancyhf{} 
\lhead{Anders St\o ttrup Larsen} \chead{Master Thesis} \rhead{\today} 
\fancyheadoffset[LE,RO]{1pt} 
}

\begin{document}

{\LARGE \bf Abstract}

The protein chemical shifts holds a large amount of information about the 3-dimensional structure 
of the protein. A number of chemical shift predictors based on the relationship between structures 
resolved with X-ray crystallography and the corresponding experimental chemical shifts have been developed.
These empirical predictors are very accurate on X-ray structures but tends to be insensitive to small 
structural changes. To overcome this limitation it has been suggested to make chemical shift predictors
based on quantum mechanical(QM) calculations. In this thesis the development of the QM derived chemical shift predictor Procs14 is presented.
 Procs14 is based on $\sim2.35$ million density functional theory(DFT) calculations on tripeptides and contains corrections
for hydrogen bonding, ring current and the effect of the previous and following residue. The hydrogen bond terms
are based on DFT calculations modeling the donor acceptor interaction and the previous/following corrections are derived from
the tripeptide calculations. Procs14 is capable at performing predictions for the \CA, \CB, \C, \N, \HN and \HA backbone atoms.
In order to benchmark Procs14, a number of QM NMR calculations
are performed on full protein structures. Of the tested empirical and QM derived predictors, Procs14 reproduced the
QM chemical shifts with the highest accuracy. A comparison with the QM derived predictor CheShift-2 on X-ray structures and
NMR ensembles with experimental chemical shift data, showed that Procs14 predicted the chemical shifts with the best accuracy. The
predictions on the NMR ensembles exhibited the best performance. This suggests that future work might benefit from
using ensemble sampling when performing simulations of protein folding with chemical shifts.  
Procs14 is implemented in the markov chain monte carlo protein folding framework PHAISTOS.
The computational efficient implementation of Procs14 allows for rapid predictions and therefore potential use 
in refinement and folding of protein structures. 
\\*
\\*
\\*
{\LARGE  Acknowledgements}

I would like to thank my two advisors Jan H. Jensen and Anders S. Christensen for guiding me in my work on
both on my bachelor and master thesis. 
Thanks to Jimmy Charnley Kromann for taking his Starcraft 2 defeats so gracefully.
Thanks to Lars Bratholm for serving me juice and finally thanks to Nina Str\o ttrup Larsen
for help with grammar.

\newpage

\tableofcontents
\thispagestyle{empty}

\newpage
\section{Introduction}

Proteins are central to the understanding of biology.
The central dogma of molecular biology states that DNA is transcribed into RNA and RNA is translated to proteins.
Proteins are involved in a huge variety of tasks including constructing the cell, enzymatic activity, acting as messenger 
molecules and providing support as structural components. The function and chemistry of a protein depends 
crucially on its 3-dimensional structure. Therefore a number of methods have been developed to infer protein structure with the most dominant being
X-ray crystallography\cite{xraycry}. The Protein Data Bank contains $\sim90,000$ protein structures solved with X-ray crystallography.
Only a limited number of proteins are suitable for this approach. Additionally a crystalised
structure might not capture the true dynamical nature of a protein in solution. This encourages the use of NMR spectroscopy which captures information about the protein in its natural environment. 
The NMR spectra contains a wealth of structural information including residual dipolar couplings(RDC)\cite{rdc}, distance restraints derived
from the Nuclear Overhauser Effect(NOE) and the chemical shift.
The NOE restraints allow the calculation of structural
ensembles that best fit the restraints\cite{noes}. Assigning the restraints can be difficult. 
Therefore the more available chemical shift is an attractive source
of structural information. The chemical shift describes the resonant frequency of an atom's magnetic moment interaction with a local magnetic field relative to a reference.
The local field is changed by the nucleus local environment and it is from this effect that the chemical shift receives its structural explanatory power.
If experimental chemical shifts are to be used to fold and refine protein structures, a method that
predicts the chemical shifts given a specific structure is needed\cite{nmrshiftscava}\cite{csfold}. Most current chemical shift predictors are derived
from the relationship between empirical data and crystal structures and as a result are insensitive to small structural changes\cite{sensitive}\cite{Vila06102009}. 
Predictors based on quantum mechanical(QM) data provides an attractive alternative and should in principle be able to 
overcome the shortcomings of the empirical methods. 

The work in this thesis documents the Procs14 QM derived
chemical shift predictor. Procs14 is based on $2.35$ million density functional theory(DFT) calculation on tripeptides and contains
corrections for hydrogen bonding, ring current and the previous/following residue. Procs14 is capable of predicting 
the chemical shift of six protein backbone atoms \CA, \CB, \C, \N, \HN and \HA.
It is implemented in the monte carlo markov chain protein folding framework PHAISTOS\cite{phaistos2}. 
This allows Procs14 to potentially be used for refinement and folding of protein structures.

Here follows a short outline of the thesis structure. 
\begin{itemize}

\item Chapter {\bf \ref{sec:Background} \chapter{Background}} The thesis' second chapter contains an introduction to the chemical shift and contains a discussion of 
current available chemical shift predictor methods. The chapter also contains a short introduction to the used computational methods and the hybrid energy approach. 

\item Chapter {\bf \ref{sec:compmethods} \chapter{Computational Methods}} Here the basics of the computational methods used in the parameterization of Procs14 is presented. 
	The first section is a short introduction to density functional theory and its use in the calculation of nuclear shielding tensors. The next section contains a test of a number of
	different methods in the optimization of tripeptides.

\item Chapter {\bf \ref{sec:Procs14} \chapter{Procs14}} Contains a detailed account of the components of
Procs14 and should in principle allow for the reproduction of Procs14 from scratch. The first section is a overview
of the terms included in Procs14. The second section and its subsection describe in detail the QM NMR calculations
on the tripeptides and the results. Next comes a section detailing the hydrogen bond term scans and some results.
The final sections documents a small hydrogen bond length correction, the scaling procedure used in Procs14 and the 
PHAISTOS implementation.

\item Chapter {\bf \ref{sec:Benchmarking} \chapter{Benchmarking}} This chapter contains benchmarks of the \HN and \HA
	hydrogen bond terms and the previous/following correction. The benchmarks includes test done on QM NMR calculations on full
	protein structures which are optimised with PM6 or a force field. The accuracy of Procs14 in reproducing 
	the QM level chemical shifts is compared with a selection of different chemical shift predictors. Finally Procs14 is compared with 
	the competing QM derived chemical shift predictor CheShift-2 on experimental data.

\item Chapter {\bf \ref{sec:Refinement} \chapter{Preliminary Results From Refinement}} This chapter contains a very preliminary test of the Procs14 PHAISTOS implementation in facilitating
		protein folding.

\item Chapter {\bf \ref{sec:theend} \chapter{Discussion and Conclusion}} This chapter starts with a discussion of the results from Procs14
	and outlines paths for future work. The thesis ends with a conclusion.
	
\item Chapter {\bf \ref{sec:append} \chapter{Appendix}} The appendix contains a number of figures and tables that could not fit in the main text. This includes 
	schematical representations of all the tripeptides used.

\end{itemize}
  
\newpage
\section{Background} \label{sec:Background}
\subsection{The Chemical Shift}
A wealth of different data can can be obtained by nuclear magnetic resonance spectroscopy. The chemical shifts
remains one of the most useful types of data generated by these experiments. They have been
used to resolve questions and structures in chemistry and biochemistry. Consider a nucleus $A$ placed in
a applied magnetic field $B_{0}$ equation (\ref{eq:naked})\cite{sauer}. 
The magnetic field will induce the nucleus magnetic moment $\mu$ to precess around the directions
of the field with the larmor frequency $\nu_{A}$. The frequency
depends on the strength of the magnetic field and the gyromagnetic ratio $\gamma_{A}$ of the nucleus type.

\begin{equation} \label{eq:naked}
	\nu_{A} = \frac{\gamma_{A}}{2\pi}B_{0}
\end{equation}  

Of great importance to NMR spectroscopy, the magnetic field interacting with $A$ can be changed by the nucleus environment equation (\ref{eq:Blocal}).
The local magnetic field is modified according to a nuclear shielding tensor $\sigma_{A}$, which is dependent on local interactions.
It is the larmor frequency of the nucleus modified by the local shielding, that is measured in the NMR experiments.

\begin{equation} \label{eq:Blocal}
	B_{A} = (1 - \sigma_{A})B_{0}
\end{equation} 

Measuring the precise strength of the magnetic field at the nucleus is infeasible in praxis. Therefore
the experiments are performed in the presence of a reference compound. The reference should contain the
same nuclei types for which the experiment is intended. The chemical shift $\delta_{A}$ can be defined using the
reference frequency $\nu_{A}^{ref}$ equation (\ref{eq:Bref}). The chemical shift is a dimensionless quantity and 
typically very small. The $10^{6}$ factor gives us the chemical shift in parts per million(\textit{ppm}).

\begin{equation} \label{eq:Bref}
	\delta_{A} = \frac{\nu_{A}-\nu_{A}^{ref}}{\nu_{A}^{ref}}10^{6}
\end{equation} 
\begin{equation} \label{eq:Bshielding}
	\delta_{A} \approx (\sigma_{A}^{ref} - \sigma_{A})10^{6}
\end{equation}

The chemical shift can be approximated as the difference between the shielding tensors of the nucleus and
the reference compound equation (\ref{eq:Bshielding}). This identity is useful, since computational chemistry
methods computes theoretical nuclear shielding tensors.
If the magnetic moment of the nucleus $k$ is placed in a magnetic field. Then the magnetic shielding tensor will take
the form shown in \eqref{eq:thematrix}, where $x$,$y$ and $z$ are cartesian components. A molecule in gas-phase or
solution will tumble and rotate, therefore only the isotropic nuclear shieldings are observed. This means that to compute the
nuclear shielding constant $\sigma^{(k)}$ we take the average of the isotropic elements of the tensor, which can be
found by taking the trace of the matrix ${\underline{\underline{\sigma}}}{}^{(k)}$. 

\begin{equation} \label{eq:thematrix}
{\underline{\underline{\sigma}}}{}^{(k)} =  
\begin{pmatrix}
\sigma_{xx}^{(k)} & \sigma_{xy}^{(k)} & \sigma_{xz}^{(k)} \\
\sigma_{yx}^{(k)} & \sigma_{yy}^{(k)} & \sigma_{yz}^{(k)} \\
\sigma_{zx}^{(k)} & \sigma_{zy}^{(k)} & \sigma_{zz}^{(k)} \\
\end{pmatrix}
\end{equation}

The individual components of the matrix is found by \eqref{eq:tensorget}. It works by taking the second order partial derivative
of the energy $E$ with respect to the magnetic field $B$ and the magnetic moment. $\alpha$ and $\beta$ are the system's cartesian components.

\begin{equation} \label{eq:tensorget}
\sigma_{\alpha\beta}^{(k)} = \frac{\partial^{2}E(\vec{B},\vec{\mu}^{(k)})}{\partial B_{\beta}\partial\mu_{\alpha}^{(k)}}\bigg|_{|\vec{B}|=0,|\vec{\mu}^{(k)}|=0}
\end{equation}

Usually the energy is found by perturbation theory in which the coupled perturbed Hartree-Fock (CPHF) equations are solved iteratively.
A problem that emerge when solving the CPHF equations is that the vector potential $\vec{A}$ depends on the gauge origin $\vec{r}_{O}$ \eqref{eq:giao}.
This would make the chemical shift dependent on the choice of the gauge origin.

\begin{equation} \label{eq:giao}
	\vec{A}(\vec{r}) = \frac{1}{2}\vec{B}\times(\vec{r} - \vec{r}_{O})
\end{equation}

The effect on the shielding from the choice of gauge origin will in principle cancel out if an infinite basis set is used. This is not 
practical instead a number of approximations have been developed. The most widely used is the Gauge-including atomic orbitals (GIAO) method.
GIAO\cite{giao} introduces a magnetic field dependence in the basis functions. The result is that integrals with the GIAO basis functions does not
depend on the gauge origin. The GIAO method is used for all QM NMR calculations in this thesis.

\subsection{Chemical Shifts on NMR Ensembles}

In order to get a chemical shift prediction for NMR ensembles one would ideally like to
compute the chemical shift by summing over boltzmann factor weighted chemical shifts equation \eqref{eq:boltz}\cite{repair}, with the condition that
$\sum_{j}^{N}\lambda_{j}=1$.

\begin{equation} \label{eq:boltz}
	\delta_{i} = \displaystyle\sum_{j=1}^{N}\lambda_{j}\delta_{ij}    
\end{equation}  

Here $\delta_{i}$ is the chemical shift of an atom that we want to compute, $N$ is the total number of conformations in the ensemble, 
$\lambda_{j}$ is the boltzmann factor of a structure $j$ in the ensemble and $\delta_{ij}$ is a chemical shift in conformation $j$. 
  Since it is not feasible to compute
QM level boltzmann factors they are approximated by assuming each that conformation contributes equally to
the chemical shift i.e. $\lambda_{j} = \frac{1}{N}$. This turns the calculation in to a simple average over the ensembles chemical shift for each atom.  

\subsection{Protein Chemical Shift Prediction Methods}

There have been developed an number of methods that predict protein chemical shift. 
Out of these, a key approach is using a semi-empirical method. SHIFTX \cite{shiftx} and 
SPARTA\cite{sparta} are among the most successful. SHIFTX uses an additive model equation (\ref{eq:shiftx}) where $i$ is the residue index, with classical or semi-classical
terms for ring current $\delta_{RC}$, electric field $\delta_{EF}$ and hydrogen bonding  $\delta_{HB}$. In addtion it uses chemical shift hypersurfaces $\delta_{HS}$ derived from emperical data.
They describe the chemical shift dependents on the backbone $\phi$ and $\psi$ torsion angles. All of these terms, are added to a random coil term $\delta_{coil}$. SHIFTX
is thereby able to predict the chemical shift of the ${}^{13}C\alpha$, ${}^{13}C\beta$, ${}^{13}C'$, ${}^{15}N^{H}$, ${}^{1}H^{N}$ and ${}^{1}H\alpha$ backbone atoms.

\begin{equation} \label{eq:shiftx}
	\delta_{calc} = \delta_{coil}+\delta_{RC}+\delta_{EF}+\delta_{HB}+\delta_{HS}  
\end{equation}

The SPARTA method searches a data base of triplets of adjacent residues for sequence homology and structural similarity. The residues are compared
using the $\phi$, $\psi$ and $\chi_{1}$ torsion angles. Each residue gets a similarity score calculated with all tripeptides in the data base. 
The $20$ best triplets are averaged and together with terms for ring current and hydrogen bonding and it results in the chemical shift prediction.
SHIFTY\cite{shifty} like SPARTA, uses sequence homology. It searches the BioMagResBank\cite{BMRB} database to produce a chemical shift prediction. 
Unlike SPARTA, SHIFTY searches for global similarity and uses only sequence information as input. Improved versions of SHIFTX and SHIFTY have 
been combined to make the SHIFT2X model\cite{shift2x}. Structure-based semi-empirical and sequence based homology
methods have been found to exhibit different strengths and weaknesses. By combining them SHIFT2X achieves better performance than either of them.
The two methods are compared on a atom by atom basis and when the difference in prediction is small, SHIFTY is weighted  higher. When the two methods disagree
above a certain threshold the SHIFTX predictions dominates. SHIFTX2 is also capable of predicting chemical shifts of certain ${}^{13}C$ and ${}^{1}H$ side chain
atoms and not just the six backbone atom types.
   
Another method distinct from the SHIFTX and SPARTA is CamShift\cite{camshift}. CamShift calculates the chemical shift of an atom by a polynomial expansion of interatomic 
distances in the vicinity of the atom, equation (\ref{eq:camshift}). 

\begin{align} \label{eq:camshift}
       \delta^{pred}_{a} &= \delta^{rc}_{a} +  \displaystyle\sum_{b,c} \alpha_{bc}d_{bc}^{\beta bc}    
\end{align}

The parameters $\alpha$ and $\beta$ are defined 
by atom and residue type. In addtion CamShift contains contributions for ring current, hydrogen bonding and a psi/psi backbone term. 
The parameters were obtained by refining CamShift predictions with a data set of experimental NMR values and PDB structures.

All of these methods use empirical data in their models. The structures are overwhelmingly PDB structures obtained by x-ray crystallography deposited in the
RSCB\cite{rscb}. There is only a limited number of good quality protein structures available. The measurement for good quality is usually an resolution of less than
$2$ \AA. In the case of SHIFTX2 197 PDB-BMRB pairs were used in the training set. The limited amount of data results in inaccurate predictions on 
structures that differs significantly from proteins in the training set. The experimental chemical shift is an ensemble average over the conformations of the 
protein during the experiment. All the previously described methods have this implicit in their models since they are fitted with experimental NMR data.
This often results in a relatively insensitivity to small structural change. To overcome these problems it has been suggested to make chemical shift predictions based on quantum mechanical calculations of chemical shifts. 
Calculations of nuclear shielding tensors have advanced considerably\cite{mulder}.
Performing a QM NMR calculation on a full protein model is computationally expensive and dependant on the size of the protein, the calculation time is 
measured in days or more. Instead of computing the chemical shifts using a single QM calculation on a full protein, CheShift-2\cite{cheshift2} predicts the ${}^{13}C\alpha$, ${}^{13}C\beta$ chemical shift using a database of DFT calculations on peptide fragments. The fragments consisted of acetyl-Gly-XXX-Gly-N-methyl model peptides where X is any of the naturally occurring amino acids.
The backbone $\phi$ and $\psi$ dihedral angles were sampled with a stepsize of $10\deg$ and the $\chi_{1}$ sampled with 30$\deg$. The $\chi_{2}$ angles were generated with
a rotamer library. In total CheShift-2 is based on $\sim 600,000$ peptide conformations. The model peptides was made using standardized bond angles and bond lengths from
the ECEPP/3 forcefield\cite{ECEPP}. CheShift-2 shows that it is more sensitive to small structural changes compared to empirically
derived methods\cite{Vila06102009}. CheShift-2 is available as a web server that can be used to refine protein structures by optimizing $\chi_{1}$/$\chi_{2}$ side chain angles. 
Unfortunately the raw data and code used in CheShift-2 is not available to the public.

Another example in SHIFTS\cite{shifts} which implements a DFT based database, which is made from calculations on small peptides. It provides predictions for ${}^{13}C$, ${}^{15}N$ atoms of the protein backbone. 
The peptides for the backbone term consisted of GAXAG where X is one of the amino acids. 
A total of 1335 model peptides were sampled from the most likely regions of the ramachandran plot.
The peptides were also used to construct corrections for the side chain and backbone angles of the previous and following residues. SHIFTS like SHIFTX uses an additive model of the chemical shift contributions. Because of the strategy of sampling from the ramachandran plot, SHIFTS can only provide good predictions in regions of the protein with well defined secondary structure.

Both CheShift-2 and SHIFTS have several limitations, most noticeably they don't provide predictions for the ${}^{1}H$ nucleus. Both methods are based on a rather limited number of samples especially
SHIFTS. CheShift-2 does not contain a hydrogen bond correction and SHIFTS implements a simple empirical model fitted with DFT data. If QM derived chemical shifts are to be used to refine and fold 
large protein structures there seems to be a need for a new and fast method.

\subsection{Introduction to the Hybrid Energy Approach}
A problem that arises when using chemical shifts as a bias in protein folding is 
how to properly include the chemical shifts as an energy term together with a force field energy.
This subsection presents the hybrid energy approach equation \eqref{eq:hybrid}, which has been developed within Jan Jensen group and
is conceptually similar to the Inferential Structure Determination(ISD) method\cite{isd}.

\begin{equation} \label{eq:hybrid}
		E_{hybrid} = w_{data}E_{data} + E_{physical}
\end{equation}  

Instead of attempting to minimize just a physical energy function $E_{physical}$ which would normaly come from a force field,
 the hybrid energy approach adds an energy term that is dependent on experimental data $w_{data}E_{data}$. Here $E_{data}$ is an
energy function which depends on the agreement between the predicted and experimental chemical shifts with a weight $w_{data}$.
The ISD method uses Bayes' theorem to represent the hybrid energy equation \eqref{eq:hybrid1}. Here the aim is to calculate 
the probability of a structure $X$ and the parameters $n$ given a set of chemical shifts $\{\delta\}$.

\begin{equation} \label{eq:hybrid1}
     p(X,n|\{\delta\}) = \frac{ p(\{\delta\}|X,n)p(X,n) }{ p( \{\delta\} ) }
\end{equation}

Since we are only interested in relative energy differences we can neglect the normalisation factor $ p( \{\delta\} )$.
If we assume that the error of our chemical shift predictor follows a gaussian distribution and $X$ and $n$ are independent, we can write the probability 
of $p(X,n|\{\delta\})$ as seen in equation \eqref{eq:hybrid2}. 

\begin{equation} \label{eq:hybrid2}
     p(X,n) = p(X)\displaystyle\prod_{j}p(\sigma_{j})
\end{equation}
 
Where $\sigma_{j}$ is the standard deviation of each Procs14 atom type which we model with Jeffreys prior $p(\sigma_{j})=\frac{1}{\sigma_{j}}$. The probability of the structure $p(X)$ is
given by the Boltzmann distribution which is a function of the physical energy of $X$ i.e. $E(X)$.

Assuming a gaussian distribution for the error of the chemical shift prediction, we can express the conditional probability of the set of chemical shifts given the structure and parameters as 
seen in equation \eqref{eq:hybrid3}. It should be understood as the product of the probability of each chemical shift where $N_{j}$ is the total number of chemical shifts for atom type $j$.
We describe the difference between predicted and experimental chemical shift with $= \Delta\delta(X)=\delta^{predicted}(X)-\delta^{experimental}$. 

\begin{equation} \label{eq:hybrid3}
\begin{split}
     p(\{\delta\}|X,n ) & = \displaystyle\prod_{j}\displaystyle\prod_{i}^{N_{j}}\frac{1}{\sigma_{j}\sqrt{2\pi}}\exp{\left(-\frac{\Delta\delta_{ij}(X)^{2}}{2\sigma_{j}^{2}}\right)} \\
                        & = \displaystyle\prod_{j} \left( \frac{1}{\sigma_{j}\sqrt{2\pi}}\right)^{N_{j}}\exp{\left(-\frac{\sum_{i}^{N_{j}}\Delta\delta_{ij}(X)^{2}}{2\sigma_{j}^{2}}\right)}  
\end{split}
\end{equation}

If we use an integrated likelihood model we can integrate out the standard deviations as seen in equation \eqref{eq:hybrid5}.

\begin{equation} \label{eq:hybrid5}
\begin{split}
	p(\{\delta\}|X) & = \displaystyle\prod_{j}\displaystyle\int_0^\infty d\sigma \frac{1}{\sigma_{j}}\left( \frac{1}{\sigma_{j}\sqrt{2\pi}}\right)^{N_{j}}\exp{\left(-\frac{\sum_{i}^{N_{j}}\Delta\delta_{ij}(X)^{2}}{2\sigma_{j}^{2}}\right)} \\     
                    & \propto \displaystyle\prod_{j}\left(\displaystyle\sum_{i}^{N_{j}}\Delta\delta_{ij}(X)^{2}\right)^{\frac{-N_{j}}{2}}  
\end{split}
\end{equation}

We can now express the hybrid energy as the logarithm of the probability of the chemical shifts given the structure multiplied by the probability of the structure, see equation \eqref{eq:hybrid7}. 
Here we set $\sum_{i}^{N_{j}}\Delta\delta_{ij}(X)^{2} = \chi_{j}^{2}(X)$. 

\begin{equation} \label{eq:hybrid7}
      E_{hybrid} = -k_{B}T \ln(p(\{\delta\}|X)p(X)) =  k_{B}T\displaystyle\sum_{j}\left(\frac{N_{j}}{2}ln(\chi_{j}^{2}(X)) \right) + E(X)
\end{equation} 

\clearpage
\section{Computational Methods}\label{sec:compmethods}
\subsection{Density functional Theory Calculations}

All calculation of the nuclear shielding tensors in this work is done with density functional theory(DFT) which provides an alternative to the 
traditional wave function approaches. DFT is based on the observation that the energy of the electronic ground state $E$ is given by 
the electron density $\rho$, equation \eqref{eq:dftenergy}\cite{dft}.

\begin{equation} \label{eq:dftenergy}
         E = E[\rho] 
\end{equation} 

The energy is calculated from the electron density with the use of functionals(functions of functions).
DFT is usually formulated in the Kohn-Sham framework. In this approach the electronic kinetic energy is expressed with Kohn-Sham orbitals instead of functionals.
This leaves the exchange-correlation energy as the only energy to be computed using functionals. 
A number of functionals have been developed to calculate the exchange-correlation energy. One of the most widely used
is the B3LYP hybrid functional\cite{b3lyp}. It has been shown to reproduce proton chemical shift with a RMSD of  $0.15$ \textit{ppm}\cite{scaling}.
In one set of tests B3LYP with a empirical scaling technique reproduced experimental ${}^{13}C$ chemical shifts with a mean absolute difference
of $2.90$ \textit{ppm}\cite{c13b3lyp} on a taxol molecule. This compared favorably $4.96$ \textit{ppm} for hartree fock calculations on the same test systems. 
In another study that included both ${}^{1}H$ and ${}^{13}C$ chemical shifts\cite{csshiftreviewc13h1}, calculations on a set of organic molecules at the B3LYP/6-31+G(d,p)//B3LYP/6-31G(d) level of theory
gave ${}^{1}H$ chemical shifts with RMSD$=0.12$ and correlation coefficient $r$ of $0.998$ and ${}^{13}C$ chemical shifts with RMSD$=2.26$ and $r=0.998$.

An attractive alternative to B3LYP is the OPBE exchange-correlation functional\cite{ opbe}.  
OPBE has been shown to produce ${}^{13}C'$ chemical shift with a mean absolute deviation of $2.3$ \textit{ppm} compared with experimental data\cite{opbe}.
In the same tests MP2 with had a mean absolute deviation $2.7$ \textit{ppm} and $3.9$ \textit{ppm} for B3LYP. For the ${}^{15}N$ the mean absolute deviation was
found to be $12.6$ \textit{ppm} compare with $10.0$ \textit{ppm} for MP2. In general the study conclude that OPBE is the superior functional 
compared with B3LYP for calculation of both magnetic shieldings and chemical shifts. OPBE also has the advantage of being less computational
intensive compared with B3LYP.  

Density functional theory methods with the Gauge including atomic orbitals(GIAO) approximation have shown to produce good results 
compared to more computational expensive \textit{ab inito} methods\cite{giaofirst}. Comparing GIAO with local gauge origin methods such as the individual gauge for localized orbitals(IGLO) and 
localized orbital/local origin(LORG) methods have shown that GIAO can achieve accurate results with smaller basis sets compared with IGLO/LORG. 
This negates the advantage in computational complexity of the local gauge origin methods. 

\subsection{Optimization}

The chemical shifts measured in an NMR experiment represent an average over conformations available to the protein during the time-scale of
the experiment. In order for the tripeptides to better represent the conformational average they were optimised with the semi-empirical method PM6\cite{pm6}.
In order to investigate the effect of the geometry optimization, tripeptides were cut from a PDB structure of ubiquitin\cite{1ubq} optimized with the AMBER forcefield\cite{amber}.
For each PDB tripeptide an NMR GIAO OPBE 6-31g(d,p) PCM calculation was done. AXA tripeptides were generated with Fragbuilder with the same central residue backbone and side chain
angles as the PDB tripeptides. The AXA tripeptides were optimised with forcefields, DFT and PM6. During the optimization all backbone and sidechain angles were kept fixed. 
The NMR, DFT, PM6 and UFF forcefield calculations were done with Gaussian09\cite{g09}. Optimization with MMFF94 was done with Fragbuilders "regularize" function.   
\tref{table:opttable} shows the results of linear regression analysis and RMSD calculation between the AXA Fragbuilder tripeptides and the PDB tripeptides.

\begin{table}[h]
\label{aggiungi}\centering 
{\scriptsize  
\begin{tabular}{lcccccc*{9}{l}}
\toprule 
 Optimization Method  &  ${}^{13}C\alpha$ & ${}^{13}C\beta$ & ${}^{13}C'$ & ${}^{15}N^{H}$ & ${}^{1}H^{N}$  & ${}^{1}H\alpha$&\\ 
                      &  $r$  RMSD        & $r$  RMSD       & $r$  RMSD   & $r$      RMSD  & $r$      RMSD  & $r$      RMSD  &\\\midrule 
 PM6                  & $0.913$ $3.84$ & $0.994$ $3.20$ & $0.550$ $2.31$ & $0.724$ $13.36$ & $0.600$ $1.42$ & $0.858$ $0.98$ &\\	
 UFF\cite{UFF}        & $0.884$ $1.96$ & $0.989$ $3.96$ & $0.320$ $6.42$ & $0.809$ $17.38$ & $0.254$ $1.36$ & $0.812$ $0.67$ &\\	
 MMFF94\cite{MMFF94}  & $0.907$ $1.74$ & $0.991$ $1.95$ & $0.398$ $4.32$ & $0.669$ $9.73$  & $0.516$ $0.95$ & $0.806$ $0.32$ &\\	 
 OPBE\cite{opbe}      & $0.904$ $1.77$ & $0.991$ $2.15$ & $0.464$ $2.18$ & $0.714$ $7.36$  & $0.603$ $1.96$ & $0.851$ $0.28$ &\\	
 OPBE PCM             & $0.914$ $1.63$ & $0.994$ $1.87$ & $0.470$ $1.83$ & $0.763$ $6.49$  & $0.601$ $1.03$ & $0.853$ $0.26$ &\\	
 B3LYP\cite{b3lyp}   & $0.907$ $1.71$ & $0.991$ $2.06$ & $0.450$ $2.17$ & $0.703$ $7.24$  & $0.589$ $1.40$ & $0.847$ $0.27$ &\\	
 B3LYP PCM            & $0.912$ $1.65$ & $0.994$ $1.77$ & $0.497$ $1.85$ & $0.744$ $6.56$  & $0.595$ $0.93$ & $0.850$ $0.25$ &\\
 PM6 vs OPBE*         & $0.980$ $3.95$ & $0.996$ $1.90$ & $0.927$ $1.14$ & $0.968$ $7.26$  & $0.991$ $0.73$ & $0.985$ $0.81$ &\\\bottomrule	
\end{tabular}}
\caption{{\bf Optimization Test.} The table shows the RMSD and correlation coefficient $r$ obtained from linear regression analysis between AXA tripeptides and 
	  tripeptides cut from an PDB structure of ubiquitin optimised with the AMBER. The NMR calculations were done at the GIAO OPBE 6-31g(d,p) PCM level of theory and
	  the AXA peptides were optimised with the methods in column $1$. In terms of correlation coefficient PM6 performs comparable with the DFT methods OPBE and B3LYP and
	  better than the two forcefields UFF and MMfF94. PM6 generally overestimates the chemical shifts and this leads to high RMSD values despite a correlation coefficient comparable
	  with the DFT methods. This highlights the importance of proper scaling the of the Procs14 chemical shifts. *shows the correlation between PM6 and the OPBE chemical shifts for the AXA
	  tripeptides. The high correlation coefficient and high RMSD implies that PM6 calculation which is scaled, will be in good agreement with the DFT methods.          	 	  
		 }
	\label{table:opttable}
\end{table}

In terms of correlation coefficient $r$ PM6 performed almost equivalent to the DFT methods and better than the forcefields. In spite of of the high correlation coefficient PM6 have a
high RMSD compared to the other methods. Inspections of the data reveals that PM6 optimization systematically overestimate the chemical shifts, especially \CA. Comparing PM6 and OPBE gives
an good average correlation coefficient of $r = 0.974$. This indicates that with prober scaling NMR calculation on tripeptides optimised with PM6, can provide chemical shifts 
comparable with tripeptides optimised with OPBE. The same relationship was found with B3LYP. This is advantageous because of PM6's much lower computational cost compared with the DFT methods.

\clearpage
\section{Procs14} \label{sec:Procs14}
\subsection{Procs14 Method Overview}

The Procs14 method is based on the assumption that the chemical shift of a backbone atom can be calculated using an additive model equation \eqref{eq:procs14}.
The chemical shift of an atom $\delta^{i}$, is a sum of the backbone $\delta^{i}_{BB}$ term, previous residue correction $\Delta\delta^{i-1}_{BB}$,
following residue correction $\Delta\delta^{i+1}_{BB}$, amide proton hydrogen bonding $\Delta\delta^{i}_{HB}$, H\(\alpha\) proton hydrogen bonding 
$\Delta\delta^{i}_{H\alpha B}$ and ring current $\Delta\delta^{i}_{RC}$.
The backbone terms are all dependent on the backbone $\phi$ and $\psi$ and $\chi_{1}$, $\chi_{2}$, $\chi_{3}$ and $\chi_{4}$ dihedral angles. The
hydrogen bond terms depend on the geometry of the acceptor donor interaction.

\begin{equation} \label{eq:procs14}
	\delta^{i} = \delta^{i}_{BB} + \Delta\delta^{i-1}_{BB} + \Delta\delta^{i+1}_{BB} + \Delta\delta^{i}_{HB} + \Delta\delta^{i}_{H\alpha B}  + \Delta\delta^{i}_{RC} 	
\end{equation}  

The backbone term $\delta^{i}_{BB}$ is parameterized on quantum mechanical calculations on AXA tripeptides, with X as any of the $20$ amino acids.
The $\phi$/$\psi$ angles of the N and C-terminus alanines are kept fixed at $-120\deg$ and $140\deg$ corresponding to a $\beta$-sheet conformation. 
Tripeptide samples were generated by scanning over the dihedral angles of the central residue. 
The N and C-terminus were capped with methyl caps to reduce terminal effects. Because of the caps, 
Procs14 is not able to provide predictions for the N and C-terminal residues. 

\begin{equation} \label{eq:pcorr}
      	\Delta\delta^{i-1}_{BB} = \delta^{ApA}(\phi^{i-1},\psi^{i-1},\chi_{n}^{i-1}) - \delta_{C}^{AAA}(\phi_{std},\psi_{std}) 
\end{equation} 
\begin{equation} \label{eq:fcorr}
      	\Delta\delta^{i+1}_{BB} = \delta^{AfA}(\phi^{i+1},\psi^{i+1},\chi_{n}^{i+1}) - \delta_{N}^{AAA}(\phi_{std},\psi_{std}) 
\end{equation} 

Interactions with the previous and following residues are modelled using data from our calculation on the tripeptides, equation \eqref{eq:pcorr}-\eqref{eq:fcorr}.
 $p$ is the previous residue type and $f$ is the following residue type. 
This model assumes that the effect from an amino acid on the previous and following residues, can be modeled by the 
tripeptides' central residue effect on the N and C-terminus alanines.  
The term $\delta^{ApA}(\phi^{i-1},\psi^{i-1},\chi_{n}^{i-1})$ is to be understood as the effect of previous residue
on its C-terminus alanine given by the dihedral angles of $p$. The term $\delta^{AfA}(\phi^{i+1},\psi^{i+1},\chi_{n}^{i+1})$ is the effect on the
N-terminus alanine. From the two terms is subtracted standard chemical shift values of alanine from the AAA tripeptide. The $\phi_{std}$/$\psi_{std}$
angles it the $-120\deg$ and $140\deg$ from the scans.

\begin{equation} \label{eq:hbond}
	\Delta\delta^{i}_{HB}=\Delta\delta_{1\deg HB}+\Delta\delta_{2\deg HB}
\end{equation}
\begin{equation} \label{eq:habond}
	\Delta\delta^{i}_{H\alpha B}=\Delta\delta_{1\deg H\alpha B}+\Delta\delta_{2\deg H\alpha B}
\end{equation}

The $ \Delta\delta^{i}_{HB}$ hydrogen bond term, consists of two individual bonding terms, equation \eqref{eq:hbond}. The primary $\Delta\delta_{1\deg HB}$ bonding term,
describes the effects of hydrogen bonding to the amide Proton \HN. The proton is a donor to either a backbone or sidechain acceptor oxygen.
The model was first used in \cite{Hbond}, but the scans have been expanded since then. Parameterization is done on a system of two N-methylacetamide molecules, which models the donor and acceptor 
atoms of the hydrogen bond. The model is constructed by scanning over the hydrogen bond length and dihedral/bond angles relevant to the hydrogen bond.
Since the hydrogen bond shows an exponential dependance on the bond length, Procs14 have a cut-off for hydrogen bonds of $4.0$ \AA.
The hydrogen bond's effect on the chemical shift, is computed by subtracting the chemical shifts from the scan from calculations on a reference formamide molecule. 
The result is stored in lookup tables for fast access.
Procs14 models also secondary hydrogen bonding $\Delta\delta_{2\deg HB}$. 
Secondary hydrogen bonding describes the effect on the acceptor oxygen residue. The secondary bonding is parameterized from the same QM calculations
as the primary bonding term.  

Procs14 also contains an \HA hydrogen bond correction $\Delta\delta^{i}_{H\alpha B}$. Both a primary and secondary correction is available and
they are constructed following the same method as the amide proton hydrogen bond. 

\begin{equation} \label{eq:rc}
	\Delta\delta_{RC}=iB\frac{1-3\cos^{2}(\theta)}{\vec{|r|}^{\,3}}
\end{equation}

The term $\Delta\delta^{i}_{RC}$ denotes the effect of ring current on the chemical shift. Usually this is only significant for proton shift and is thus only
calculated for the \HA and \HN protons. The ring current is calculated by a simple point-dipole model equation \eqref{eq:rc} described in \cite{ringcurrent}.
The model depends on two parameters $i$ and $B$ and the vector $r$, which is the vector from the proton to the center of the aromatic ring. $\theta$ is the angle between
$r$ and the vector normal to the aromatic ring system. The cut-off for calculating ring current is $8$ \AA\xspace in Procs14.

\subsection{Protein Backbone Tripeptide Scans}
\subsubsection{Backbone Scans}

The strategy for constructing the backbone terms are DFT calculations on tripeptide model systems. 
The nuclear shielding tensors were calculated at the NMR GIAO OPBE 6-31g(d,p) PCM level of theory. Calculations were done on the PM6 geometry optimized AXA tripeptides.
To make the tripeptide the FragBuilder Python module\cite{FragBuilder} was chosen. It allows for the generation of the large number of samples
required. The model tripeptides consisted of AXA triples where X is one of the $20$ amino acids. The N and C-terminus was
capped with methyl caps to eliminate terminal effects. Phaistos only models aspartic and glutamic acids in their deprotonated
state and lysine, arginine and histidine in their protonated state. The tripeptide with these amino acids were either protonated
or deprotonated accordingly. Cysteine is only modeled and not the disulfide bonded cystine. Therefore Procs14 only models cysteine residues.
For each tripeptide, a scan on the central residue's backbone and sidechain dihedral angles $\phi$, $\psi$, $\chi_{1}$, $\chi_{2}$, $\chi_{3}$, $\chi_{4}$ 
was carried out. The $\omega$ dihedral angle that runs over the peptide double-bond C-N bond rarely deviate from $180\deg$. 
This allowed the scans over the tripeptides to fix $\omega$ at this value. The $\phi$/$\psi$ backbone angles 
on the N and C-terminues alanine residues were fixed at $-140\deg$ and $120\deg$ corresponding to a typical $\beta$-sheets residue backbone angles. See \fref{fig:ADA}
for an example ADA tripeptide and see appendix {\bf \ref{tripep} \chapter{Tripeptides}} for schematic representations of the tripeptides and the dihedral angles modeled in 
Procs14.

\begin{figure}[h!]
	\centering
	  \begin{subfigure}[b]{0.8\textwidth}
                \centering
                \includegraphics[width=\textwidth]{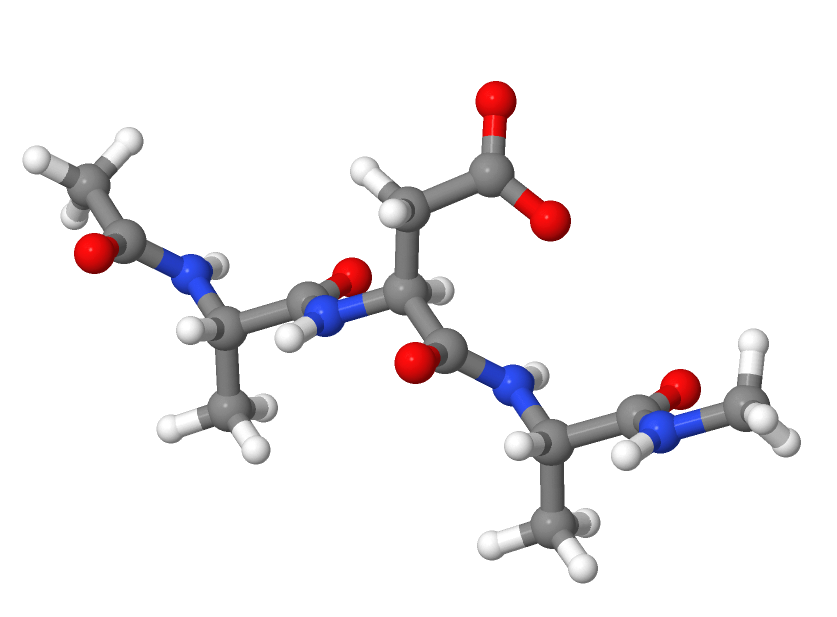}
                \caption{\small{ADA Tripeptide }}
                \label{fig:ADA}
           \end{subfigure}
	   \caption{ {\bf Tripeptide Example.} This figure shows an ADA tripeptide generated with the Fragbuilder Python module. The $\phi$, $\psi$, $\chi_{1}$ and $\chi_{2}$ angles are $-140\deg$, $120\deg$, $-160\deg$ and $140\deg$. The aspartic acid is deprotonated and the N and C-terminus alanine residues is capped with methyl groups.
				}
	\label{fig:ADA}
\end{figure}
 
\begin{figure}[h!]
	\centering
	  \begin{subfigure}[b]{0.455\textwidth}
                \centering
                \includegraphics[width=\textwidth]{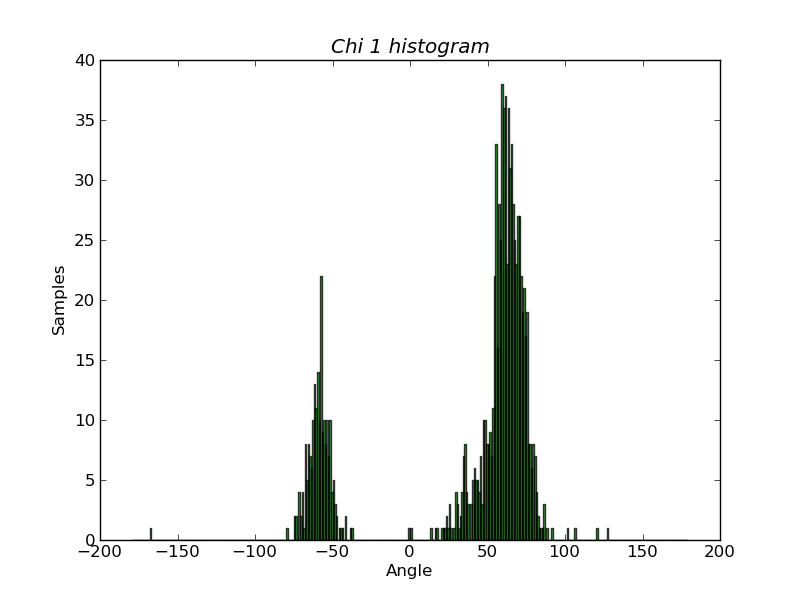}
                \caption{\small{Chi 1 samples}}
                \label{fig:1a}
     \end{subfigure}
	\begin{subfigure}[b]{0.455\textwidth}
                \centering
                \includegraphics[width=\textwidth]{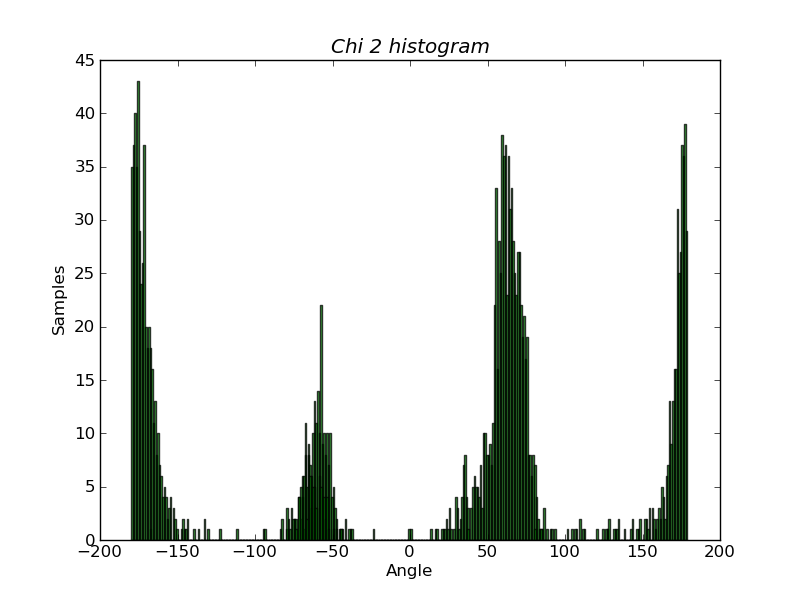}
                \caption{\small{Chi 2 samples}}
                \label{fig:1b}
     \end{subfigure}
	 \\
	\begin{subfigure}[b]{0.455\textwidth}
                \centering
                \includegraphics[width=\textwidth]{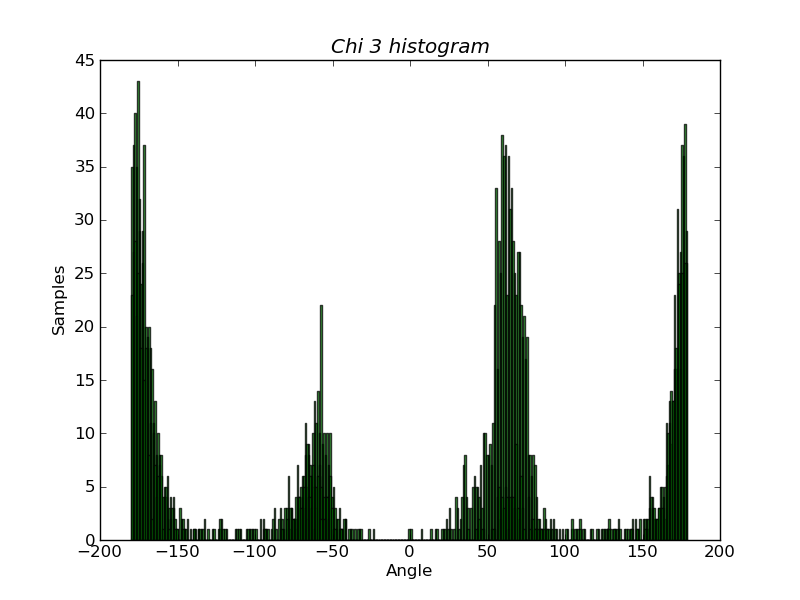}
                \caption{\small{Chi 3 samples}}
                \label{fig:1c}
     \end{subfigure}
	\begin{subfigure}[b]{0.455\textwidth}
                \centering
                \includegraphics[width=\textwidth]{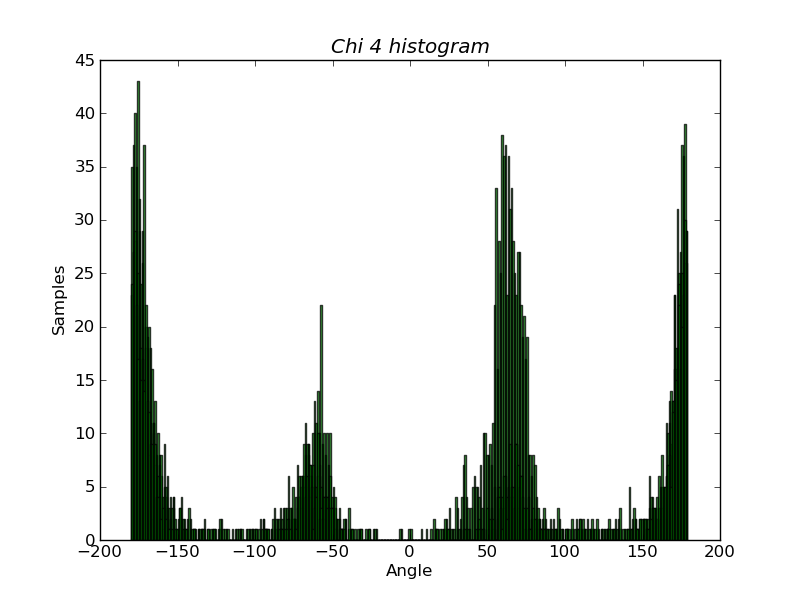}
                \caption{\small{Chi 4 samples}}
                \label{fig:1d}
     \end{subfigure}
	\caption{ {\bf BASILISK Torsion Angle Sampling.} The Figures {\bf (a)}, {\bf (b)}, {\bf (c)} and {\bf (c)} shows histograms of the $\chi_{1}$, $\chi_{2}$, $\chi_{3}$, $\chi_{4}$ side chain angles generated with BASILISK for 
	         a single backbone angle pair. The bin size of the histograms is $1\deg$ and contains angles from $1000$ samples.
				}
	\label{fig:basi}
\end{figure}

The scans on the central residues $\phi$/$\psi$ backbone angles was carried out in the range  $-180\deg$ to $180\deg$, with a step-size of $20\deg$.  
For amino acids with less than three side chain angles, scans were done in the same range and step-size as with the backbone angles.
Scanning in the same way on amino acids with three and four side chain angles would lead to an unpractically large amount of samples. For lysine/arginine this would result
in $19^{6}=\sim 47$ million samples. Amino acid side chain adopts favorable conformations that usually is modeled by rotamer libraries. This fact can be used
to, instead of scanning over the entire $-180$ to $180\deg$ range, the scan can be done with samples generated with
favourable side chain dihedral angles. Instead of using rotamer libraries that only possess discrete side chain conformations the BASILISK model\cite{BASILISK} was used.
It impliments an probabilistic model that allows sampling in continuous conformational space. For each backbone angle pair $\sim1000$ samples were generated with BASILISK.
\fref{fig:basi} shows the histograms of the distribution of lysine $\chi_{1}$, $\chi_{2}$, $\chi_{3}$, $\chi_{4}$ angles for a single backbone angle pair. 

\subsubsection{Interpolation and Data Files}

When scanning over the AXA tripeptides for Procs14, the backbone and side chain angles were kept fixed to ensure that our calculations represented a specific angle conformation. This resulted 
in some optimisations failing to converge. See appendix {\bf \ref{appendpro} \chapter{Procs14}}, \fref{fig:AGAopt} for a plot showing which tripeptides that converged in the case of glycine. The failed structures tended to be in unfavorable conformations with steric hindrance.
The nuclear shielding tensor values for the missing structures needed to be interpolated. For amino acids with $0$-$1$ side chain angles cubic interpolation was used and for $2$-$4$
side chain angles nearest neighbour interpolation. See Appendix {\bf \ref{appendpro} \chapter{Procs14}}, \fref{table:6dlysineinterpoaltion} for an comparison 
of the two interpolation schemes. 
For amino acids with $0$ side chain angles, the data was interpolated to a grid with $1\deg$ grid spacing,
$1$ side chain angles was to a grid of $5\deg$ and the rest of the amino acids $20\deg$. The interpolation was done with the Python package SciPy\cite{scipy}.
The grids were saved in the $.npy$ compressed file format from the Numpy Python package. In the compressed state on the hard disk the data size is $\sim 16$ GB 
and when loaded in to random access memory(RAM) $\sim 32$ GB. 

\subsubsection{Results From Backbone Scans}

\begin{figure}[h]
   \hspace{-15 mm}
  \begin{minipage}[b]{0.54\linewidth}
    \centering
    \includegraphics[width=\linewidth]{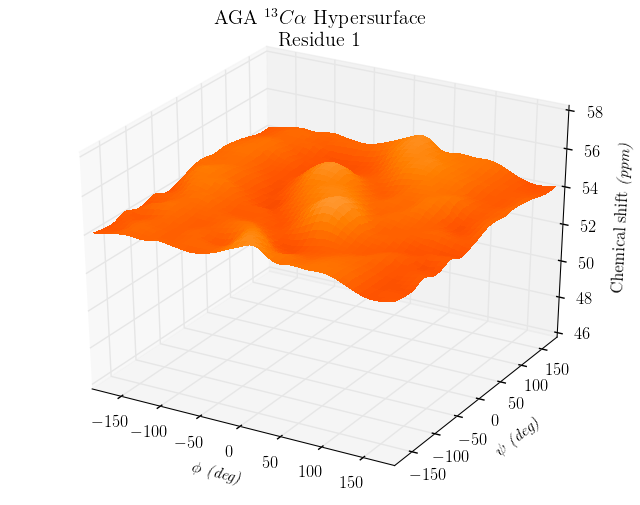}
  \end{minipage}
  \begin{minipage}[b]{0.54\linewidth}
    \centering
    \includegraphics[width=\linewidth]{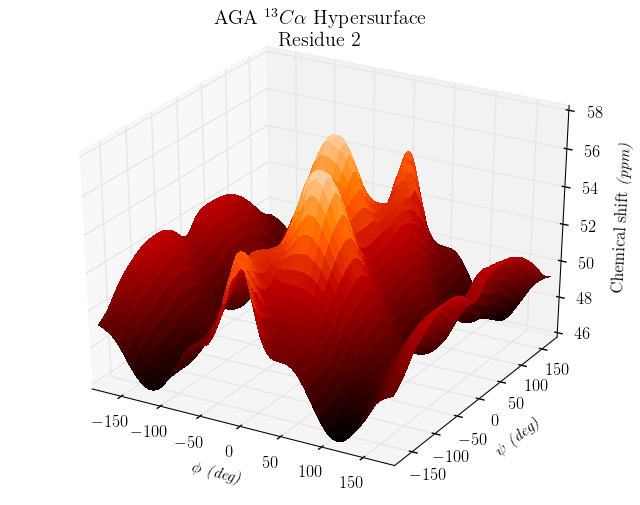}
  \end{minipage} \\
    \centering
  \begin{minipage}[b]{0.54\linewidth}
    \centering
    \includegraphics[width=\linewidth]{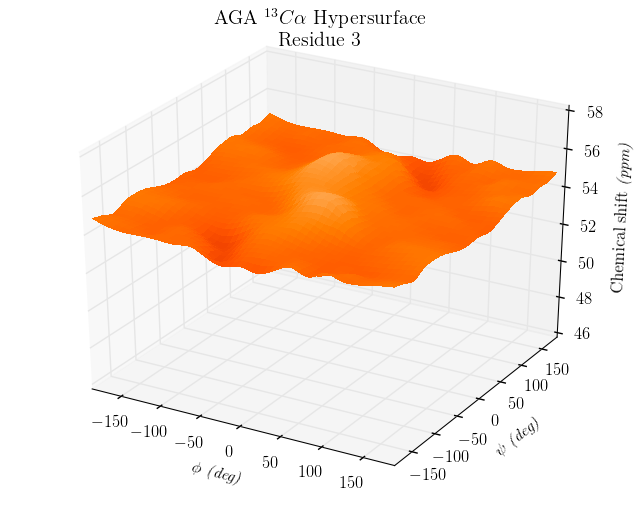}
  \end{minipage}

\caption{ {\bf \CA Glycine Hypersurfaces.} The figure $C\alpha$ shows chemical shift hypersurfaces for the AGA tripeptide. Residue 1 is the N-terminus alanine, residue 2 is the central
	   glycine residue and residue 2 is the C-terminus alanine residue. The x and y axis are the $\phi$ and $\psi$ angles on the central glycine residue.
	   The chemical shifts is calculated with TMS as a reference.
                 }
        \label{fig:AGAsurface}
\end{figure}

A hypersurface example of the \CA interpolated data can be seen in \fref{fig:AGAsurface} for the AGA tripeptides. It shows the effect of the backbone angles
on the \CA  chemical shifts and the effect of the central residue on the N and C-terminus alanines. The change on the two terminal alanines is used in the previous/following correction.
From the figure it can be deduced that the effect on previous/following residue from glycine is marginal. 
See appendix {\bf \ref{appendhyper} \chapter{Glycine Hypersurfaces}}, {\bf Figure \ref{fig:agahyper2}-\ref{fig:agahyper4}} for hypersurfaces for the other atom types.
See {\bf Table 2} for an overview of all the data used in the backbone term in Procs14.

\begin{table}[h]
\label{aggiungi}\centering
{\scriptsize
\begin{tabular}{llccccc*{9}{l}}
\toprule %
 Amino Acid       &  Spacing             & Data file size   & Samples  & Data points & Interpolation  &  Side chain angles  &\\\midrule
 Glycine          & $1\,^{\deg}$        & 3.0 MB           & 361      &  344        & Cubic          &  $0$                &\\
 Alanine          & $1\,^{\deg}$        & 3.0 MB           & 361      &  343        & Cubic          &  $0$                &\\
 Proline          & $1\,^{\deg}$        & 3.0 MB           & 361      &  246        & Cubic          &  $0$                &\\
 Serine           & $5\,^{\deg}$        & 9.0 MB           & 6859     &  6259       & Cubic          &  $1$                &\\
 Cysteine         & $5\,^{\deg}$        & 9.0 MB           & 6859     &  6326       & Cubic          &  $1$                &\\
 Valine           & $5\,^{\deg}$        & 9.0 MB           & 6859     &  5861       & Cubic          &  $1$                &\\
 Threonine        & $20\,^{\deg}$       & 3.0 MB           & 130321   &  114464     & Nearest        &  $2$                &\\
 Asparagine       & $20\,^{\deg}$       & 3.0 MB           & 130321   &  113566     & Nearest        &  $2$                &\\
 Aspartic Acid    & $20\,^{\deg}$       & 3.0 MB           & 130321   &  113790     & Nearest        &  $2$                &\\
 Histidine        & $20\,^{\deg}$       & 3.0 MB           & 130321   &  110787     & Nearest        &  $2$                &\\
 Isoleucine       & $20\,^{\deg}$       & 3.0 MB           & 130321   &  93722      & Nearest        &  $2$                &\\
 Leucine          & $20\,^{\deg}$       & 3.0 MB           & 130321   &  97803      & Nearest        &  $2$                &\\
 Phenylalanine    & $20\,^{\deg}$       & 3.0 MB           & 130321   &  107570     & Nearest        &  $2$                &\\
 Tryptophan       & $20\,^{\deg}$       & 3.0 MB           & 130321   &  101471     & Nearest        &  $2$                &\\
 Tyrosine         & $20\,^{\deg}$       & 3.0 MB           & 130321   &  111975     & Nearest        &  $2$                &\\
 Glutamine        & $20\,^{\deg}$       & 57.0 MB           & 143769   &  130134     & Nearest        &  $3$                &\\
 Glutamic Acid    & $20\,^{\deg}$       & 57.0 MB           & 144360   &  129638     & Nearest        &  $3$                &\\
 Methionine       & $20\,^{\deg}$       & 57.0 MB           & 144341   &  129019     & Nearest        &  $3$                &\\
 Arginine         & $20\,^{\deg}$       & 1.0GB             & 360909   &  327057     & Nearest        &  $4$                &\\
 Lysine           & $20\,^{\deg}$       & 1.0GB             & 360909   &  326607     & Nearest        &  $4$                &\\\bottomrule
\end{tabular}}
\caption{ {\bf Overview Table.} Column $0$ is the central residue type in the tripeptide. Column $1$ contains the grid spacing in the datafile. Column $2$ 
	   is the size of the data files for a single atom type after data compression. Column $3$ is the amount of initial generated samples. 
          Column $4$ is number of chemical shift data points after the geometry optimization and NMR calculations. Column $5$ is the interpolation method
           used to interpolate the missing data points. Column $6$ is the amino acid's number of side chain angles in Procs14.
                }
        \label{table:overview}
\end{table}

\clearpage
\subsection{Hydrogen Bond Scans}
\subsubsection{${}^{1}H^{N}$ Hydrogen Bond Scans}
\begin{figure}[h!]
	\centering
	  \begin{subfigure}[b]{0.455\textwidth}
                \centering
                \includegraphics[width=0.5\textwidth]{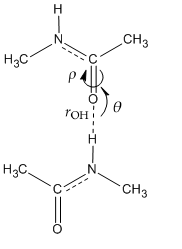}
                \caption{\small{N-methylacetamide}}
                \label{fig:hba}
     \end{subfigure}
	\begin{subfigure}[b]{0.455\textwidth}
                \centering
                \includegraphics[width=0.5\textwidth]{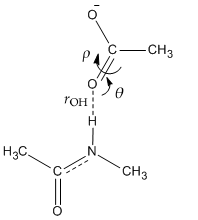}
                \caption{\small{acetate anion}}
                \label{fig:hbb}
     \end{subfigure}
	 \\
	\begin{subfigure}[b]{0.455\textwidth}
                \centering
                \includegraphics[width=0.5\textwidth]{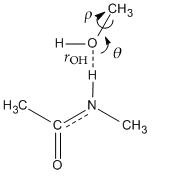}
                \caption{\small{methanol}}
                \label{fig:hbc}
     \end{subfigure}
	\caption{ {\bf ${}^{1}H^{N}$ Hydrogen Bond Model Systems.} The figure shows the three systems used to parametrize the $\Delta\delta_{1\deg HB}$ and $\Delta\delta_{2\deg HB}$ terms. The sytems consist
		of a N-methylacetamide molecule used to represent the protein backbone and an second molecule with an hydrogen bond acceptor group. {\bf (a)}
		  is the system with another N-methylacetamide for backbone-backbone hydrogen bonding. {\bf (b)} is the system for hydrogen bonding with residues with
		  carboxylate groups. {\bf (c)} is the system for residues with alcohol groups. The systems are scanned over the bond angle $\theta$, dihedral angle $\rho$ and
                  the hydrogen bond distance $r_{OH}$.     
				}
	\label{fig:hbondscans}
\end{figure}

Three systems with an N-methylacetamide and a hydrogen bond acceptor were the basis of the \HN hydrogen bond parameterization.
The hydrogen bonding system for the negatively charged aspartic and glutamic acids side chains were modeled with an acetate anion.
The system for bonding with the alcohol groups of threonine and serine was done with a methanol molecule. The system for bond
with another backbone segment was modeled with an additional N-methylacetamide. See \fref{fig:hbondscans} for representations of 
the three systems. 

The parameterization was done by scanning over the hydrogen bond length $r_{OH}$, the bond angle  $\theta$ and a dihedral angle
$\rho$. The bond lenth was scanned from $1.5$ \AA\xspace to $4.0$ \AA\xspace in $0.125$ \AA\xspace steps. The H..O=C, H..O=C and H..O-C bond angles were scanned from $180$ to $90\deg$ in $10\deg$ steps. The H..O=C-N, 
H..O=C-C and H..O-C(..)H\textsuperscript{O} dihedral angle was scanned in the entire $-180$ to $180\deg$ interval with a step size of $15\deg$. The NMR calculations were done at the
GIAO OPBE/6-311++g(2d,p) level of theory. To get the $\Delta\delta$ change in chemical shift an NMR calculation was done on a N-methylacetamide system with no hydrogen bonding. The results from
the scan were subtracted from this reference system.

A special case that needs to be treated in Procs14 is solvent exposed amide protons. 
In a protein almost all of the amide protons inside the protein is part of a hydrogen bond network. This allows Procs14
to treat amide protons with no hydrogen bonds as surface protons and therefore bound to a water molecule. In the old Procs
model the hydrogen bond with a water molecule was found to give a correction on the \HN chemical shift of $2.07$ \textit{ppm}\cite{Hbond}.
The same model is reused in Procs14.
\subsubsection{${}^{1}H\alpha$ Hydrogen Bond Scans}

\begin{figure}[h!]
	\centering
	  \begin{subfigure}[b]{0.49\textwidth}
                \centering
                \includegraphics[width=0.5\textwidth]{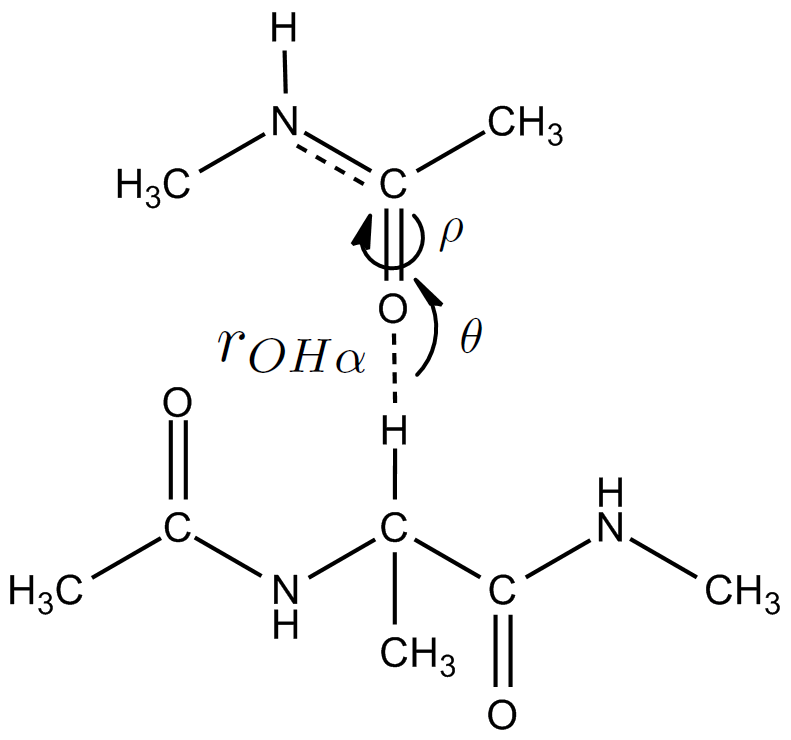}
                \caption{\small{N-methylacetamide}}
                \label{fig:haba}
     \end{subfigure}
	\begin{subfigure}[b]{0.49\textwidth}
                \centering
                \includegraphics[width=0.5\textwidth]{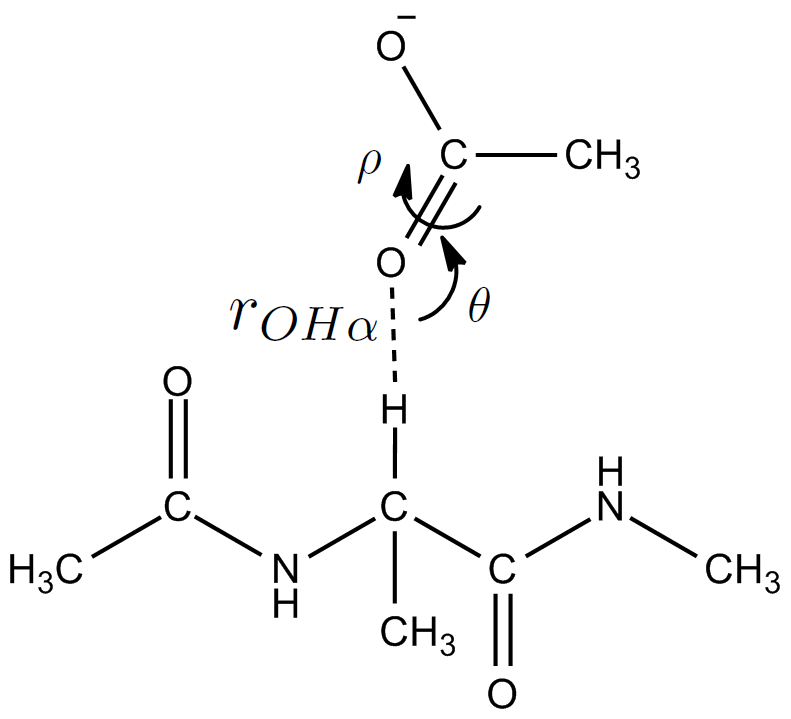}
                \caption{\small{acetate anion}}
                \label{fig:habb}
     \end{subfigure}
	 \\
	\begin{subfigure}[b]{0.49\textwidth}
                \centering
                \includegraphics[width=0.5\textwidth]{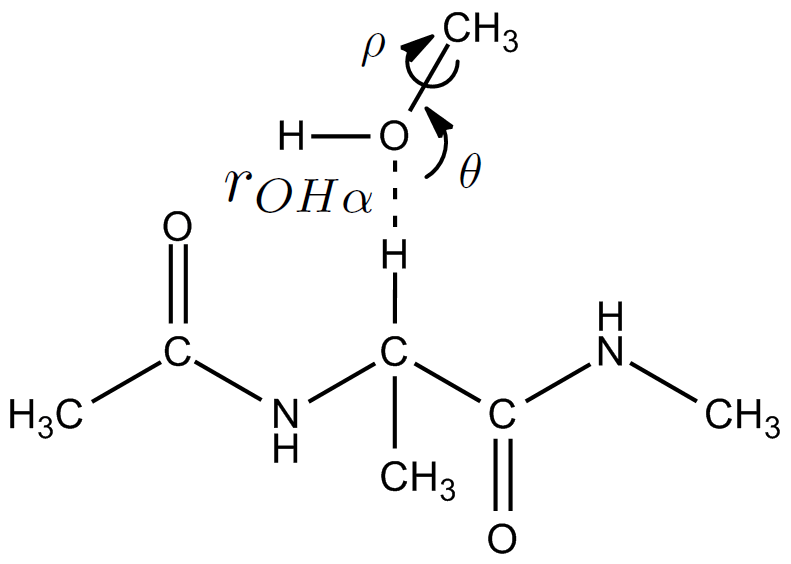}
                \caption{\small{methanol}}
                \label{fig:hbc}
     \end{subfigure}
	\caption{ {\bf ${}^{1}H\alpha$ Hydrogen Bond Model System.} The model system used to calculate the contribution to the chemical shift from hydrogen bonding 
		with ${}^{1}H\alpha$ as the donor. {\bf (a)} The scan is over the hydrogen bond distance $r_{OH\alpha}$, the H\(\alpha\)..O=C' bond angle and the dihedral angle 
		defined by H\(\alpha\)..O=C'N. {\bf (b)} Acetate anion, the scan is over $r_{OH\alpha}$, H\(\alpha\)..O=C bond angle and H\(\alpha\)..O=C-C torsion angle. {\bf (c)}
		Methanol, the scan is over $r_{OH\alpha}$ and H\(\alpha\)..O-C bond angle and H\(\alpha\)..O=C(..)H\textsuperscript{O}.
}
	\label{fig:ALAscans}
\end{figure}

In addition to hydrogen bonding between amide protons and acceptor groups, Procs14 contains a term for calculating interaction between
the C\(\alpha\)-H\(\alpha\) donor sytems and an acceptor oxygen.
Like the amide proton hydrogen bond scans the parameterization is done on model systems. The residue with the donor hydrogen is treated as
an alanine, see \fref{fig:ALAscans}. The acceptor molecule is an N-methylacetamide that models the 
bonding with an backbone oxygen. The $\phi$ and $\psi$ backbone angles of the alanine is both set to $45\deg$ corresponding to a left handed 
alpha helix conformation. This was done to reduce steric effects and minimize unwanted interactions between the acceptor oxygen and the two amide protons.
The scan was over the H\(\alpha\)..OC' bond angle from $90\deg$ to $180\deg$ in steps of $10\deg$. The H\(\alpha\)..O=C'N. dihedral was scanned in
steps of $15\deg$ over the entire range. The hydrogen bond distance $r_{OH\alpha}$ was scanned from $1.8$ to $4.0$ \AA\xspace in steps of $0.2$ \AA.
The C\(\alpha\)-H\(\alpha\)..O bond angle was fixed at $130\deg$, since an investigation of protein structures found this to be a common value.  
The same principles were used to scan over model systems with an acetate anion and methanol to model alpha hydrogen bonding with carboxylate 
and alcohol oxygens.

\clearpage
\subsubsection{Results From Hydrogen Bond Scans}

\begin{figure}[h!]
	\centering
        \includegraphics[width=0.5\textwidth]{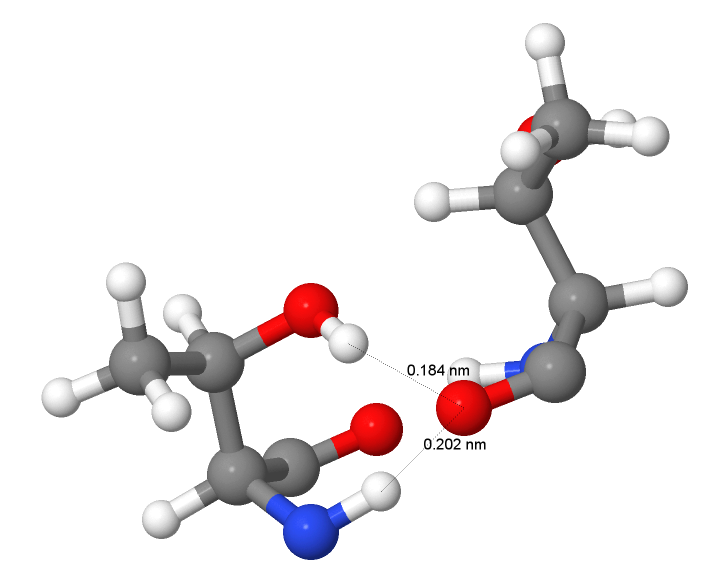}
	\caption{{\bf Multiple Hydrogen Bonding}. 2OED residue 53(left) threonine and 44(right) threonine. The acceptor oxygen bonds with both the alcohol proton and amide proton.}
	\label{fig:THRout}
\end{figure}

When investigating the ${}^{15}N^{H}$ predictions a significant number of outliers was found. Analysis indicated that the amino acids with 
hydroxyl groups in their side chain serine and threonine, systematically had its chemical shift overestimated. The protein structures
showed that the hydrogon bond acceptor formed an hydrogen bond both the amide proton and the hydroxyl group. An example is showed
in \fref{fig:THRout}. It shows residue 53 threonine and 44 threonine from an PM6 optimised protein g sturcture.

Addtional outliers were found, when a single amide proton bonded to both a carboxylate and carbonyl oxygen and an amide proton bonding with both an alcohol and carbonyl oxygen.
The examples highlights the importance of hydrogen bonding networks in proteins. Although inherently difficult to model, small test systems with the three described cases were made.  
The systems consisted of the same N-methylacetamide, acetate anion and methanol molecules from the hydrogen bond scans.  
By comparing the system with the multiple hydrogen bonding to the system without, small corrections to the Procs14 chemical shifts can be constructed. See appendix {\bf \ref{appendpro} \chapter{Procs14}}, \tref{table:hbondcorr} for the results.

\clearpage
\subsection{Hydrogen Bond Length Correction}

The average bond length of the H\(\alpha\)-C\(\alpha\) and H-N bond in the PM6 optimized tripeptides differs a lot from the same bond lengths in forcefield optimized proteins.
For example, the average H\(\alpha\)-C\(\alpha\) bond length in the PM6 tripeptides is $1.131$ \AA\xspace while a protein optimized with the CHARMM forcefield have bond lengths of $1.082$ \AA\xspace.
To investigate the effect of bond length on the chemical shift two H\(\alpha\)-C\(\alpha\) and H-N bond length scans were performed on a Ala-Ala-Ala tripeptide. The result is seen
in \fref{fig:hbondcorrshow}. Using linear regression analysis it is clear that the dependence of the bond length on the chemical shift adequately modelled  by a linear relationship.
 
\begin{equation} \label{eq:hacorr}
      	\Delta\delta^{b}_{{}^{1}H\alpha} = (<b>-<Procs14>)\cdot20.24
\end{equation} 
\begin{equation} \label{eq:hbcorr}
      	\Delta\delta^{b}_{{}^{1}H^{N}} = (<b>-<Procs14>)\cdot22.67 
\end{equation} 
\begin{equation} \label{eq:cacorr}
      	\Delta\delta^{b}_{{}^{13}C\alpha{}} = (<b>-<Procs14>)\cdot55.56 
\end{equation} 

Using the slopes found from regression analysis a correction from the bond length can be made, see equation \eqref{eq:hacorr}-\eqref{eq:cacorr}.
The effect of the H\(\alpha\)-C\(\alpha\) on the \CA chemical shifts is also modeled in equation \eqref{eq:cacorr}.
$<b>$ is the average bond length of the structure for which the prediction is to be performed and $<Procs14>$ is the average H\(\alpha\)-C\(\alpha\) or H-N bond length
from the PM6 optimized tripeptides. For the bondlengths from the CHARMM forcefield the $\Delta\delta^{b}_{{}^{1}H\alpha}$ correction will be $-1.001$ \textit{ppm}.
One should be careful with using the corrections on structures with hydrogen bond lengths that differs a lot from the optimal. For example the crystal structure of cutinase(1CEX)
 has an H-N bond length of $0.85$ \AA\xspace. The H-N bond is typically close to $1$\AA\xspace. For this reason the two $\Delta\delta^{b}_{{}^{1}H\alpha}$ $\Delta\delta^{b}_{{}^{1}H^{N}}$ corrections 
can be turned off if necessary.

\begin{figure}[h!]
	\centering
	  \begin{subfigure}[b]{0.60\textwidth}
                \centering
                \includegraphics[width=1.0\textwidth]{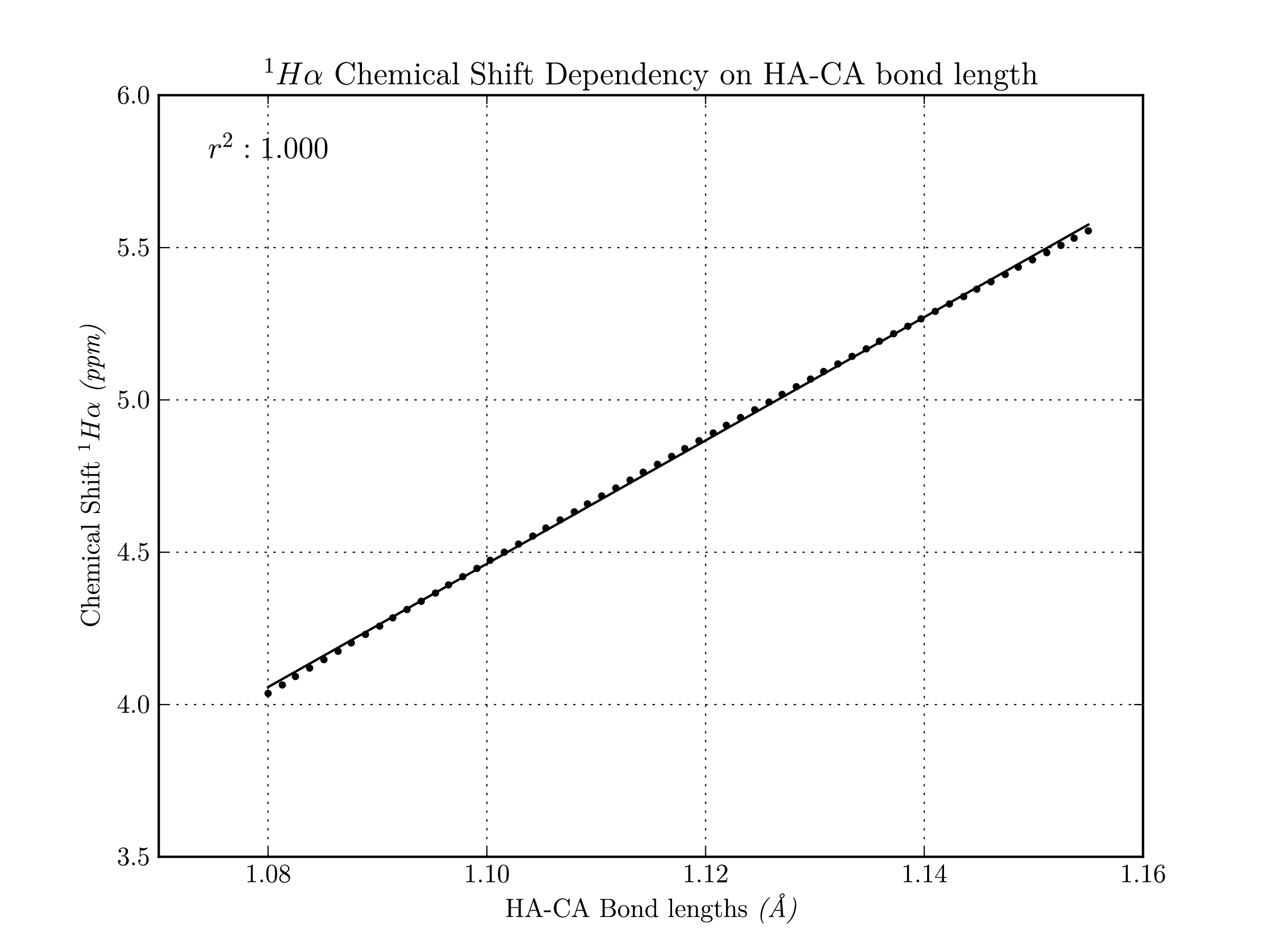} 
                \caption{\small{${}^{1}H\alpha$ Bond Length Scan }}
                \label{fig:hbcorrha}
     \end{subfigure} \\
	\begin{subfigure}[b]{0.60\textwidth}
                \centering
                \includegraphics[width=1.0\textwidth]{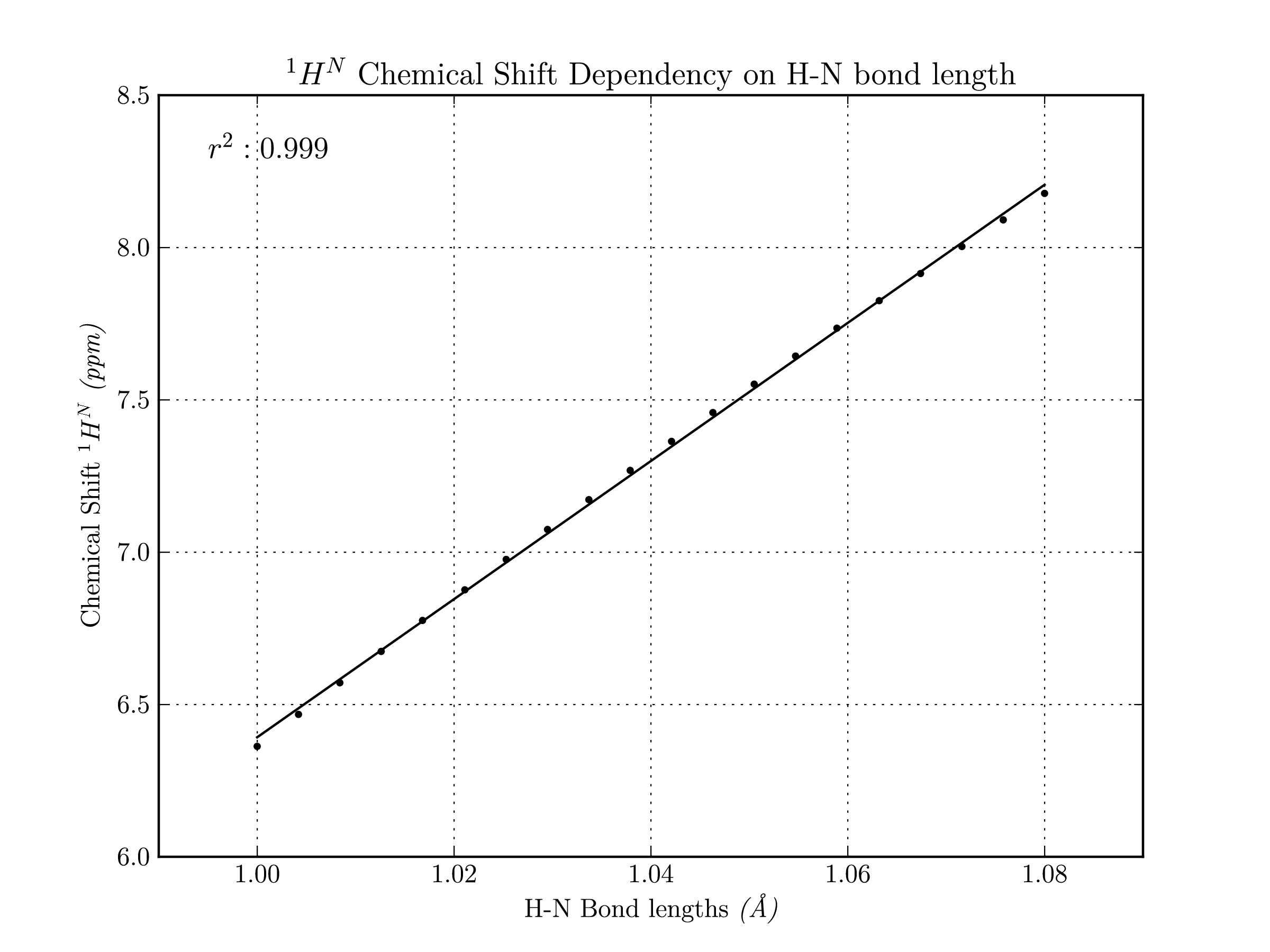}
                \caption{\small{${}^{1}H^{N}$ Bond Length Scan}}
                \label{fig:hbcorrhn}
     \end{subfigure} 

	\caption{ {\bf The Effect of Bond length on ${}^{1}H^{N}$ and ${}^{1}H\alpha$ Chemical Shift.} {\bf (a)} shows the chemical shift of ${}^{1}H\alpha$ from 
                   an scan over the H\(\alpha\)-C\(\alpha\) bond length. The scan is performed on an Ala-Ala-Ala tripeptide. The data point is fitted with an
		           linear regression line with an slope of $22.67$. {\bf (b)} shows the chemical shift of ${}^{1}H^{N}$ from 
                   a scan over the H-N bond length. The linear regression line has an slope of $20.24$.
         }\label{fig:hbondcorrshow}
\end{figure}

\clearpage
\subsection{Scaling}

The NMR calculations give the theoretical nuclear shieldings and have to be converted to actual chemical shifts.
The first inclination is to use the same basis set and method to calculate the chemical shift on a reference compound.
Basis set dependant errors may persist in spite of this approach. Instead of using a reference compound a method using 
linear regression analysis on theoretical shieldings and experimental values\cite{scaling} is used. The slope $a$ and intercept $b$
from the regression analysis is used with equation \eqref{eq:scaling} to make the final chemical shift prediction. 

\begin{equation} \label{eq:scaling}
	\delta^{prediction} = \frac{(b-\sigma^{calculated})}{a} 
\end{equation}

For Procs14 we have the additional problem of NMR calculations on PM6 optimized tripeptides systematically 
over or under estimating the chemical shift, dependent on the atom type. By scaling Procs14 predictions on crystal
structures with their experimental value this systemtic error can be corrected. The scaling procedure is performed on
crystal structures of cutinase(1CEX), CheY(1CHN)\cite{1chn}, RNase H(2RN2)\cite{2rn2} and experimental chemical shifts cutinase\cite{1cexexp}, CheY\cite{chyexp} and RNase H\cite{rnaseexp}, this yielded the average scaling factors 
seen in \tref{table:scaling}.

\begin{table}[h]
\label{aggiungi}\centering 
{\scriptsize  
\begin{tabular}{lcccccc*{9}{l}}
\toprule %
 Atom type            &  ${}^{13}C\alpha$ & ${}^{13}C\beta$   &  ${}^{13}C'$     & ${}^{15}N^{H}$ & ${}^{1}H^{N}$  & ${}^{1}H\alpha$  &\\\midrule
 Slope $a$            & $0.92$             & $1.04$           &  $0.68$          & $1.12$         & $1.04$         & $1.05$          &\\	
 Intercept $b$        & $188.17$           & $196.35$         &  $155.74$        & $264.36$       & $32.03$        & $30.89$          &\\\bottomrule
\end{tabular}}
\caption{ {\bf Scaling Factors.} The table shows the scaling factors used in Procs14 to calculate the chemical shifts from nuclear shieldings. The scaling factor values
	  are an average from linear regression analysis on three proteins 1CEX, 1CHN and 2RN2.    
		 }
	\label{table:scaling}
\end{table}

\clearpage
\subsection{Phaistos implementation}

The Procs14 method is implemented as an energy term in Phaistos and as a part of the Chemshift module. During each monte carlo iteration a Procs14 chemical shift prediction is done.  
These shifts are compared to a user provide input file containing experimental chemical shifts. For each atom type an energy is computed. The energy is used to evaluate 
each monte carlo move. Phaistos and Procs14 are both coded in C++.
The Procs14 datafiles are loaded as simple lookup tables. Each data point can be accessed by converting dihedral angles to indices matching the grid of the lookup table.
For residues with $1$ or more sidechain angles, the final chemical shifts predictions are found by linear interpolation. 

\begin{equation} \label{eq:linear}
	\delta = \delta_{0}+(\delta_{1}-\delta_{0})\frac{d - d_{0}}{d_{1}-d_{0}}
\end{equation}

Equation \eqref{eq:linear} shows the case of $1$-dimensional linear interpolation\cite{linear}. The chemical shift $\delta$ at the dihedral angle $d$ is interpolated
using the grid points $(d_{0},\delta_{0})$ and $(d_{1},\delta_{1})$. Where the grid points are the closest dihedrals on the grid, i.e. $d_{0}$ is closest dihedral
that is smaller than $d$ and vice versa for $d_{1}$.
The multidimensional interpolation case can be treated as a succession of $1$D interpolations on each variable.
In addition to the backbone terms the hydrogen bonding and ring current terms are also computed.  The interatomic distance is computed for each hydrogen bond acceptor and donator atom type. If 
the distance is less than a cutoff of $4.0$ \AA\xspace the hydrogen bond terms are calculated. The same approach is used for the ring current term $\Delta\delta^{i}_{RC}$. Where the cutoff distance between the aromatic ring system and the hydrogen is $8$ \AA.

\begin{table}[h!]
\label{aggiungi}\centering
{\scriptsize
\begin{tabular}{llcccc*{9}{l}}
\toprule %
                         & 1UBQ               & 1CEX       &\\
                         & $<ms>$             & $<ms>$     &\\\midrule
 Procs14                 & 3.96               & 15.38      &\\
 Camshift                & 19.0               & 113.07     &\\
 Procs14 Cached          & 0.59               & 1.32       &\\
 Camshift Cached         & 3.54               & 13.48      &\\\bottomrule
\end{tabular}}

\caption{ {\bf Phaistos Speed Test.} This table shows a comparison between the Procs14 and Camshift Phaistos implementation. Column $2$ and $3$ contains
	  the average amount of milliseconds to complete $1000$ iterations. The test is done on the proteins ubiquitin(1UBQ) or cutinase(1CEX)\cite{1cex}.
	  The cached version is significantly faster $\sim11.6$X for Procs14 on cutinase. The cached Procs14 version is also $\sim10.2$X times faster than Camshift Cached on 
          cutinase. Both methods were noticeably slower on the 197 residue cutinase compared to the 76 residue ubiquitin. The exact speed of both methods
	  will depend on the amino acid composition and the types of monte carlo moves used in Phaistos.
                 }
        \label{table:phaistosspeed}
\end{table}

During the monte carlo simulation the protein is sampled using monte carlo moves.
For each iteration only a segment of the protein is changed and it is therefore advantageous to only re-calculate the chemical shift for the modified region. This concept is called caching and
provides a significant, speedup see \tref{table:phaistosspeed}. For the backbone term the only modified region $\mp1$ needs to be re-calculated.
For hydrogen bonding each donor/acceptor in the modified region gets it's interaction recalculated with donor/acceptors inside and outside the modified region. 
The same method is used for the ring current term, each aromatic ring gets it interaction recalculated with ${}^{1}H^{N}$ and ${}^{1}H\alpha$ hydrogens inside and outside the modified 
region. The speedup gained by using the cached version generally increase with protein size. For cutinase it is $\sim11.6$X compared to $\sim6.7$X in ubiquitin.
The concept of storing the backbone term chemical shifts in lookup tables in RAM is vindicated by Procs14 cached $\sim10.2$X speedup compared to Camshift cached.
The energy contribution from Procs14 is calculated with \eqref{eq:energyphaistos}. 

The energy contribution from Procs14 is calculated with the relation seen in \eqref{eq:energyphaistos}. $N_{j}$ is the number of chemical shifts for atom type $j$ and
$\chi$ is the difference between the experimental and predicted chemical shift. This is derived from the expression for the hybrid energy equation \eqref{eq:hybrid7}.
 
\begin{equation} \label{eq:energyphaistos}
	E_{Procs14} = k_{B}T\displaystyle\sum_{j}\left(\frac{N_{j}}{2}ln(\chi_{j}^{2}(X)) \right)
\end{equation}

\clearpage
\section{Benchmarking} \label{sec:Benchmarking}
\subsection{NMR Calculations on Full Proteins}
In the following sections Procs14 is benchmarked on QM NMR calculations on full protein models.
The proteins are a crystal structure of ubiquitin(1UBQ) with 76 residues and two models of
Protein G, 2OED\cite{2oed} which is a crystal structure refined with dipolar couplings and
1IGD\cite{1igd} the crystal structure of protein G. The protonation states of
LYS, ARG, HIS, GLU and ASP is if necessary changed to the same as in Procs14. The QM NMR calculations
is done on both the unmodified crystal structures and optimized versions.
PM6 was used in order to get an optimised structure with approximately the same bond angles and bond lengths as
the tripeptides. The PM6-D3H+\cite{dh3plus} with a PCM model was used and PM6-DH\cite{pm6dh} with the COSMO solvation model.
Pure PM6 was not able to converge most likely because it lacks the dispersion and hydrogen bonding corrections of PM6-D3H+ and PM6-DH.
In addition to PM6 the proteins was also optimised with a selection of forcefields, AMBER\cite{amber}, CHARMM22/CMAP\cite{CHARMM2009} and
AMOEBAPRO13\cite{AMOEBA}. The NMR calculations were done at the GIAO OPBE 6-31g(d,p) PCM level of theory in Gaussian 09.

\subsection{Benchmarking the Hydrogen Bond Terms}

\begin{figure}[h!]
	\centering
	  \begin{subfigure}[b]{0.49\textwidth}
                \centering
                \includegraphics[width=1.0\textwidth]{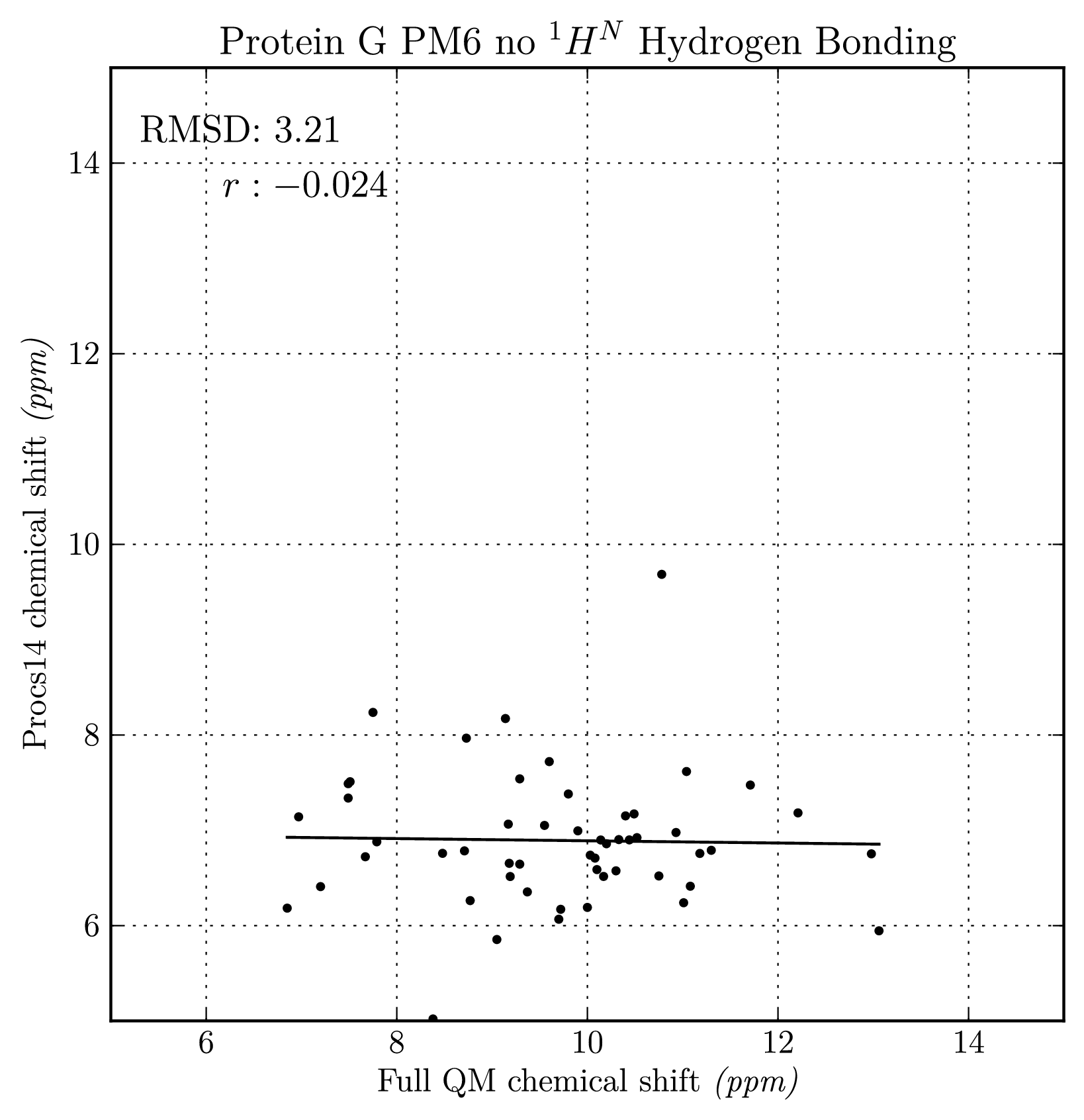}
                \caption{\small{${}^{1}H^{N}$ Without hydrogen bonding}}
                \label{fig:hnnohb}
     \end{subfigure}
	\begin{subfigure}[b]{0.49\textwidth}
                \centering
                \includegraphics[width=1.0\textwidth]{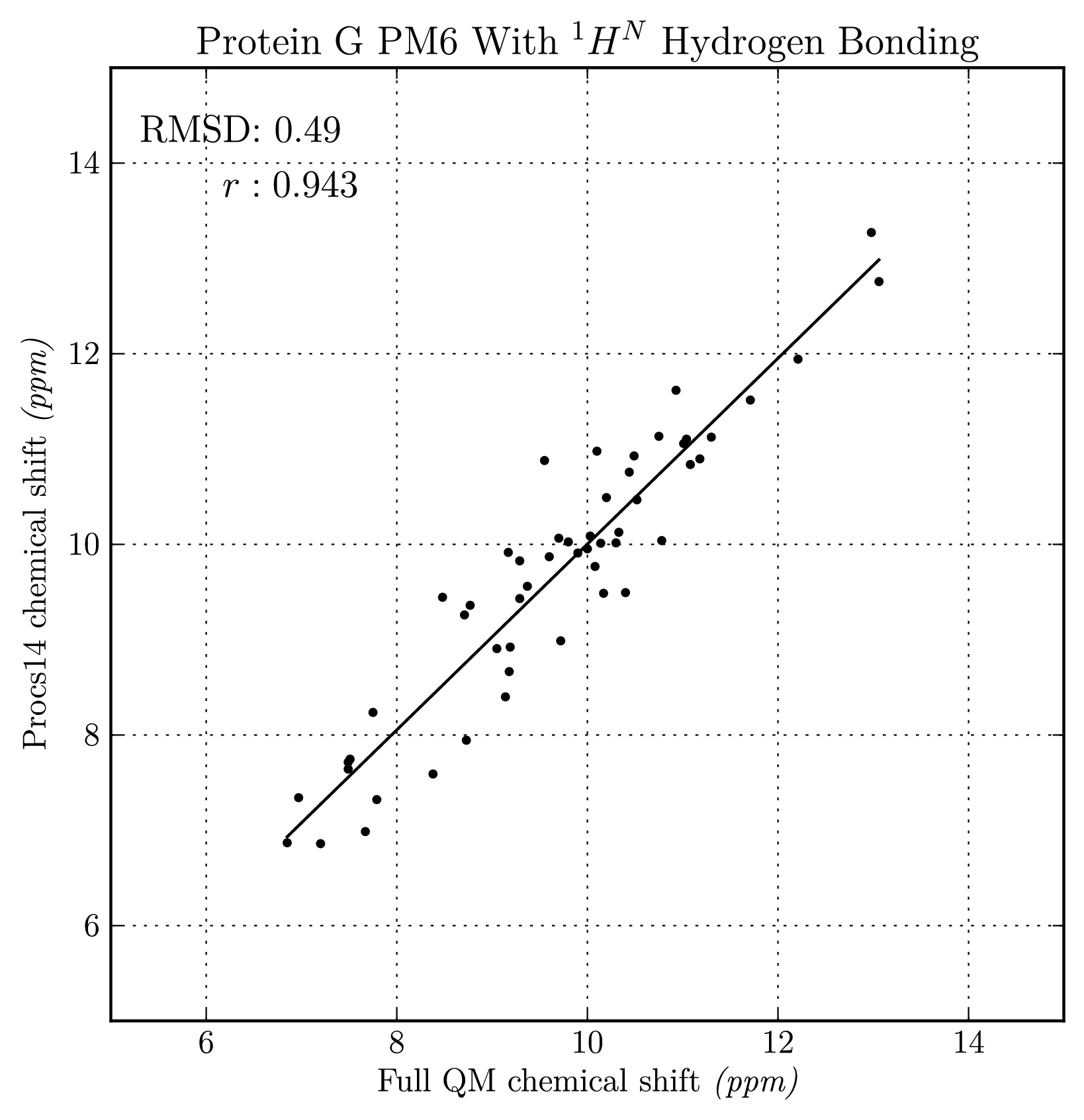}
                \caption{\small{${}^{1}H^{N}$ With hydrogen bonding}}
                \label{fig:hnwithhb}
     \end{subfigure} \\
     \begin{subfigure}[b]{0.49\textwidth}
                \centering
                \includegraphics[width=1.0\textwidth]{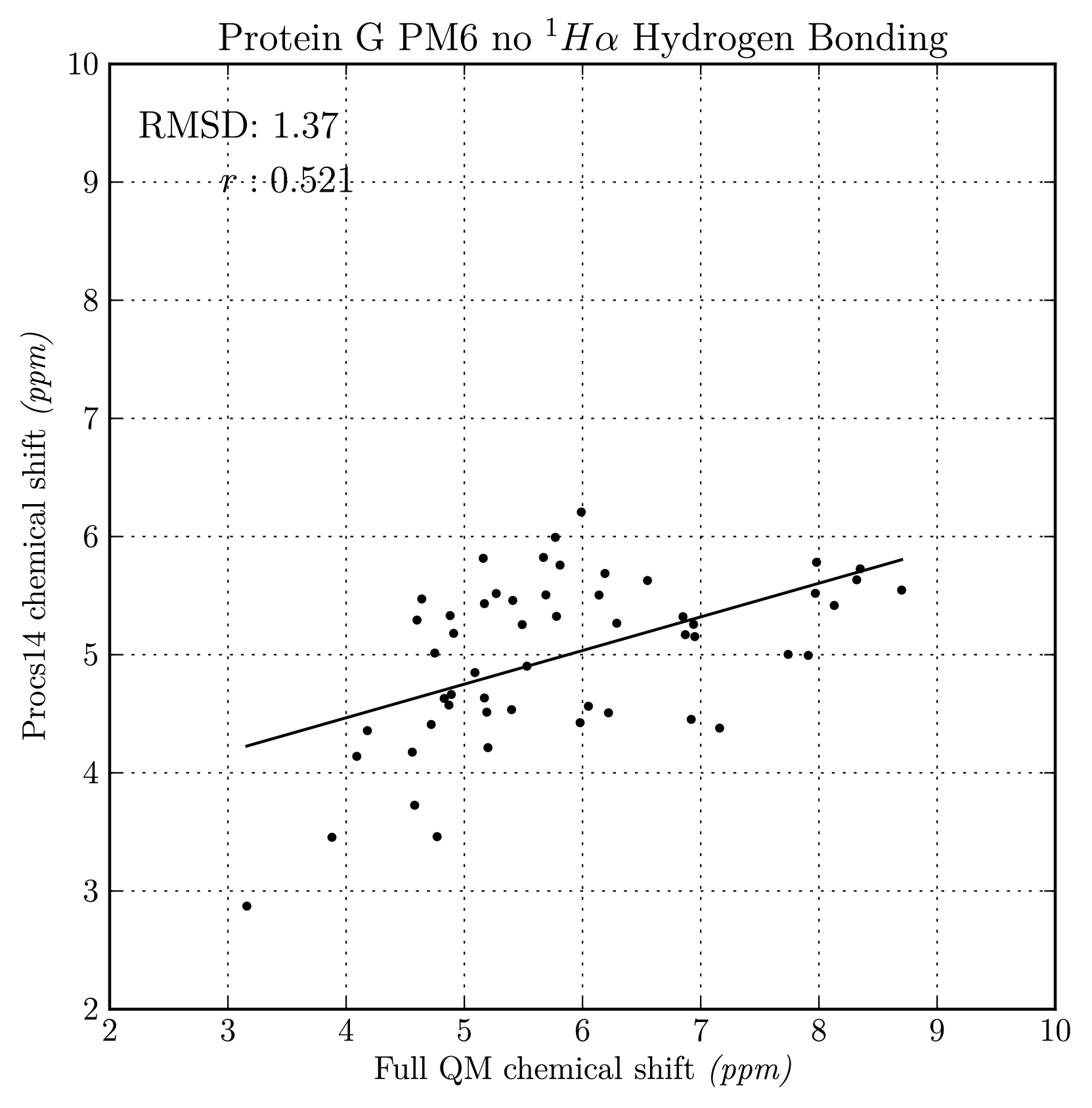}
                \caption{\small{${}^{1}H\alpha$ Without hydrogen bonding}}
                \label{fig:nohab}
     \end{subfigure}
	\begin{subfigure}[b]{0.49\textwidth}
                \centering
                \includegraphics[width=1.0\textwidth]{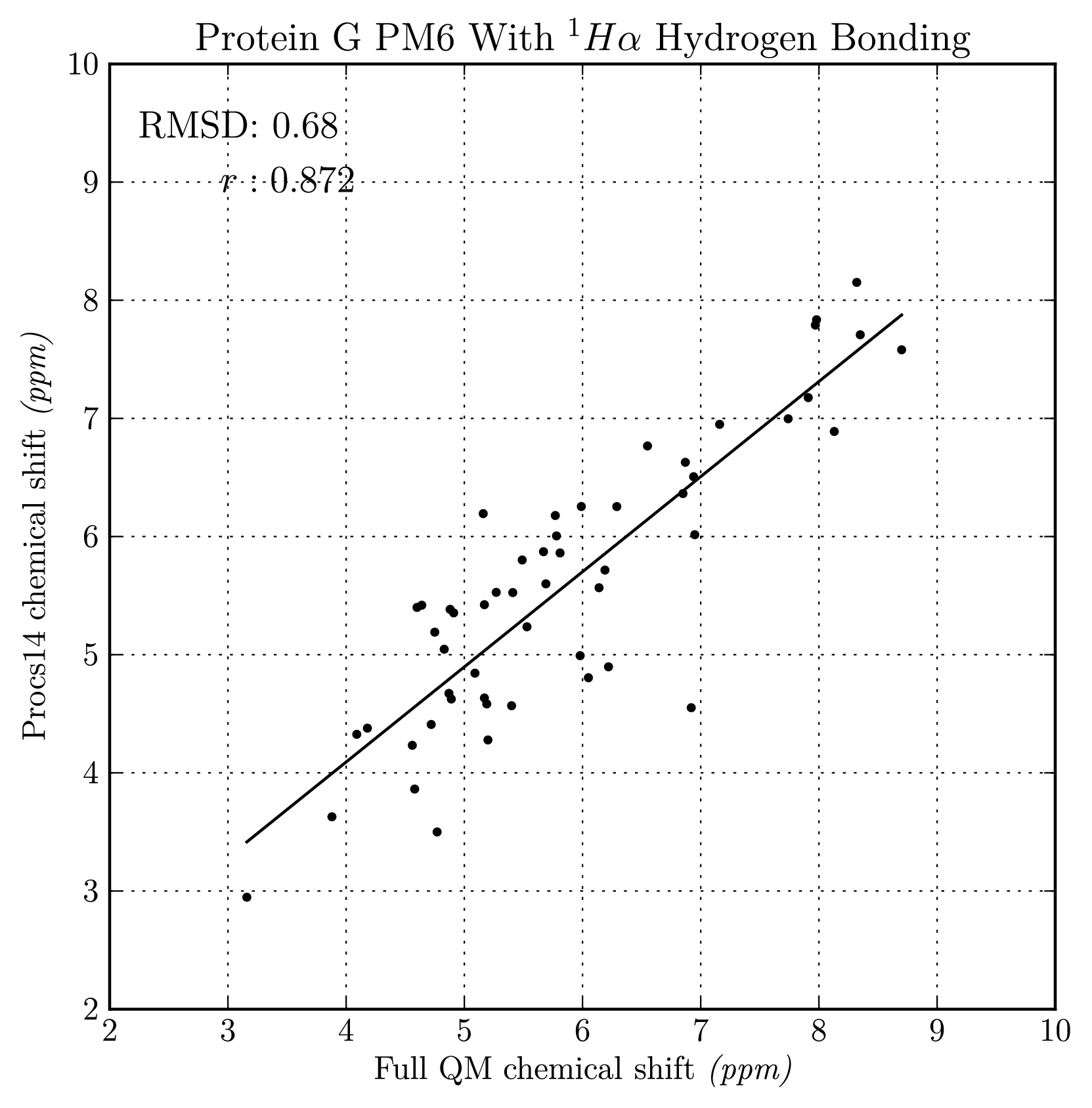}
                \caption{\small{${}^{1}H\alpha$ With hydrogen bonding}}
                \label{fig:withab}
     \end{subfigure}

	\caption{ {\bf The Effect of ${}^{1}H^{N}$ and ${}^{1}H\alpha$ Hydrogen Bonding.} The figures shows Procs14 predictions with and without hydrogen bonding terms and 
		chemical shifts calculated on a full protein. The protein is protein G optimised with PM6-DH. {\bf (a)} and {\bf (b)} shows the effect of the ${}^{1}H^{N}$
		hydrogen bonding term. {\bf (c)} and {\bf (d)} shows the effect of ${}^{1}H\alpha$ hydrogen bonding. Both bonding terms add a significant improvement to the 
correlation factor $r$ and the RMSD. Although both still contains a number of outliers.}

	\label{fig:hbondtest}
\end{figure}

To test the effect of the hydrogen bonding terms in Procs14, chemical shift predictions are performed on PM6-DH and PM6-DH3+ optimized structures.
Since the proteins are optimized with PM6 like the tripeptides the scaling procedure and hydrogen bond length correction is not used.
Instead the chemical shifts are found with \eqref{eq:Bshielding} and NMR calculations at the same level of theory on reference compounds
TMS and ammonia. For the ${}^{1}H^{N}$ chemical shift the $\Delta\delta^{i}_{HB}$ term significantly improve the linear correlation factor $r$ and RMSD compared
to a control prediction \fref{fig:hbondtest} {\bf (a)}-{\bf (b)}. The prediction without the hydrogen bonding term have an RMSD of $3.21$ \textit{ppm}
and $r=-0.024$ compared to an RMSD of $0.49$ and $r=0.943$ with the $\Delta\delta^{i}_{HB}$. This highlights the importance of proper treatment of hydrogen
bonding for the ${}^{1}H^{N}$ chemical shift. The value of the contribution from hydrogen bonding shows a range from close to zero and up to $\sim 6.5$ \textit{ppm}.

\clearpage

In order to test the performance of the primary $\Delta\delta_{1\deg HB}$, secondary $\Delta\delta_{2\deg HB}$ and the combination of the two $\Delta\delta^{i}_{HB}$ hydrogen bond terms, 
Procs14 predictions were carried out and compared with the QM NMR calculations, see \tref{table:hydrogenbond}. The table shows a benchmark of the \HN hydrogen bond
terms on protein structures optimised with PM6. The hydrogen bond terms are compared on all the six atom types. The predictions were improved with
the combination of both the primary and secondary terms on ${}^{1}H^{N}$ and ${}^{15}N^{H}$ chemical shifts. This improved the average \HN RMSD from $3.22$ \textit{ppm} to $0.66$ 
\textit{ppm} and the average correlation coefficient from $-0.042$ to $0.917$. The ${}^{13}C'$ and ${}^{1}H\alpha$ shifts showed an improvement only with the 
secondary term. For \C the RMSD went from $3.44$ to $2.36$ \textit{ppm} and $r$ from $0.495$ to $0.705$. For \CA only the secondary bonding term provided a modest 
improvement in chemical shift. Since the hydrogen bond model systems does not contain any \CB atom no correction is available for this atom type.

The same kind of test of the ${}^{1}H\alpha$ hydrogen bond terms is shown in \tref{table:hydrogenalphabond}. The table shows a comparison between Procs14 predictions with
the \HA hydrogen bond terms and the QM NMR calculations on full protein structures. The \HA chemical shifts are improved significantly with the RMSD improving from $1.38$ 
to $0.75$ \textit{ppm} and $r$ from $0.485$ to $0.842$. The secondary term $\Delta\delta_{2\deg H\alpha B}$ does not seem to be of much use, offering only a small 
improvement in RMSD on ${}^{13}C'$ and ${}^{15}N^{H}$ chemical shift. The hydrogen bond terms that do not seem to improve the chemical shift predictions is turned off, 
see appendix {\bf \ref{appendbench} \chapter{Benchmarking}}, \tref{table:hbondtermsused} for an overview of the terms used.

In general the correction on the atoms directly involved with the hydrogen bond performed very good. 
It is possible that the hydrogen bond scans lacks degrees of freedom crucial to the hydrogen bond correction
on the not atoms directly involved with the hydrogen bonding. One possibility is that the N-H..O and C\(\alpha\)-H\(\alpha\)..O bond angles play an important role in the correction. For example
in the crystal structure 1UBQ the average N-H..O bond angle was $123.72\deg$, while the angle is fixed at $180\deg$ in the Procs14 model systems. To investigate this, a scan over the two bond angles was performed. The scans were done on N-methylacetamide N-methylacetamide and N-methylacetamide alanine model systems. See appendix {\bf \ref{appendbench} \chapter{Benchmarking}}, \fref{fig:theta1} for the result. The change in chemical shift is small on the ${}^{1}H$ atoms, while the change can be quite large on the rest of the atom types. It is possible that the effect comes not just from the hydrogen bond interaction but also other types of interactions with other atoms in the model system.

\renewcommand\tabcolsep{2pt}
\begin{table}[h]
\hspace*{-0.30in}
\label{aggiungi}\centering 
{\scriptsize
\begin{tabular}{ lcccccccccccc }
\hline
\multicolumn{1}{l}{} & \multicolumn{2}{l}{\hspace{14pt}${}^{13}C\alpha{}$} & \multicolumn{2}{l}{\hspace{14pt}${}^{13}C\beta$} & \multicolumn{2}{l}{\hspace{14pt}${}^{13}C'$} & \multicolumn{2}{l}{\hspace{14pt}${}^{15}N^{H}$} & \multicolumn{2}{l}{\hspace{14pt}${}^{1}H^{N}$} & \multicolumn{2}{l}{\hspace{14pt}${}^{1}H\alpha$} \\
\hline
\multicolumn{1}{l}{}  & $<r>$   & $<$RMSD$>$ & $<r>$   & $<$RMSD$>$ & $<r>$   & $<$RMSD$>$ & $<r>$    & $<$RMSD$>$ & $<r>$   & $<$RMSD$>$ & $<r>$   & $<$RMSD$>$ \\
Without                                       & $0.923$ & $2.03$     & $0.985$ & $2.54$     & $0.495$ & $3.44$     & $0.761$ & $9.66$     & $-0.042$ & $3.22$    & $0.485$ & $1.38$    \\
$\Delta\delta_{2\deg HB}$                     & $0.919$ & $1.98$     & $0.985$ & $2.54$     & $0.705$ & $2.36$     & $0.844$ & $5.82$     & $0.009$ & $3.01$     & $0.527$ & $1.28$    \\
$\Delta\delta_{1\deg HB}$                     & $0.914$ & $2.42$     & $0.985$ & $2.54$     & $0.471$ & $3.48$     & $0.802$ & $6.02$     & $0.908$ & $0.73$     & $0.460$ & $1.41$    \\
$\Delta\delta^{i}_{HB}$                       & $0.912$ & $2.29$     & $0.985$ & $2.54$     & $0.683$ & $2.48$     & $0.878$ & $5.16$     & $0.917$ & $0.66$     & $0.505$ & $1.31$    \\
\hline
\end{tabular}
}
\caption{ {\bf Benchmarking the Primary and Secondary ${}^{1}H^{N}$ Hydrogen Bond Terms.} This table shows a comparison of the primary $\Delta\delta_{1\deg HB}$ and secondary $\Delta\delta_{2\deg HB}$ ${}^{1}H^{N}$ hydrogen bond terms. They are compared with Procs14 without the hydrogen bond terms and both primary and secondary at the same time $\Delta\delta^{i}_{HB}$. The
${}^{1}H^{N}$ and ${}^{15}N^{H}$ chemical shift are improved with both primary and secondary bonding terms. For the ${}^{13}C'$ and ${}^{1}H\alpha$ chemical shift only the secondary bonding 
alone was better. None of the bonding terms seems to improve the ${}^{13}C\alpha{}$ chemical shift. For these comparisons the ${}^{1}H\alpha$ hydrogen bond terms are turned off. 
		 }
	\label{table:hydrogenbond}
\end{table}
\renewcommand\tabcolsep{6pt}

\renewcommand\tabcolsep{2pt}
\begin{table}[h]
\hspace*{-0.30in}
\label{aggiungi}\centering 
{\scriptsize
\begin{tabular}{ lcccccccccccc }
\hline
\multicolumn{1}{l}{} & \multicolumn{2}{l}{\hspace{14pt}${}^{13}C\alpha{}$} & \multicolumn{2}{l}{\hspace{14pt}${}^{13}C\beta$} & \multicolumn{2}{l}{\hspace{14pt}${}^{13}C'$} & \multicolumn{2}{l}{\hspace{14pt}${}^{15}N^{H}$} & \multicolumn{2}{l}{\hspace{14pt}${}^{1}H^{N}$} & \multicolumn{2}{l}{\hspace{14pt}${}^{1}H\alpha$} \\
\hline
\multicolumn{1}{l}{}  & $<r>$   & $<$RMSD$>$ & $<r>$   & $<$RMSD$>$ & $<r>$   & $<$RMSD$>$ & $<r>$    & $<$RMSD$>$ & $<r>$   & $<$RMSD$>$ & $<r>$   & $<$RMSD$>$ \\
Without                                              & $0.923$ & $2.03$     & $0.985$ & $2.54$     & $0.495$ & $3.44$     & $0.761$ & $9.66$     & $-0.042$ & $3.22$     & $0.485$ & $1.38$    \\
$\Delta\delta_{2\deg H\alpha B}$                     & $0.921$ & $2.08$     & $0.985$ & $2.54$     & $0.498$ & $3.40$     & $0.760$ & $9.58$     & $-0.039$ & $3.21$     & $0.472$ & $1.39$  \\
$\Delta\delta_{1\deg H\alpha B}$                     & $0.908$ & $2.07$     & $0.982$ & $3.04$     & $0.474$ & $3.12$     & $0.729$ & $8.85$     & $-0.045$ & $3.23$     & $0.848$ & $0.73$    \\
$\Delta\delta^{i}_{H\alpha B}$                       & $0.907$ & $2.11$     & $0.982$ & $3.04$     & $0.475$ & $3.10$     & $0.729$ & $8.80$     & $-0.043$ & $3.22$     & $0.842$ & $0.75$    \\

\hline
\end{tabular}
}

\caption{ {\bf Benchmarking the Primary and Secondary ${}^{1}H\alpha$ Hydrogen Bond Terms.} This table shows a comparison of the primary $\Delta\delta_{1\deg H\alpha B}$ and secondary $\Delta\delta_{2\deg H\alpha B}$ ${}^{1}H\alpha$  hydrogen bond terms. They are compared with Procs14 without the hydrogen bond terms and both primary and secondary at the same time $\Delta\delta^{i}_{H\alpha}$. Only the ${}^{1}H\alpha$ chemical shifts shows a clear improvement. ${}^{13}C'$ and ${}^{15}N^{H}$ showed a better RMSD although the correlation coefficient got worse. In these comparisons the ${}^{1}H^{N}$ hydrogen bond terms are turned off.}
	\label{table:hydrogenalphabond}
\end{table}
\renewcommand\tabcolsep{6pt}

\subsection{Benchmarking the Corrections for the Previous and Following Residue }

In order to test the correction for previous and following residue small methyl capped tripeptides were cut out from the PM6 optimized structures. An NMR
calculation was performed at the same level of theory as the full protein calculations. This approach allows for isolation of the chemical shift from
hydrogen bonding, ring current and other non-local interactions. The tripeptides are cut from the same structures as used in the test of the hydrogen bond terms.
The chemical shifts are computed with TMS and ammonia as reference compounds. Procs14 predictions are performed on the PM6 structures with the hydrogen bond and
ring current terms turned off. See \tref{table:PFcorr} for the results. 
The chemical shifts are in general improved by the corrections, the RMSD in the \CA chemical shifts went from $1.5$ to $1.42$ \textit{ppm} and 
correlation coefficient from $0.922$ to $0.934$. The strongest affected atom types were clearly the \C, \N and \HN. The correlation coefficient 
went from $0.511$, $0.629$ and $0.439$ to $0.656$, $0.852$ and $0.736$. The RMSD on \C, \N and \HN went from $3.56$, $5.69$ and $0.77$ to $1.94$, $4.48$ and $0.51$.
The previous/following correction terms only made the predictions \CB worse.  

\renewcommand\tabcolsep{2pt}
\begin{table}[h]
\hspace*{-0.30in}
\label{aggiungi}\centering 
{\scriptsize
\begin{tabular}{ lcccccccccccc }
\hline
\multicolumn{1}{l}{} & \multicolumn{2}{l}{\hspace{14pt}${}^{13}C\alpha{}$} & \multicolumn{2}{l}{\hspace{14pt}${}^{13}C\beta$} & \multicolumn{2}{l}{\hspace{14pt}${}^{13}C'$} & \multicolumn{2}{l}{\hspace{14pt}${}^{15}N^{H}$} & \multicolumn{2}{l}{\hspace{14pt}${}^{1}H^{N}$} & \multicolumn{2}{l}{\hspace{14pt}${}^{1}H\alpha$} \\
\hline
\multicolumn{1}{l}{}  & $<r>$   & $<$RMSD$>$ & $<r>$   & $<$RMSD$>$ & $<r>$   & $<$RMSD$>$ & $<r>$    & $<$RMSD$>$ & $<r>$   & $<$RMSD$>$ & $<r>$   & $<$RMSD$>$ \\
Without                                            & $0.922$ & $1.50$     & $0.994$ & $1.60$     & $0.511$ & $3.56$     & $0.629$ & $5.69$     & $0.439$ & $0.77$     & $0.803$ & $0.34$    \\
$\Delta\delta^{i-1}_{BB}$                          & $0.921$ & $1.45$     & $0.994$ & $1.69$     & $0.609$ & $2.83$     & $0.853$ & $4.67$     & $0.751$ & $0.56$     & $0.814$ & $0.33$    \\
$\Delta\delta^{i+1}_{BB}$                          & $0.933$ & $1.50$     & $0.994$ & $1.62$     & $0.530$ & $2.18$     & $0.626$ & $5.64$     & $0.415$ & $0.79$     & $0.798$ & $0.34$    \\
$\Delta\delta^{i-1}_{BB}+\Delta\delta^{i+1}_{BB}$  & $0.934$ & $1.42$     & $0.993$ & $1.72$     & $0.645$ & $1.94$     & $0.852$ & $4.48$     & $0.736$ & $0.51$     & $0.811$ & $0.32$    \\
\hline
\end{tabular}
}

\caption{ {\bf Benchmarking the Correction for the Previous and Following Residue.} This table shows the result of benchmarking of the correction for previous $\Delta\delta^{i-1}_{BB}$ and following $\Delta\delta^{i+1}_{BB}$ residues and the combination of them. The RMSD and linear correlation factor $r$ is computed on predictions from Procs14 without hydrogen bonding and ring current and tripeptides cut from the PM6 optimized structures. The NMR calculation on the cut peptides is calculated at the same level of theory as the full protein calculations.             
		 }
	\label{table:PFcorr}
\end{table}
\renewcommand\tabcolsep{6pt}
\subsection{Benchmarking on Forcefield Optimised Structures}

\begin{figure}[h!]
  \hspace{-15 mm}
  \centering
  \begin{minipage}[b]{0.45\linewidth}
    \includegraphics[width=\linewidth]{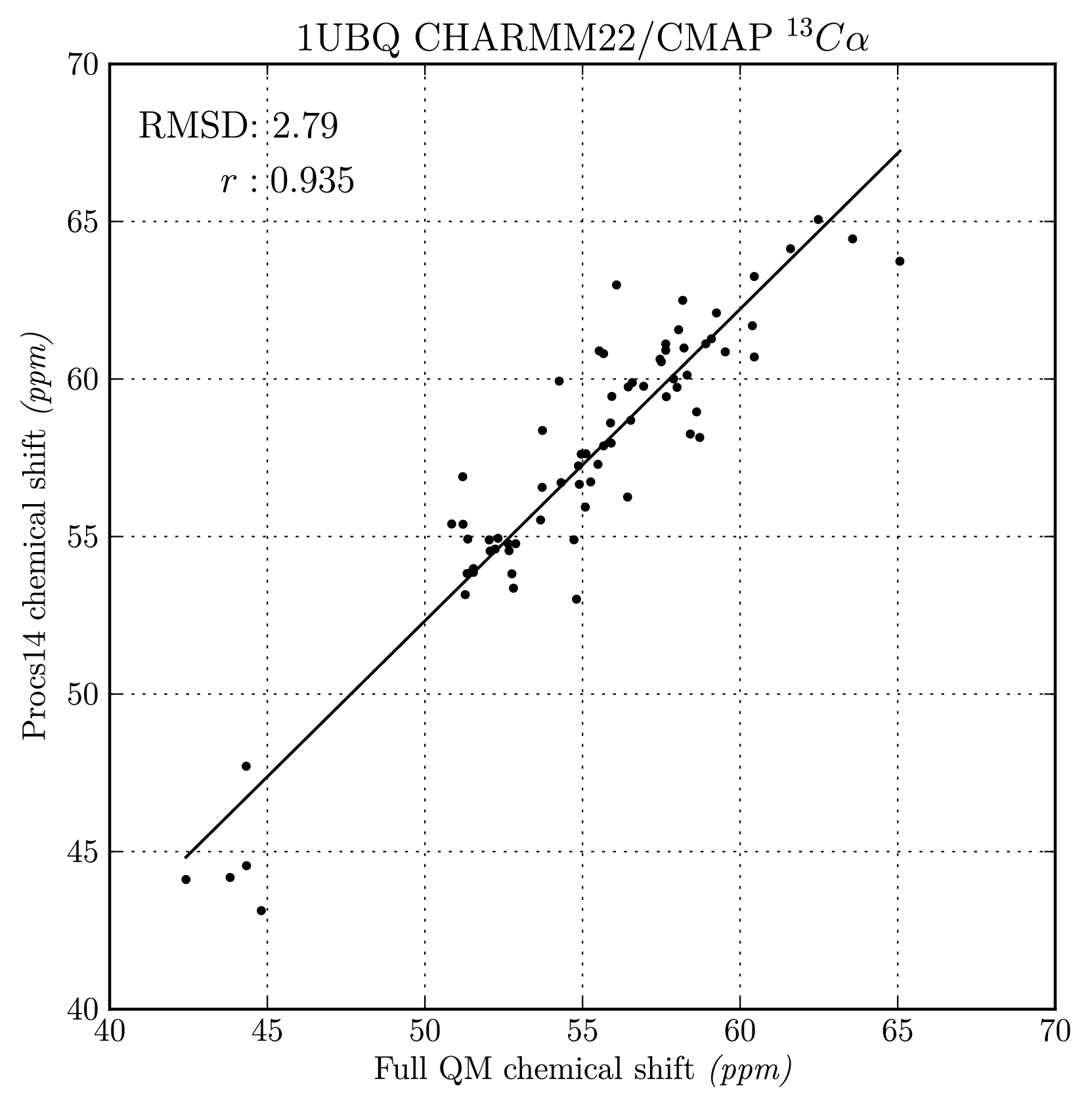}
  \end{minipage}
  \begin{minipage}[b]{0.45\linewidth}
    \includegraphics[width=\linewidth]{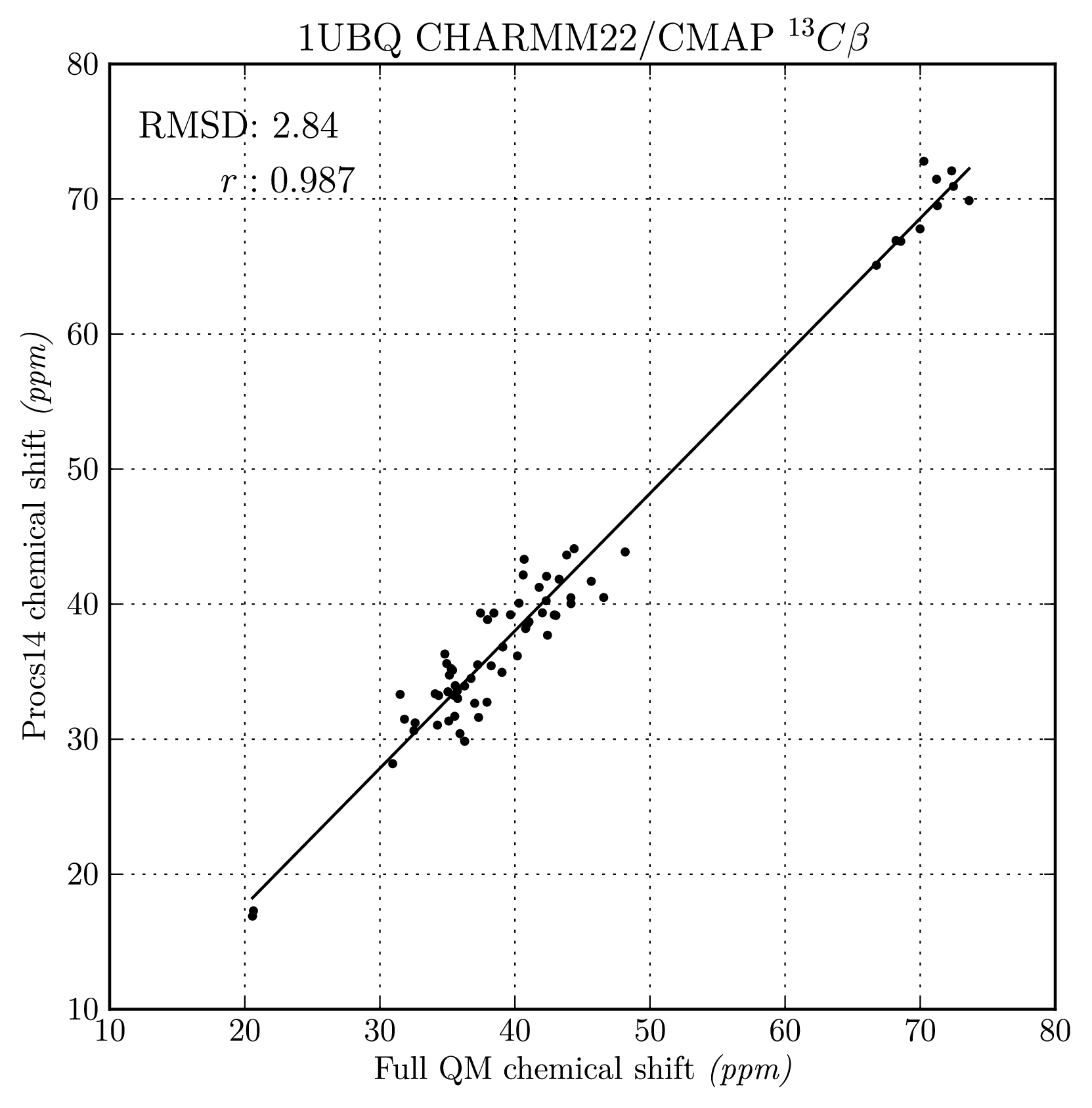}
  \end{minipage} \\
  \hspace{-15 mm}
  \begin{minipage}[b]{0.45\linewidth}
    \includegraphics[width=\linewidth]{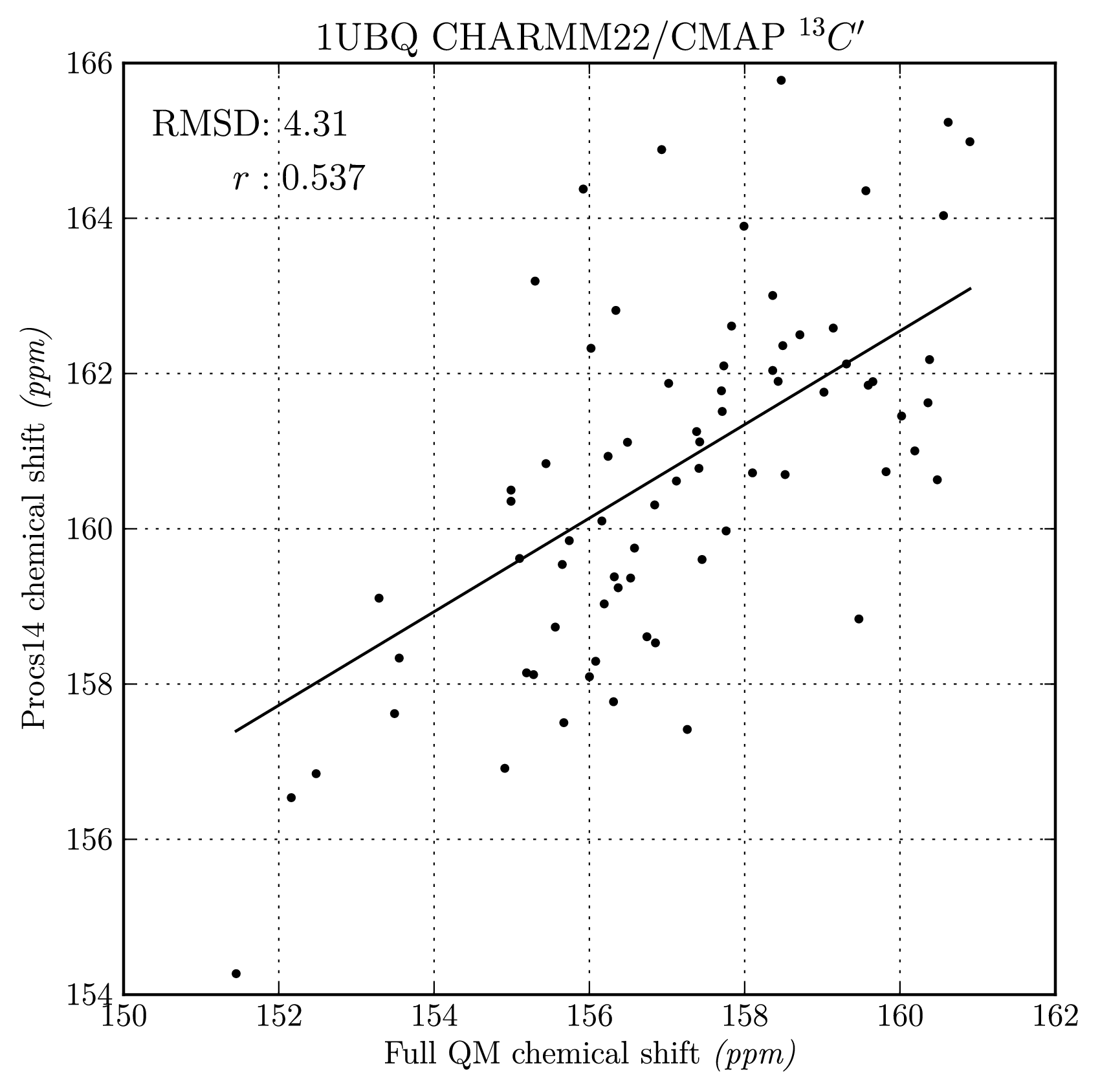}
  \end{minipage}
  \begin{minipage}[b]{0.45\linewidth}
    \includegraphics[width=\linewidth]{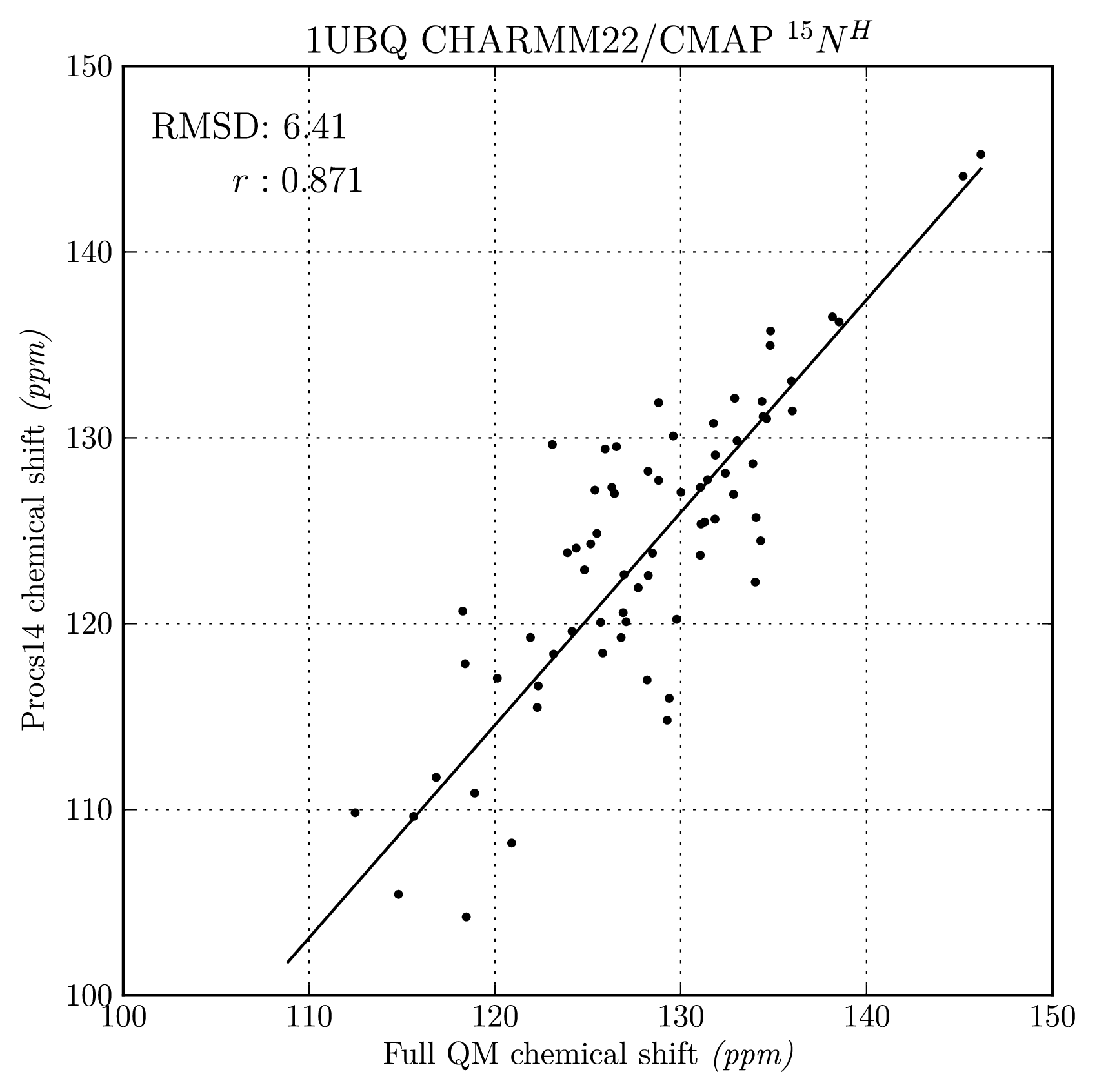}
  \end{minipage}\\
  \hspace{-15 mm}
  \begin{minipage}[b]{0.45\linewidth}
    \includegraphics[width=\linewidth]{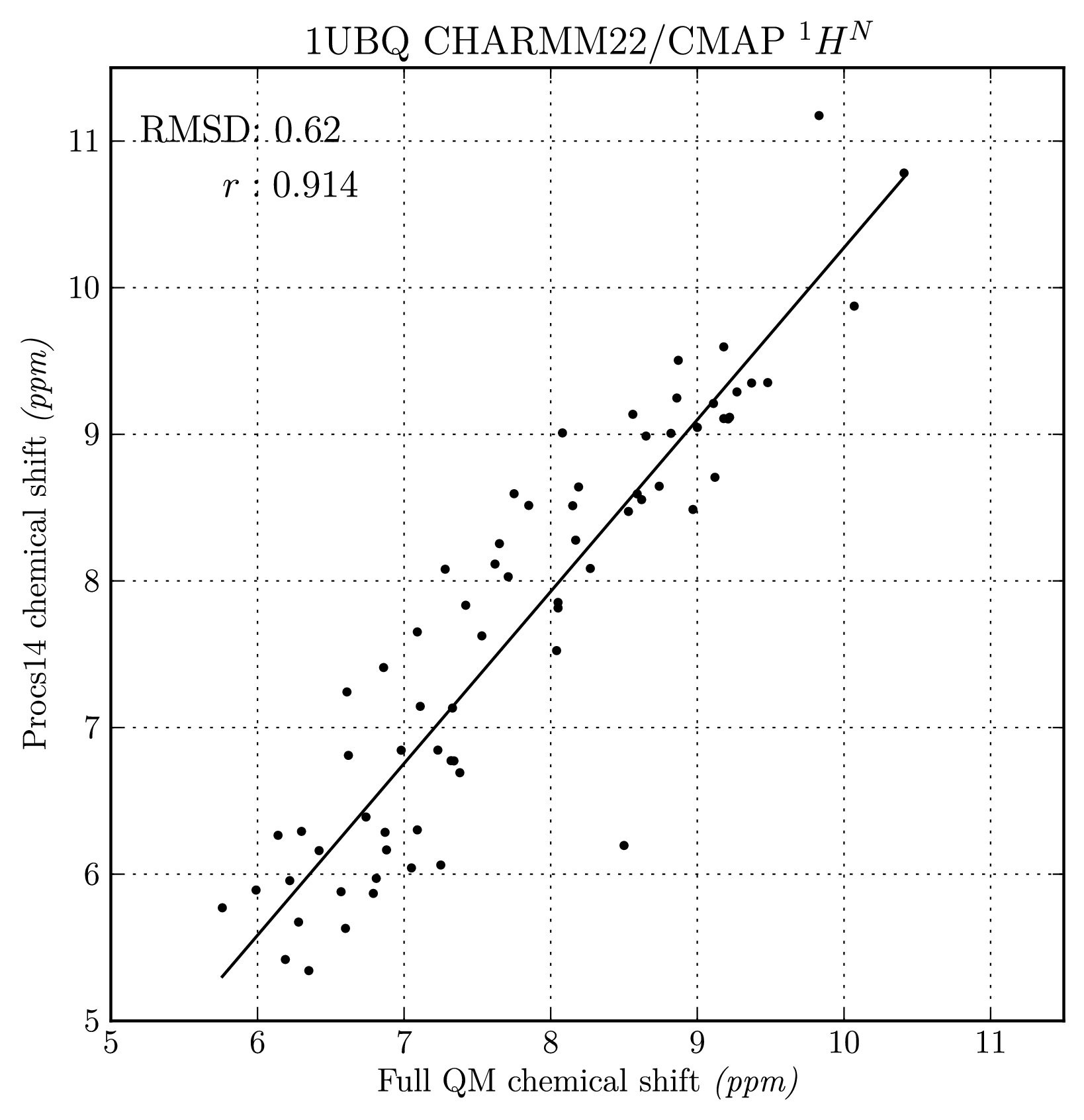}
  \end{minipage}
  \begin{minipage}[b]{0.45\linewidth}
    \includegraphics[width=\linewidth]{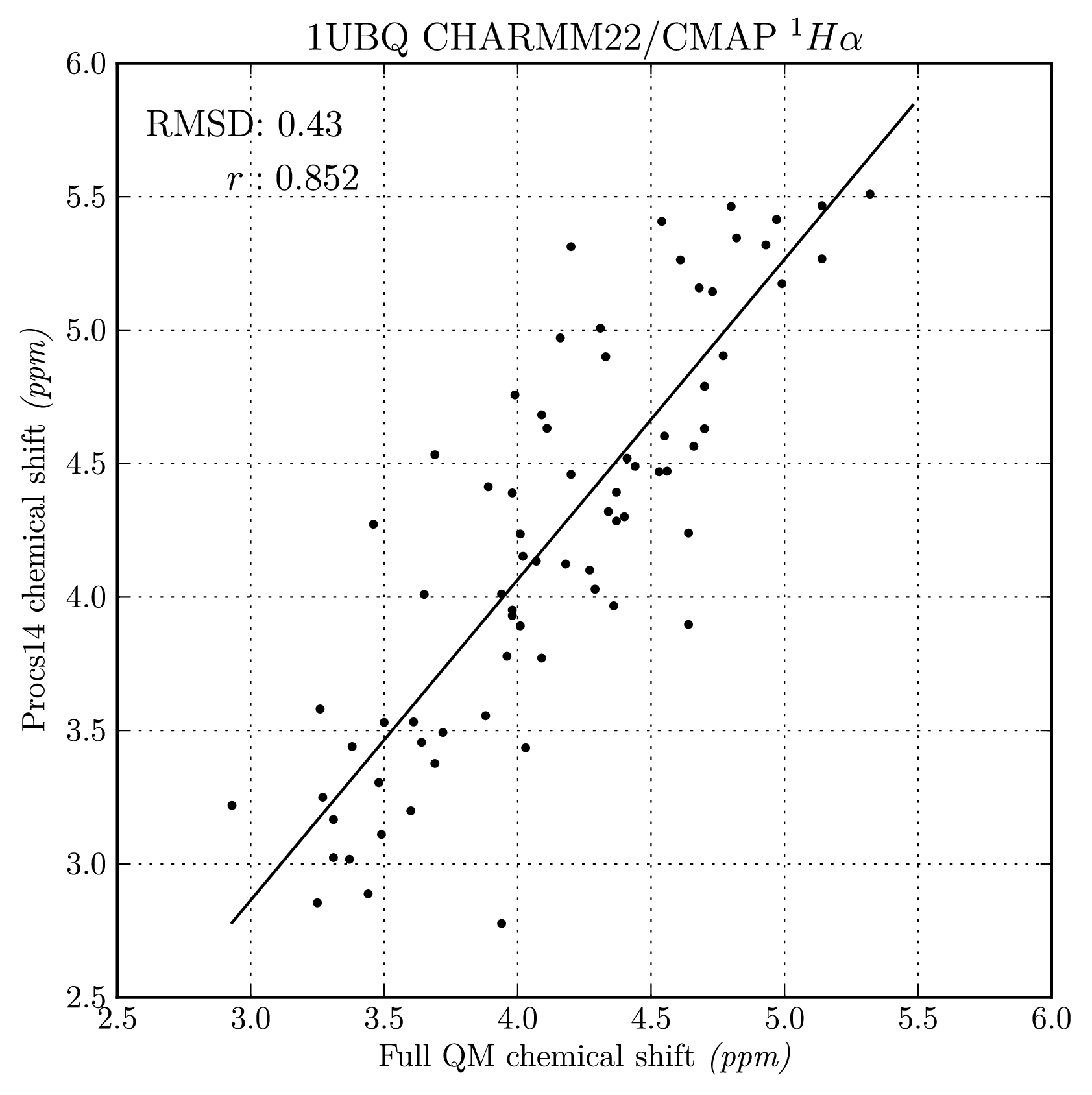}
  \end{minipage}
  
  \caption{ {\bf Procs14 Predictions Compared With QM NMR Calculations on Ubiquitin.}
		 }
	\label{fig:charmpredic}
\end{figure}

In addition to benchmarking on PM6 structures, Procs14 is also tested on NMR calculation on proteins optimised with forcefields.
The tests is done with the hydrogen bond length correction and the scaling procedure. The results are compared with \CA and \CB predictions 
from the CheShift-2 web server. See \tref{table:procscheshift} for the results. Procs14 have better correlation coefficient than
CheShift in almost all cases and a better RMSD for most structures. For Procs14 the \CA RMSD has a range of $1.62$ to $2.39$ \textit{ppm} and correlation coefficient range of $0.892$ to $0.948$.
This is significantly better than CheShift-2's RMSD from $1.67$ to $3.94$ \textit{ppm} and $r$ from $0.840$ to $0.940$. The \CB chemical shifts
are also in favor of Procs14. Procs14 has a \CB RMSD from $2.05$ to $3.87$ \textit{ppm} and $r$ from $0.970$ to $0.995$. This is better than
CheShift-2's RMSD of $2.20$ to $10.21$ and $r$ of $0.972$ to $0.992$. For the \C chemical shifts the Procs14 predictions are quite bad, they
show an average correlation coefficient of $0.512$ and average RMSD of $5.61$ \textit{ppm}. For the \HN chemical shift the predictions shows
a large variance with $r$ up to $0.936$ and the lowest RMSD equal to $0.51$ \textit{ppm}. The \HA chemical shifts have a RMSD range of
$0.38$ to $0.49$ and a $r$ range of $0.799$ to $0.878$. The \N Procs14 prediction have a quite high average RMSD of $7.26$ and average $r$ equal to
$0.812$.
It is important to be careful when interpreting the RMSD 
since it was shown that the chemical shifts are very sensitive to bond lengths. This can
be seen in \fref{fig:charmpredic}, which shows Procs14 prediction of ubiquitin optimised with CHARMM22-CMAP.  For \CA, \CB and \C the chemical shifts is systematically too high or low, dependent on the atom type.  
The hydrogen bond length correction on the \HN and \HA chemical shifts are quite successful in correction for this error. The 
C\(\alpha\)-H\(\alpha\) bond length correction on \CA  is only moderately successful. This indicates that the bond length corrections
should be expanded to the atom types with multiple covalent bonds, namely \CA, \CB, \C   and \N. For the \HA and \HN chemical shifts,
the correlation coefficient and RMSD is often worsened by a small number of outliers. For example in the case of protein G optimised
with AMOEBAPRO14, the glutamine residue 2 have an hydrogen bond with it own side chain oxygen. This is a case presently not treated explicitly
in Procs14.

\begin{table}[h]
\label{aggiungi}\centering 
{\scriptsize  
\begin{tabular}{lcccccc*{9}{l}}
\toprule %
 Procs14              &  ${}^{13}C\alpha{}$     & ${}^{13}C\beta$       & ${}^{13}C'$            & ${}^{15}N^{H}$            & ${}^{1}H^{N}$        & ${}^{1}H\alpha$      &\\
 Structures           &  $r$  RMSD     & $r$  RMSD      & $r$  RMSD       & $r$      RMSD  & $r$      RMSD  & $r$      RMSD  &\\\midrule 
 1UBQ:AMBER           & $0.924$ $1.79$ & $0.990$ $2.05$ & $0.348$ $5.99$  & $0.775$ $7.02$ & $0.837$ $0.79$ & $0.844$ $0.51$  &\\ 	
 1UBQ:CHARMM22/CMAP   & $0.935$ $2.79$ & $0.987$ $2.84$ & $0.537$ $4.31$  & $0.871$ $6.41$ & $0.914$ $0.62$ & $0.852$ $0.43$ &\\
 1UBQ:AMOEBAPRO14     & $0.928$ $2.70$ & $0.982$ $3.87$ & $0.520$ $4.32$  & $0.790$ $9.73$ & $0.866$ $1.01$ & $0.846$ $0.49$ &\\
 1UBQ:Crystal         & $0.892$ $2.30$ & $0.970$ $3.57$ & $0.201$ $12.28$ & $0.769$ $8.46$ & $0.307$ $2.30$ & $0.832$ $0.46$ &\\
 2OED:AMBER           & $0.926$ $1.86$ & $0.990$ $3.04$ & $0.500$ $5.53$  & $0.781$ $6.75$ & $0.896$ $0.66$ & $0.799$ $0.49$ &\\	
 2OED:CHARMM22/CMAP   & $0.948$ $1.62$ & $0.994$ $2.46$ & $0.676$ $4.42$  & $0.897$ $6.08$ & $0.927$ $0.52$ & $0.868$ $0.43$ &\\
 2OED:AMOEBAPRO14     & $0.943$ $2.25$ & $0.989$ $3.74$ & $0.537$ $4.34$  & $0.719$ $8.59$ & $0.625$ $1.05$ & $0.855$ $0.43$ &\\
 2OED:Original        & $0.947$ $2.39$ & $0.995$ $3.32$ & $0.657$ $6.17$  & $0.869$ $4.31$ & $0.902$ $0.51$ & $0.860$ $0.45$ &\\
 1IGD:AMBER           & $0.928$ $1.83$ & $0.994$ $2.63$ & $0.425$ $5.63$  & $0.792$ $6.99$ & $0.936$ $0.55$ & $0.833$ $0.47$ &\\
 1IGD:CHARMM22/CMAP   & $0.948$ $2.78$ & $0.995$ $2.39$ & $0.654$ $4.06$  & $0.877$ $6.38$ & $0.933$ $0.52$ & $0.878$ $0.38$ &\\
 1IGD:AMOEBAPRO14     & $0.928$ $2.16$ & $0.982$ $3.87$ & $0.520$ $4.32$  & $0.790$ $9.73$ & $0.866$ $1.01$ & $0.846$ $0.49$ &\\
 1IGD:Crystal         & $0.947$ $2.39$ & $0.995$ $3.32$ & $0.657$ $6.17$  & $0.869$ $4.31$ & $0.902$ $0.51$ & $0.860$ $0.45$ &\\\midrule
 CheShift-2           &  ${}^{13}C\alpha{}$     & ${}^{13}C\beta$       & ${}^{13}C'$            & ${}^{15}N^{H}$            & ${}^{1}H^{N}$        & ${}^{1}H\alpha$      &\\
 Structures           &  $r$  RMSD     & $r$  RMSD      & $r$  RMSD       & $r$      RMSD  & $r$      RMSD  & $r$      RMSD  &\\\midrule
 1UBQ:AMBER           & $0.888$ $3.19$ & $0.982$ $2.85$ &  & &  &  &\\	
 1UBQ:CHARMM22/CMAP   & $0.880$ $2.12$ & $0.986$ $5.22$ &  & &  &  &\\
 1UBQ:AMOEBAPRO14     & $0.922$ $3.94$ & $0.978$ $6.34$ &  & &  &  &\\
 1UBQ:Crystal         & $0.892$ $2.31$ & $0.972$ $3.57$ &  & &  &  &\\
 2OED:AMBER           & $0.908$ $3.02$ & $0.987$ $2.90$ &  & &  &  &\\	
 2OED:CHARMM22/CMAP   & $0.902$ $2.13$ & $0.990$ $4.95$ &  & &  &  &\\
 2OED:AMOEBAPRO14     & $0.931$ $3.81$ & $0.988$ $6.29$ &  & &  &  &\\
 2OED:Original        & $0.940$ $1.67$ & $0.992$ $2.20$ &  & &  &  &\\
 1IGD:AMBER           & $0.854$ $3.02$ & $0.989$ $2.75$ &  & &  &  &\\
 1IGD:CHARMM22/CMAP   & $0.840$ $2.10$ & $0.991$ $4.69$ &  & &  &  &\\
 1IGD:AMOEBAPRO14     & $0.860$ $3.54$ & $0.990$ $5.95$ &  & &  &  &\\
 1IGD:Crystal         & $0.871$ $3.19$ & $0.988$ $10.21$ &  & &  &  &\\\bottomrule
\end{tabular}}
\caption{ {\bf Procs14 and Cheshift-2 Comparison With Full Protein NMR Calculations.} The table shows predictions from Procs14 and CheShift-2 compared to QM NMR calculations 
	on full protein structures. The proteins are a crystal structure of ubiquitin, a crystal structure of protein G refined with Residual Dipolar Couplings(RDC) and a crystal structure of protein G 1IGD. Next to the names of the proteins are the forcefields used in the geometry optimization. For each atom types the RMSD and coefficient of determination $r$ from linear regression analysis are displayed.
          CheShift-2 only preforms predictions for $C\alpha$ and $C\beta$ carbons and therefore
          lacks results for the remaining atom types. Note that the \C chemical shifts are not scaled predictions, but chemical shifts calculated with TMS as reference. The correction for surface amide protons is turned off.      	  
		 }
	\label{table:procscheshift}
\end{table}

\clearpage
 \vspace{-15 mm}
\clearpage

\renewcommand\tabcolsep{2pt}
\begin{table}[h]
\hspace*{-0.30in}
\label{aggiungi}\centering 
{\scriptsize
\begin{tabular}{ lcccccccccccc }
  \hline
  \multicolumn{1}{l}{Method} & \multicolumn{2}{c}{\CA} & \multicolumn{2}{c}{\CB} & \multicolumn{2}{c}{\C} & \multicolumn{2}{c}{\N} & \multicolumn{2}{c}{\HN} & \multicolumn{2}{c}{\HA} \\
\hline
\multicolumn{1}{l}{}   & $<r>$   & $<$RMSD$>$ & $<r>$   & $<$RMSD$>$ & $<r>$   & $<$RMSD$>$ & $<r>$   & $<$RMSD$>$ & $<r>$   & $<$RMSD$>$ & $<r>$   & $<$RMSD$>$ \\
Procs14                & $0.932$ & $2.21$     & $0.989$ & $3.05$     & $0.512$ & $5.61$     & $0.812$ & $7.26$     & $0.818$ & $0.83$     & $0.840$ & $0.46$    \\
SHIFTX                 & $0.896$ & $2.52$     & $0.985$ & $3.22$     & $0.501$ & $17.59$    & $0.750$ & $10.04$    & $0.437$ & $1.41$     & $0.810$ & $0.51$    \\
SPARTA                 & $0.903$ & $2.46$     & $0.985$ & $3.22$     & $0.545$ & $17.69$     & $0.754$ & $9.54$     & $0.375$ & $1.45$    & $0.791$ & $0.54$    \\
SPARTA+                & $0.910$ & $2.41$     & $0.985$ & $3.24$     & $0.545$ & $17.65$     & $0.768$ & $9.50$     & $0.468$ & $1.34$    & $0.843$ & $0.50$    \\
CamShift               & $0.901$ & $2.51$     & $0.984$ & $3.33$     & $0.528$ & $17.64$     & $0.719$ & $9.93$     & $0.482$ & $1.37$     & $0.820$ & $0.51$    \\
CheShift-2             & $0.891$ & $2.84$     & $0.986$ & $4.83$     &         &            &         &            &         &            &         &           \\

\hline
\end{tabular}
}

\caption{ {\bf A Comparison of Chemical Shift Predictors on Full QM NMR Calculations.}
 	   This table shows correlation coefficient $r$ and RMSD averages for Procs14 and five other chemical shift predictors. The chemical shift predictors are compared with 
          full QM NMR calculations on full protein structures. The proteins are ubiquitin and protein G. In addition to the crystal structure, the structures are optimised 
	  with one of the forcefields AMBER, CHARMM22/CMAP or AMOEBAPRO14. The RMSD and correlation coefficient reported, is an average over 12 chemical shift predictions.
		 }
	\label{table:predictors}
\end{table}
\renewcommand\tabcolsep{6pt}

The results from the full QM NMR calculations on proteins are compared with Procs14 and five other chemical shift predictors in \tref{table:predictors}. Procs14 is better or 
comparable on all six atom types. On \CA Procs14 has an average correlation coefficien $<r>$ of $0.932$ and average $<$RMSD$>$ of $2.21$ \textit{ppm}, this compares
favorably with the best empirical predictor SPARTA+'s $<r>$ of $0.910$ and $<$RMSD$>$ of $2.41$ \textit{ppm}. For \CB Procs14 has $<r>$ of $0.989$ and $<$RMSD$>$ of $3.05$ \textit{ppm} compared
with SHIFTX $0.985$ and $3.22$ \textit{ppm}. The \C chemical shift predictions are very bad for all methods with a $<r>$ of $\sim0.52$. Procs14 is again best on the \N
chemical shifts with a $<r>$ of $0.812$ and $<$RMSD$>$ of $7.26$ \textit{ppm} versus SPARTA+'s $<r>$ of $0.768$ and $<$RMSD$>$ of $9.50$ \textit{ppm}.
The empirical predictors exhibit very bad results on the \HN chemical shifts with a $<r>$ in the range $0.375$-$0.482$ and $<$RMSD$>$ of $1.37$-$1.45$ \textit{ppm}. Compared to this,
Procs14 is much better with $<r>$ of $0.818$ and $<$RMSD$>$ of $0.83$ \textit{ppm}.
This is most likely the result of the empirical methods being relativity insensitive to the geometry of hydrogen bonding. For example on a protein structure of protein G optimised with CHARMM22/CMAP the Procs14
\HN chemical shifts have a range of $\sim6$-$10$ {\textit{ppm}}, while the empirical methods are between $7$ and $9$ {\textit{ppm}}. For the \HA chemical shift all the predictors 
are comparable. Procs14 only have a slightly better $<$RMSD$>$ of $0.46$ \textit{ppm} compared with the best empirical SPARTA+'s $0.50$ \textit{ppm}.

\subsection{Comparison with Experimental Chemical Shift}

In this section Procs14 predictions are compared with experimental determined chemical shift data.
Since Procs14 is expected to be very sensitive to the input structure, a number of different structures 
are used. They are structures of ubiquitin, derived using either X-ray crystallography or they are NMR structural ensembles.
The experimental chemical shifts come from the model in \cite{1d3z} BMRB code 17769.
In \tref{table:expshift} the comparison of Procs14 and CheShift-2 predictions with experimental chemical shifts is shown.
The predictions are clearly more accurate on the NMR ensembles than the crystal structure. On the 2KOX ensemble, Procs14 achieved
a \CA RMSD of $1.04$ \textit{ppm} and correlation coefficient $r$ of $0.974$. This is much better than the 1UBQ crystal structure which
had an \CA RMSD of $1.66$ \textit{ppm} and $r$ of $0.932$. The predictions are improved on all atom types by using the NMR ensembles.
The improvement in RMSD on \CA is up to $\sim0.62$ \textit{ppm,} \CB $\sim 0.41$ \textit{ppm}, \C $\sim 0.98$ \text{ppm}, \N $\sim 0.9$ \textit{ppm}, \HN $\sim 0.34$ \textit{ppm}
and \HA $\sim 0.17$ \textit{ppm}. Procs14 also provides better prediction than CheShift-2 on all structures. Although
I suspect that Cheshift-2's poor predictions are partly a result of systematic errors. For \CB the best Procs14 prediction 
was on the 1D3Z ensembles with $r$ of $0.992$ and RMSD of $1.71$ \textit{ppm}. For \CB Cheshift-2 gives a $r$ of $0.955$ and RMSD of
$2.58$ \textit{ppm}.
Unfortunately the Procs14 RMSD and correlation coefficient
are quite bad on \HN. If the correction for the solvent exposed amide protons is turned off and surface amide protons excluded, a much better
result can be achieved. This gives a RMSD of $0.58$ \textit{ppm} and $r=0.864$ for the 2K39 ensemble. This shows that Procs14 models the protein's internal hydrogen bond 
network with decent accuracy but not the surface amide protons. 

\begin{table}[h]
\label{aggiungi}\centering 
{\scriptsize  
\begin{tabular}{lccccccc*{10}{l}}
\toprule %
Procs14              &  ${}^{13}C\alpha{}$     & ${}^{13}C\beta$       & ${}^{13}C'$            & ${}^{15}N^{H}$            & ${}^{1}H^{N}$        & ${}^{1}H\alpha$    &  models  &\\
Structures           &  $r$  RMSD     & $r$  RMSD      & $r$  RMSD       & $r$      RMSD  & $r$      RMSD  & $r$      RMSD  &          &\\\midrule 
2KOX:NMR\cite{2kox}  & $0.974$ $1.04$ & $0.991$ $1.77$ & $0.502$ $2.53$ & $0.819$ $5.72$ & $0.609$ $0.85$ & $0.870$ $0.32$ &  $640$ &\\
2LJ5:NMR\cite{2lj5}  & $0.968$ $1.14$ & $0.991$ $1.72$ & $0.535$ $2.44$ & $0.818$ $5.53$ & $0.556$ $0.72$ & $0.836$ $0.38$ &  $301$ &\\
1XQQ:NMR\cite{1xqq}  & $0.964$ $1.22$ & $0.990$ $1.80$ & $0.496$ $2.63$ & $0.809$ $5.10$ & $0.582$ $0.75$ & $0.848$ $0.32$ &  $128$ &\\
2K39:NMR\cite{2k39}  & $0.967$ $1.16$ & $0.991$ $1.76$ & $0.463$ $2.49$ & $0.819$ $5.13$ & $0.550$ $0.75$ & $0.835$ $0.31$ &  $116$ &\\              
2K5N:NMR\cite{2kn5}  & $0.963$ $1.23$ & $0.991$ $1.76$ & $0.468$ $2.54$ & $0.859$ $4.52$ & $0.510$ $0.64$ & $0.821$ $0.32$ &  $50$ &\\
1D3Z:NMR\cite{1d3z}  & $0.959$ $1.33$ & $0.992$ $1.71$ & $0.466$ $2.67$ & $0.852$ $5.72$ & $0.580$ $0.97$ & $0.794$ $0.40$ &  $10$ &\\
1UBQ:Crystal         & $0.932$ $1.66$ & $0.987$ $2.12$ & $0.391$ $3.42$ & $0.809$ $5.42$ & $0.282$ $0.98$ & $0.790$ $0.48$ &  $1$  &\\\midrule
CheShift-2           &  ${}^{13}C\alpha{}$     & ${}^{13}C\beta$       & ${}^{13}C'$            & ${}^{15}N^{H}$            & ${}^{1}H^{N}$        & ${}^{1}H\alpha$    &  models  &\\
Structures           &  $r$  RMSD     & $r$  RMSD      & $r$  RMSD       & $r$      RMSD  & $r$      RMSD  & $r$      RMSD  &          &\\\midrule 
2KOX:NMR             & $0.961$ $2.16$ & $0.984$ $2.62$ &  & &  &  & $640$ &\\	 
2LJ5:NMR             & $0.956$ $2.29$ & $0.994$ $2.63$ &  & &  &  & $301$ &\\		
1XQQ:NMR             & $0.952$ $2.27$ & $0.993$ $2.79$ &  & &  &  & $128$ &\\	
2K39:NMR             & $0.956$ $2.28$ & $0.994$ $2.55$ &  & &  &  & $116$ &\\	
2K5N:NMR             & $0.955$ $2.34$ & $0.994$ $2.58$ &  & &  &  & $50$  &\\	
1D3Z:NMR             & $0.954$ $2.15$ & $0.995$ $2.38$ &  & &  &  & $10$  &\\	
1UBQ:Crystal         & $0.919$ $2.73$ & $0.989$ $3.00$ &  & &  &  & $1$   &\\\bottomrule	
\end{tabular}}
\caption{ {\bf Procs14 and Cheshift-2 on Ubiquitin Crystal Structures and NMR Ensembles With Experimental Data.} The table shows Procs14 and CheShift-2 prediction on crystal structures
          and NMR ensembles of ubiquitin. The table contains RMSD values computed between the predictions and experimental chemical shifts and a correlation coefficient $r$ between them. The chemical shifts are calculated with \refeq{eq:boltz} for both Procs14 and CheShift-2. The preditions on the NMR ensembles are better than 
	  The crystal structures for all atom types. The RMSD on \CA is up to $\sim0.62$ \textit{ppm}, \CB $\sim 0.41$ \textit{ppm}, \C $\sim 0.98$ \textit{ppm}, \N $\sim 0.9$ \textit{ppm}, \HN $\sim 0.34$ \textit{ppm} and \HA $\sim 0.17$ \textit{ppm}.   
          The Procs14 predictions are much better than CheShift-2 with a much better RMSD and slightly better correlation coefficient. 
		 }
	\label{table:expshift}
\end{table}

\clearpage
\section{Preliminary Results From Refinement} \label{sec:Refinement}
\subsection{Refinement}

The section presents a very preliminary test of the Procs14 implementation in PHAISTOS by refining a protein structure.
The simulation was run on protein G(2OED) with the PROFASI forcefield\cite{profasi}.
Because of time constraints the protein simulation was started from a partly unfolded
state with a $C\alpha$-RMSD of $3.42$ \AA. The code used was also not completely final.
Since the cached version of Procs14 provides significantly faster predictions compared to the non-cached version, the non-cached
version was not tested. The \C chemical shift predictions were excluded since Procs14 does not
seem to provide robust predictions for this atom type. 
It all ran for 96 hours on a 12 core Intel Xeon CPU X5675 3.07GHz machine with 36 GB of RAM.
The weight of the Procs14 energy term was kept fixed at $0.4$. A control
refinement with just PROFASI was run on the same machine for the same amount of time. The result of the energy scoring 
can be seen in \fref{fig:hbondcorrshow}. Both with and without Procs14 the simulations were able to refine the structures
closer to the experimental structure. The simulation with Procs14 achieves a $C\alpha$-RMSD of $1.47$ \AA\xspace and the simulation
with just PROFASI get a $C\alpha$-RMSD of $1.37$ \AA. This is slightly better than Procs14, but in a real world application
the correct experimental structure would not be available as a comparison. If we select the sample with the lowest energy,
Procs14 achieves a minimal $C\alpha$-RMSD of $2.09$ \AA\xspace which is better than $2.56$ \AA\xspace from the simulation with just PROFASI.      

\begin{figure}[h!]
	\centering
	  \begin{subfigure}[b]{0.70\textwidth}
                \centering
                \includegraphics[width=1.0\textwidth]{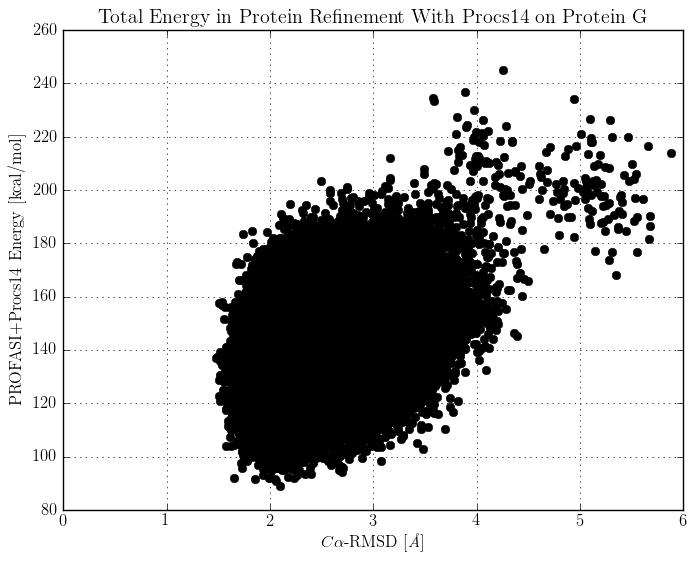} 
                \caption{\small{Refinement Energy Scoring with Procs14 and PROFASI}}
                \label{fig:refprocs}
     \end{subfigure} \\
	\begin{subfigure}[b]{0.70\textwidth}
                \centering
                \includegraphics[width=1.0\textwidth]{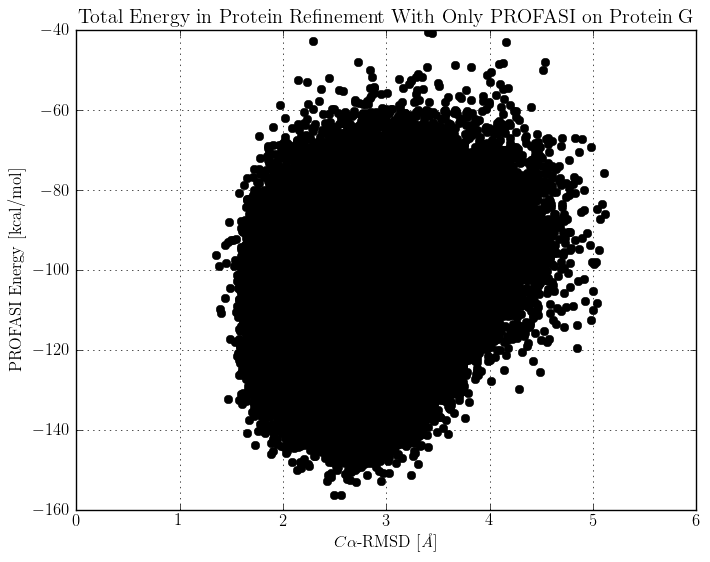}
                \caption{\small{Refinement Energy Scoring with PROFASI}}
                \label{fig:refnocs}
     \end{subfigure} 

	\caption{ {\bf Refinement of Protein G.} {\bf (a)} shows the energy scoring during a refinement of protein G with the cached Procs14 term.
                   The RMSD is calculated on $C\alpha$ atoms on all residues. {\bf (b)} shows the energy scoring during a refinement of protein G with
                   just the PROFASI force field. 
                   }\label{fig:hbondcorrshow}
\end{figure}

\clearpage
\section{Discussion and Conclusion} \label{sec:theend}
\subsection{Discussion and Future Work}

Procs14 beat the \CA and \CB predictions from CheShift-2 in all measurements. This is likely results from the 
larger amount of samples and better representation of the side chain angles in the Procs14 model.  
When comparing Procs14 with the empirical predictors on the full protein QM NMR calculations it had better predictions
on the \CA, \CB, \C, \N and \HN atom types. 
For \HA Procs14 provided only slightly better predictions.
The failure of the empirical methods to reproduce the QM NMR calculations shows the limitations of the empirical approach.
For the \N chemical shifts, Procs14 had a quite high RMSD. This might result from the hydrogen bond scans, not taking into account 
enough bond angles and dihedrals of hydrogen bond system. 
The \HN hydrogen bond terms showed a good accuracy when comparing them to the full protein QM NMR calculations. 
When comparing them to the experimental chemical shifts, the correlation coefficient and RMSD worsened considerably. This
is largely a result of the ineptitude of the model for surface-exposed amide protons. Using a solvent accessible surface area
calculation might improve this, but will most likely be too slow if Procs14 is to be used to in PHAISTOS. Instead a method that calculates the number
of contact atoms in the vicinity of the amide protons could be investigated. This might be a fast enough model to be useful.

Investigation of the effect of bond length, showed that the chemical shift exhibited a strong dependence on the bond lengths.   
Currently the bond length corrections depend only on a average of bond length, therefore it would be interesting to expand the bond length
correction to depend on the individual residues.

Presently the interaction between side chain donors and backbone oxygens is always modelled with two N-methylacetamide molecules.
The negative charge on lysine, arginine and histidine will surely alter the strength of the hydrogen bond interaction.
Therefore another pathway to improving Procs14 would be to perform scans with model systems representing hydrogen bonding between
backbone and the three negatively charged amino acids. This might improve the overall bad Procs14 predictions on \C. 

An interesting approach to study the quality of Procs14 predictions would be to investigate the accuracy of Procs14
on each amino acid type. The standard for reporting chemical shift prediction in the literature, is to report the  
correlation coefficient and RMSD for all amino acids types on a single atom type. This would reveal whether or not Procs14
is able to distinguish different states of the same amino acid type.

Procs14 showed improved predictions on the NMR ensembles compared to the x-ray crystal structures.
This highlights the chemical shifts dependence on the different conformational states of the protein in solution. The empirical
chemical shift predictor methods have this effect implicit in their models. If QM derived chemical shifts is to be used 
to predict experimental chemical shifts with good accuracy, this effect has to be taken into account. This suggest that 
it would be interesting to use ensemble sampling in the refinement/folding of protein structures in PHAISTOS.  

\subsection{Conclusion}

The Procs14 chemical shift predictor decidedly beat the other QM derived predictor CheShift-2 in all tests. 
When comparing the empirical methods with Procs14 in their ability to reproduce full protein QM NMR chemical shifts, Procs14
also performed better.
The RMSD of the chemical shifts versus the experimental data, significantly improved on
NMR ensembles. This suggest that investigating ensemble sampling might be fruitful.
The Procs14 model still does not provide robust predictions on all atom types and there is therefore still much work to be done to make a truly accurate QM chemical shift predictor.         

\clearpage
\bibliographystyle{plain}
\bibliographystyle{ieeetr}
\bibliography{biblografi}

\clearpage
\section{Appendix} \label{sec:append}
\appendix
\section{Procs14}\label{appendpro}

\begin{figure}[h!]
	\centering   
                \includegraphics[width=0.7\textwidth]{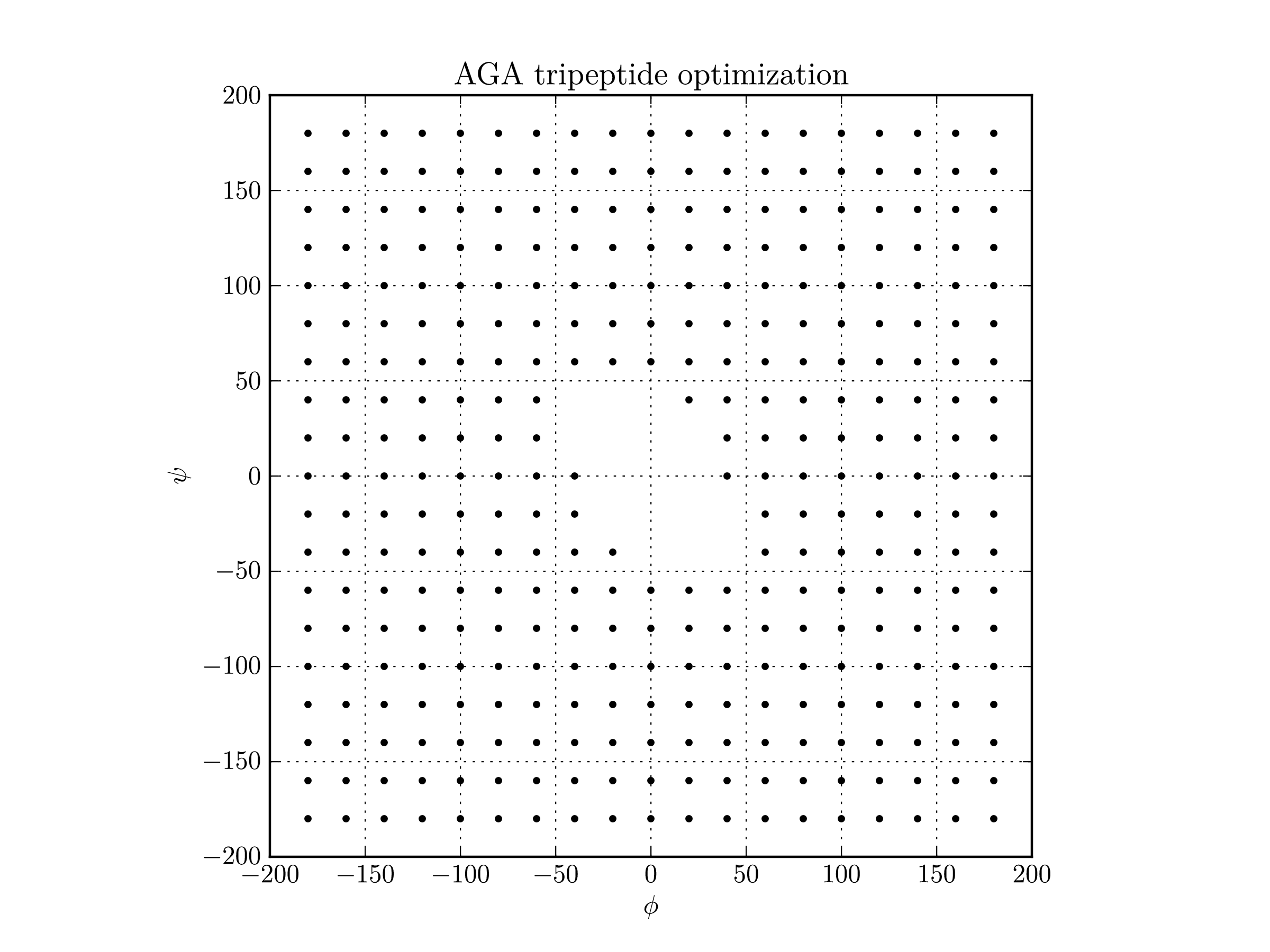}
	\label{AGAopt}
	\caption{ This figure shows which AGA tripeptides structures failed in the geometry optimization with PM6.
			  each dot represents the structures that converge successfully. The missing structures are centerd around
			  $\phi$ and $\psi$ equal to zero. This region is it self unfavored acording to typical ramachandran plots
			  for glycine.
		 }
	\label{fig:AGAopt}
	
\end{figure}	

\begin{table}[h!]
\label{aggiungi}\centering 
{\scriptsize  
\begin{tabular}{llcccc*{9}{l}}
\toprule %
                               & Raw data alpha helix   & New samples alpha helix     & Raw data Beta sheet     &  New samples Beta sheet  &\\
						       & RMSD ppm              & RMSD ppm                    & RMSD ppm               &  RMSD ppm                &\\\midrule 
 Linear           & 1.694                 & 1.629                       & 1.417                  &  1.489                   &\\		   
 Nearest-neighbor & 0.146                 & 0.964                       & 0.149                  &  0.822                  &\\\bottomrule

\end{tabular}}
\caption{ The table show a comparison between of the accuracy of linear interpolation versus nearest-neighbor interpolation. Two sets of backbone angles
          corresponding to a alpha helix and a beta sheet is used. $1000$ samples is created using BASILISK. From the samples a grid is interpolated using either
          nearest-neighbor or linear Interpolation. From the grids linear interpolation is used to give the final chemical shift prediction. The grids are compared to the raw data used in the interpolation, column $1$ and $3$. Column $2$ and $4$ is the comparison between
		  the grids and $1000$ new samples made with BASILISK.		  
		  The values of the columns are the RMSD in ppm between the grid and the samples. The nearest-neighbor interpolation is by far the most accurate.  
		 }
	\label{table:6dlysineinterpoaltion}
\end{table}

\begin{table}[h!]
\label{aggiungi}\centering 
{\scriptsize  
\begin{tabular}{lcccccc*{9}{l}}
\toprule %
                        &  ${}^{13}C\alpha$ & ${}^{13}C\beta$   &  ${}^{13}C'$     & ${}^{15}N^{H}$ & ${}^{1}H^{N}$  & ${}^{1}H\alpha$  &\\  
Model system            & (\textit{ppm})     & (\textit{ppm})   &  (\textit{ppm})  & (\textit{ppm}) & (\textit{ppm}) & (\textit{ppm})   &\\\midrule	
 NMA NMA COH            & $0.31$             & $0.0$            &  $-0.6$          & $-4.48$        & $-0.57$        & $-0.49$          &\\	
 NMA COO NMA            & $0.29$             & $0.0$            &  $-1.16$         & $-2.51$        & $-0.54$        & $0.03 $          &\\
 NMA COO COH            & $0.0 $             & $0.0$            &  $0.15$          & $-5.38$        & $-0.36$        & $0.0$            &\\\bottomrule
\end{tabular}}
\caption{ Shows the corrections found by modeling hydrogen bonding with multiple acceptor/donors. NMA NMA COH is the system with an alcohol hydrogen bonding
	  in addtional to the amide proton. NMA COO NMA is bonding to two acceptors carbonyl oxygen and carboxylate oxygen. NMA COO COH is bonding to two acceptors
	  carbonyl oxygen and alcohol oxygen.		 }
	\label{table:hbondcorr}
\end{table}

\clearpage
\section{Benchmarking}\label{appendbench}

\begin{figure}[h!]
	\centering
	  \begin{subfigure}[b]{0.75\textwidth}
                \centering
                \includegraphics[width=1.0\textwidth]{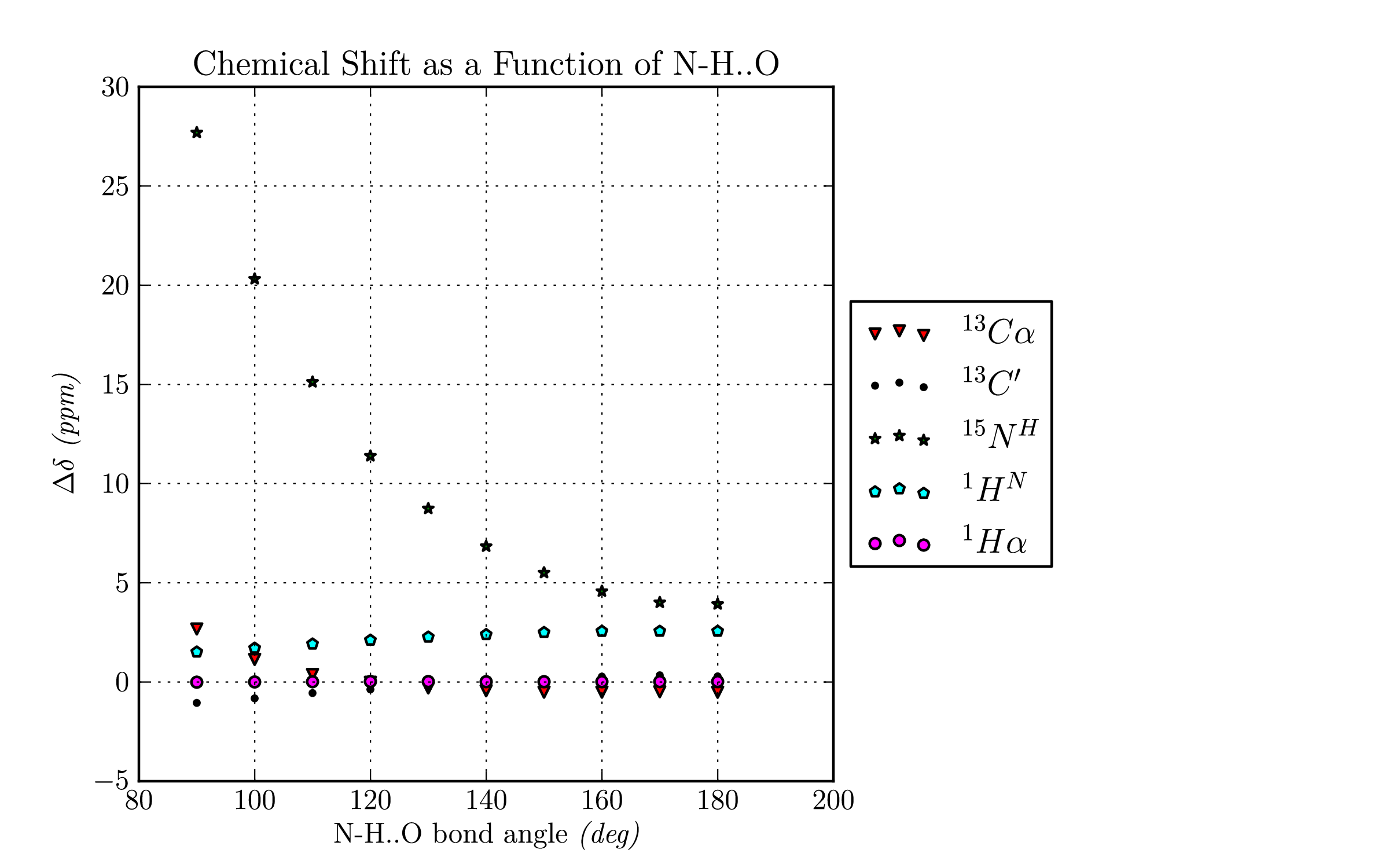}
                \caption{\small{${}^{1}H^{N}$ Chemical Shift N-H..O Dependency}}
                \label{fig:theta1hn}
     \end{subfigure} \\
	\begin{subfigure}[b]{0.75\textwidth}
                \centering
                \includegraphics[width=1.0\textwidth]{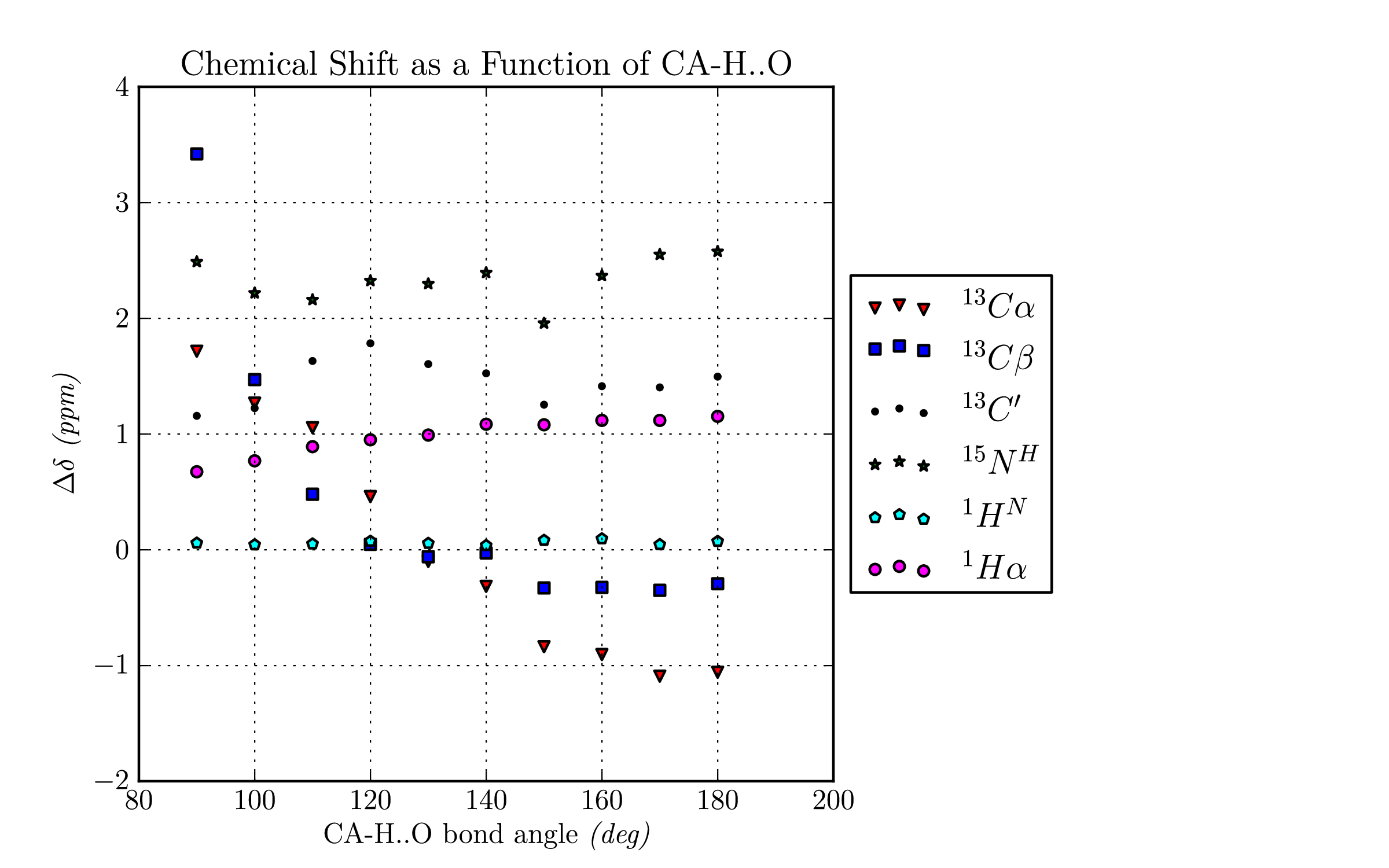}
                \caption{\small{${}^{1}H\alpha$ Chemical Shift C\(\alpha\)-H\(\alpha\)..O  Dependency}}
                \label{fig:theta1ha}
     \end{subfigure} 

	\caption{ {\bf Chemical Shift Dependence on the N-H..O and C\(\alpha\)-H\(\alpha\)..O Bond Angles.} The figure shows the change in chemical shift as an function of the 
                  N-H..O and C\(\alpha\)-H\(\alpha\)..O bond angles. {\bf (a)} shows the results of an scan over the N-H..O bond angle, in the model system seen in
                  \fref{fig:hba}. Especially ${}^{15}N^{H}$ and ${}^{13}C\alpha$ exhibits the largest change in chemical shift. {\bf (b)} is an scan over the C\(\alpha\)-H\(\alpha\)..O Bond Angle on the model system in \fref{fig:haba}.
                  ${}^{13}C\alpha$ and ${}^{13}C\beta$ shows strong dependence on the bond angle. In both scans the model systems had the remaining bond angles and lengths fixed. The $\Delta\delta$ chemical shifts
				  is the change from a model system without hydrogen bonding.
                  } 

	\label{fig:theta1}
\end{figure}

\begin{table}[h]
\label{aggiungi}\centering 
{\scriptsize  
\begin{tabular}{lcccccc*{9}{l}}
\toprule %
                                          &  ${}^{13}C\alpha$ & ${}^{13}C\beta$   &  ${}^{13}C'$     & ${}^{15}N^{H}$ & ${}^{1}H^{N}$  & ${}^{1}H\alpha$  &\\\midrule
$\Delta\delta_{2\deg H\alpha}$            &                    &                  &  X               & X              & X              &                  &\\	
$\Delta\delta_{1\deg H\alpha}$            &                    &                  &  X               & X              &                & X                &\\	
$\Delta\delta_{2\deg HB}$                 &                    &                  &  X               & X              & X              & X                 &\\	
$\Delta\delta_{1\deg HB}$                 &                    &                  &                  & X              & X              &                  &\\\bottomrule
\end{tabular}}
\caption{ {\bf Hydrogen Bond Terms Used in Procs14.} This tables shows for which atom types the hydrogen bond terms is used. An X indicate that the term is used for the
	  specific atom type.
		 }
	\label{table:hbondtermsused}
\end{table}

\clearpage
\section{Glycine Hypersurfaces}\label{appendhyper}

\begin{figure}[h!]
  \hspace{-15 mm}
  \centering
  \begin{minipage}[b]{0.49\linewidth}
    \includegraphics[width=\linewidth]{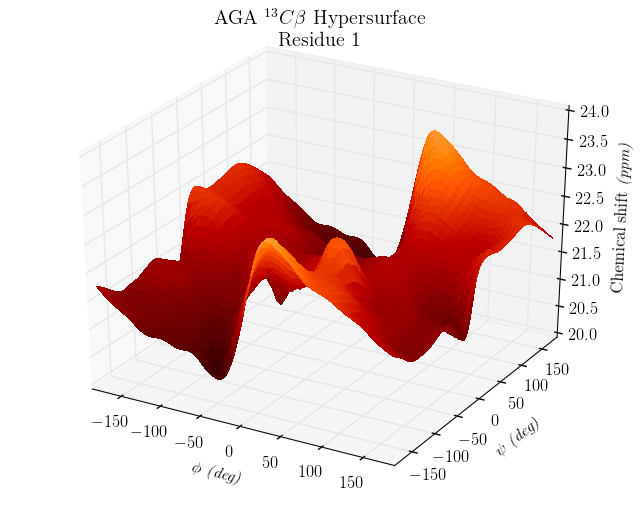}
  \end{minipage}
  \begin{minipage}[b]{0.49\linewidth}
    \includegraphics[width=\linewidth]{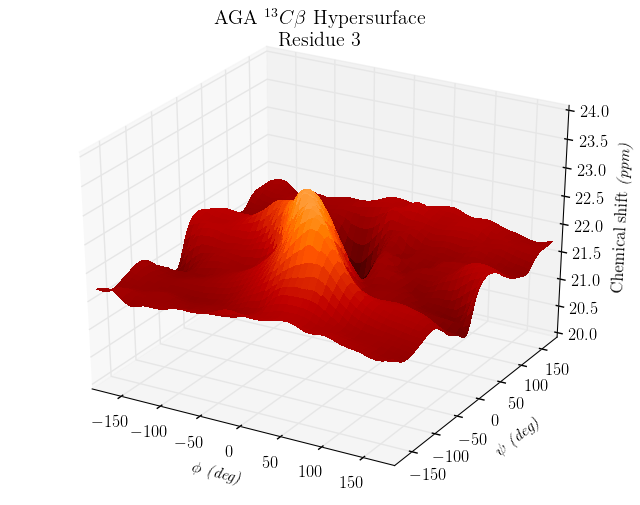}
  \end{minipage} \\
  \hspace{-15 mm}
  \begin{minipage}[b]{0.49\linewidth}
    \includegraphics[width=\linewidth]{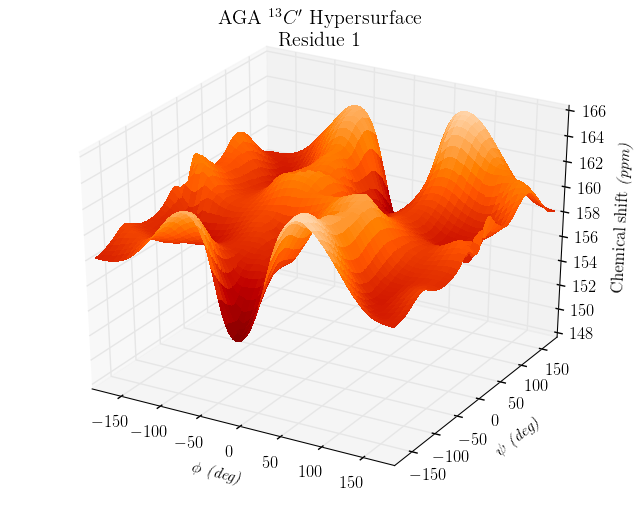}
  \end{minipage}
  \begin{minipage}[b]{0.49\linewidth}
    \includegraphics[width=\linewidth]{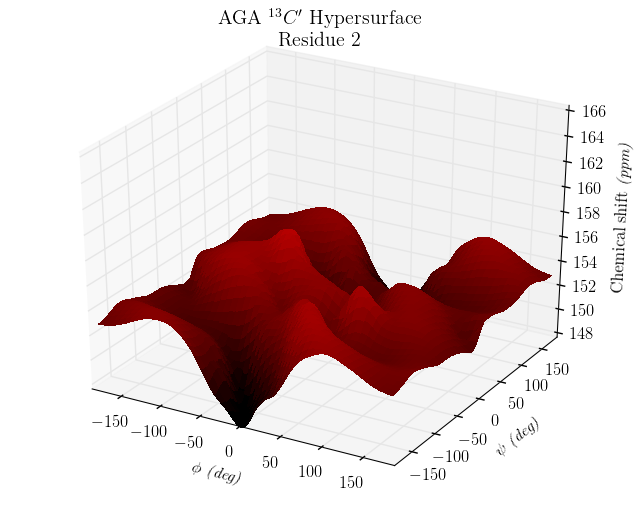}
  \end{minipage}\\

  \begin{minipage}[b]{0.49\linewidth}
    \includegraphics[width=\linewidth]{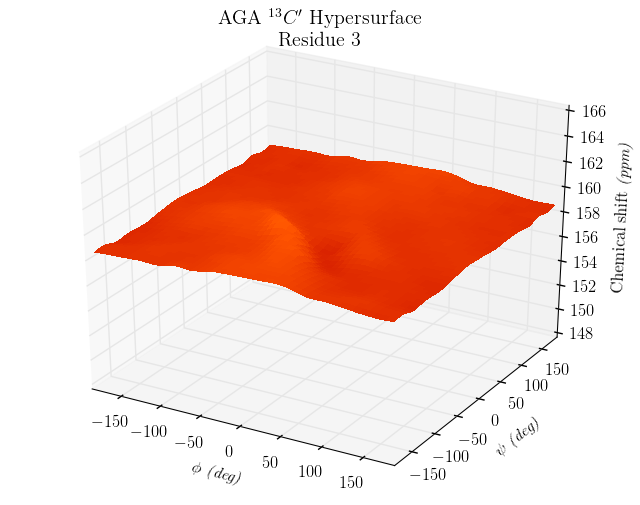}
  \end{minipage}
\caption{{\bf \CB   and \C   Glycine Hypersurfaces.}
		 }
	\label{fig:agahyper2}
\end{figure}

\begin{figure}[h!]
   \hspace{-15 mm}
   \centering
  \begin{minipage}[b]{0.48\linewidth}
    \includegraphics[width=\linewidth]{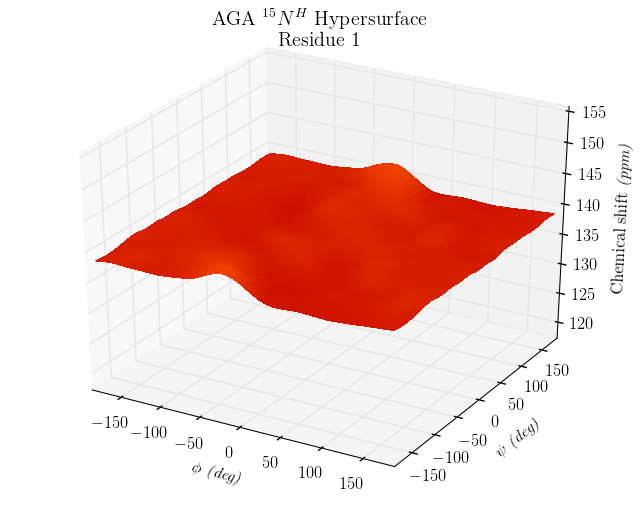}
  \end{minipage}
  \begin{minipage}[b]{0.48\linewidth}
    \includegraphics[width=\linewidth]{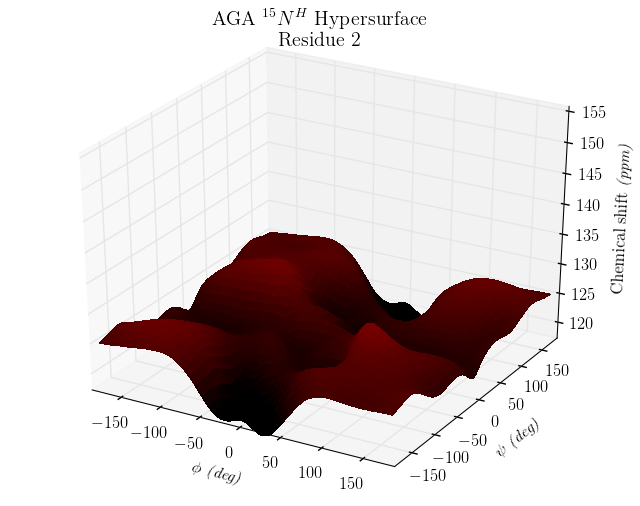}
  \end{minipage} \\
  \hspace{-15 mm}
  \begin{minipage}[b]{0.48\linewidth}
    \includegraphics[width=\linewidth]{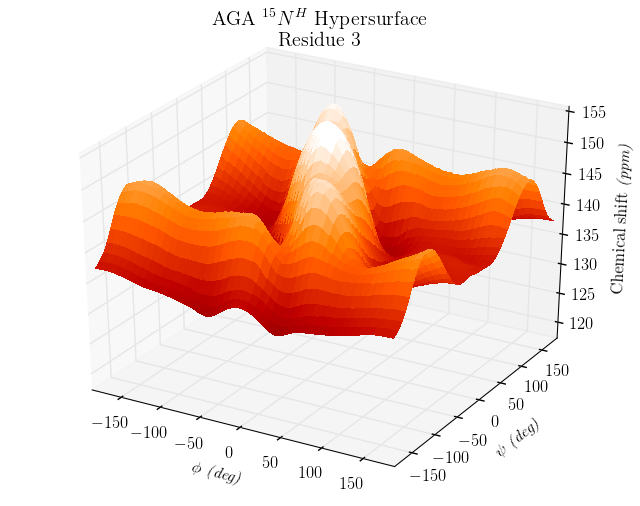}
  \end{minipage}
  \begin{minipage}[b]{0.48\linewidth}
    \includegraphics[width=\linewidth]{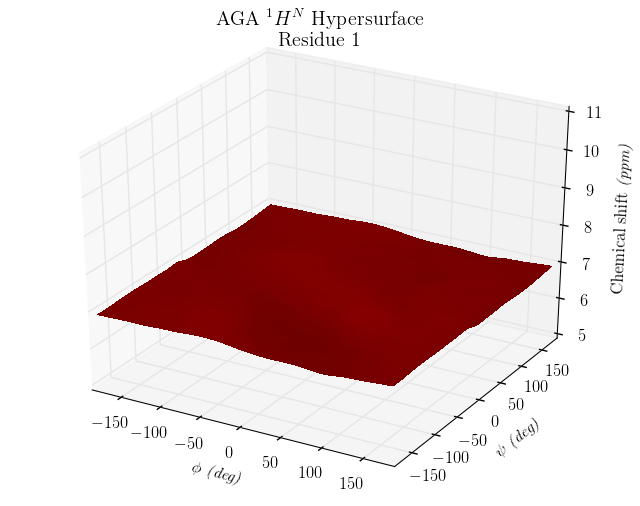}
  \end{minipage}\\
  \hspace{-15 mm}
  \begin{minipage}[b]{0.48\linewidth}
    \includegraphics[width=\linewidth]{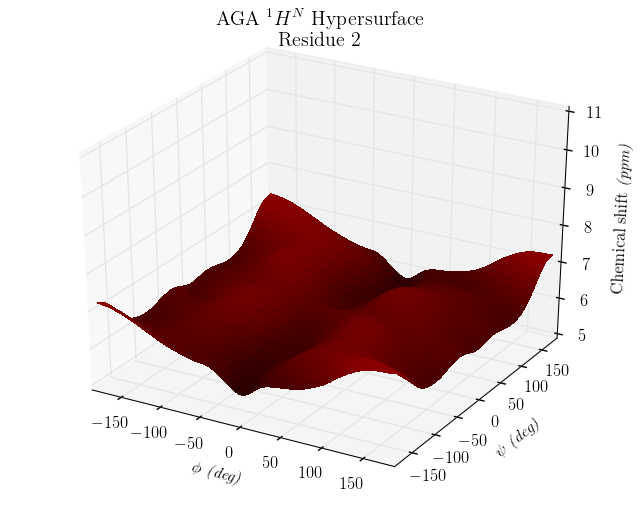}
  \end{minipage}
  \begin{minipage}[b]{0.48\linewidth}
    \includegraphics[width=\linewidth]{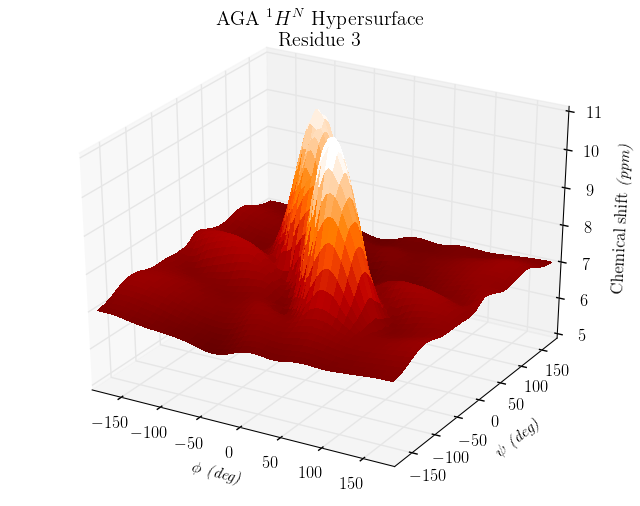}
  \end{minipage}
\caption{{\bf \N   and \HN   Glycine Hypersurfaces.}
		 }
	\label{fig:agahyper3}
\end{figure}

\newpage

\begin{figure}[h!]
   \hspace{-15 mm}
  \begin{minipage}[b]{0.48\linewidth}
    \centering
    \includegraphics[width=\linewidth]{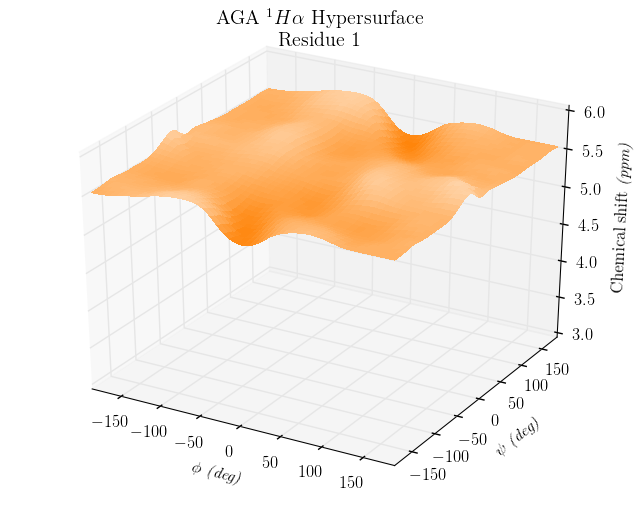}
  \end{minipage}
  \begin{minipage}[b]{0.48\linewidth}
    \centering
    \includegraphics[width=\linewidth]{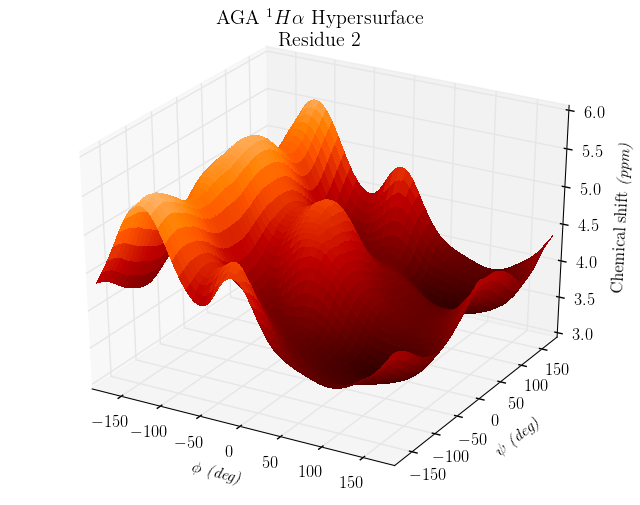}
  \end{minipage} \\
  \centering
  \begin{minipage}[b]{0.48\linewidth}    
     \includegraphics[width=\linewidth]{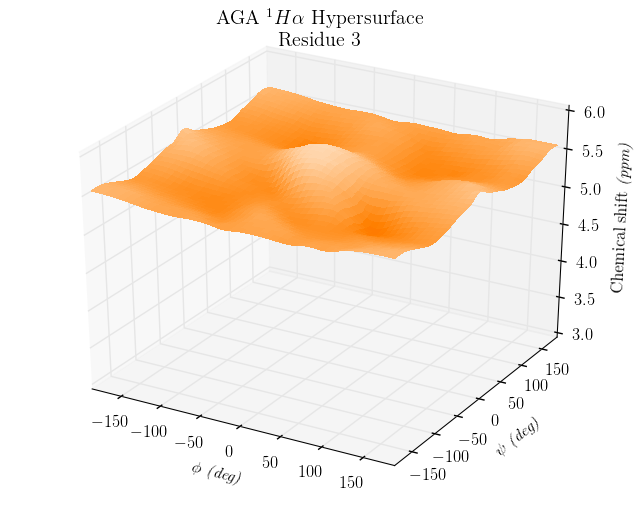}
  \end{minipage}
\caption{ {\bf \HA   Glycine Hypersurfaces.}
		 }
	\label{fig:agahyper4}
\end{figure}
\newpage
\clearpage
\section{Tripeptides}\label{tripep}

\begin{figure}[h!]
	\centering \hspace{-8 mm}
	  \begin{subfigure}[b]{0.49\textwidth}
                \centering
                \includegraphics[width=\textwidth]{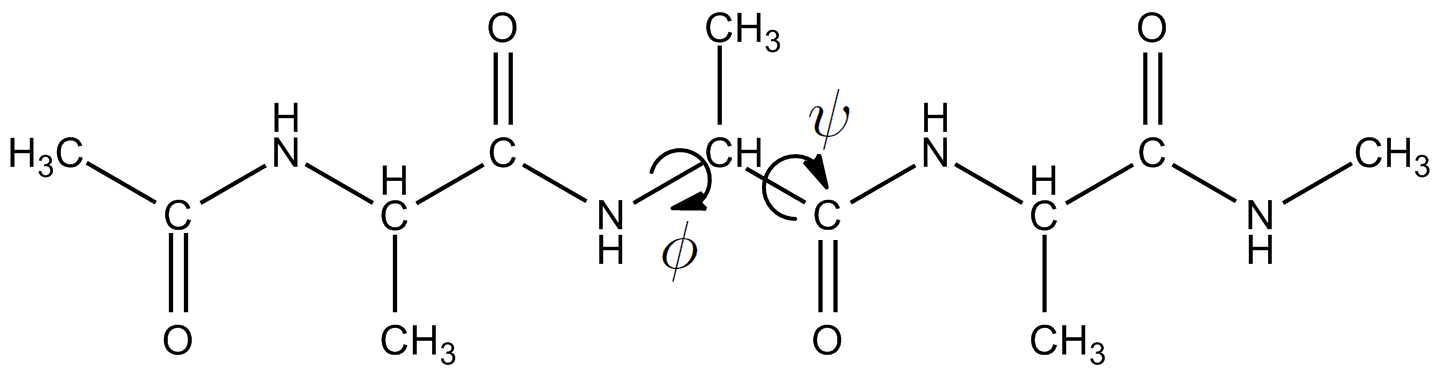}
                \caption{\small{AAA Alanine }}
                \label{fig:1a}
     \end{subfigure} \hspace{4 mm}
	\begin{subfigure}[b]{0.49\textwidth}
                \centering
                \includegraphics[width=\textwidth]{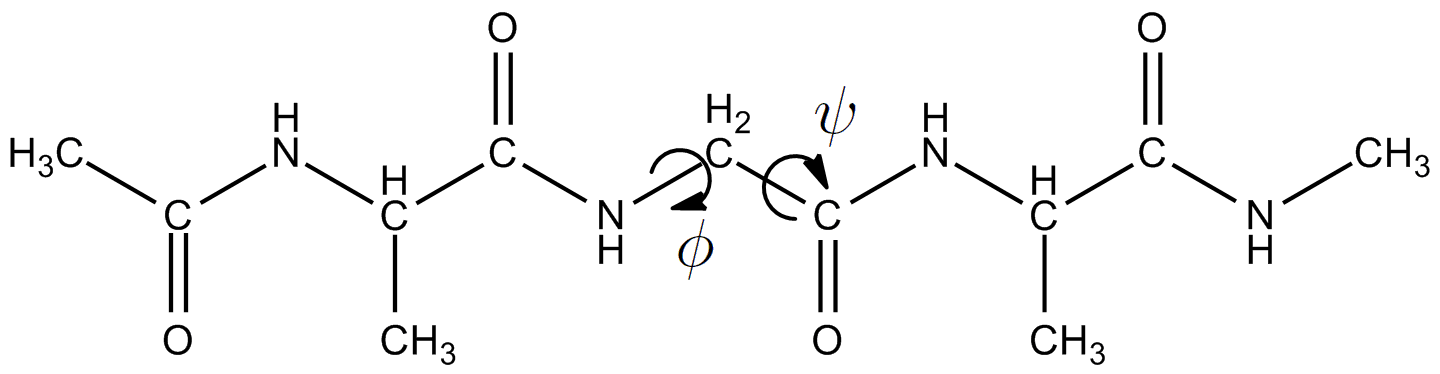}
                \caption{\small{AGA Glycine}}
                \label{fig:1b}
     \end{subfigure} \\ 
	 \vspace{9 mm}
	 \centering \hspace{-8 mm}\begin{subfigure}[b]{0.49\textwidth}
                \centering
                \includegraphics[width=\textwidth]{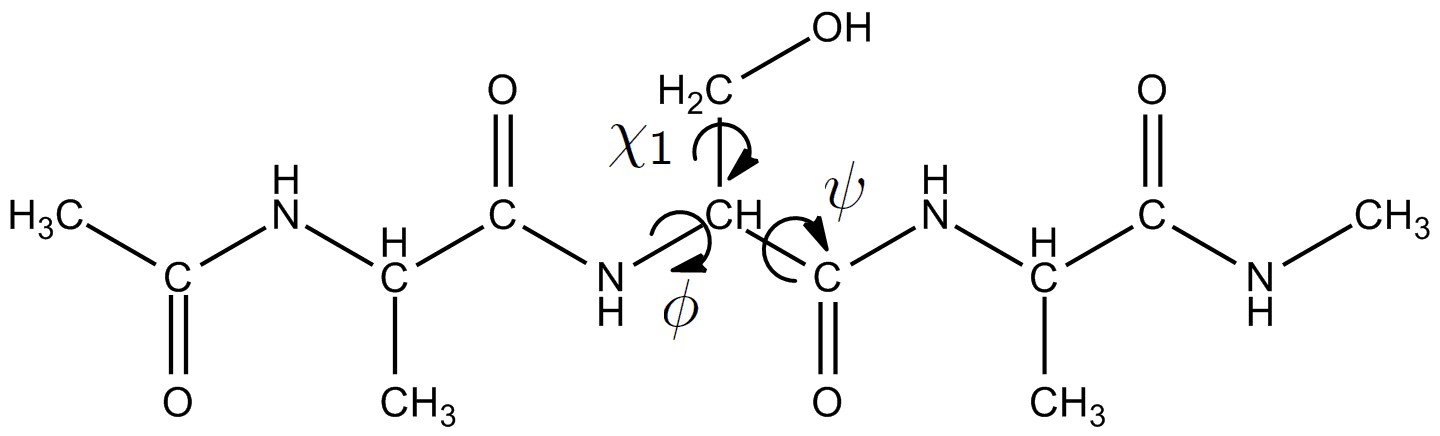}
                \caption{\small{ASA Serine}}
                \label{fig:1c}
     \end{subfigure} \hspace{4 mm}
	 \begin{subfigure}[b]{0.49\textwidth}
                \centering
                \includegraphics[width=\textwidth]{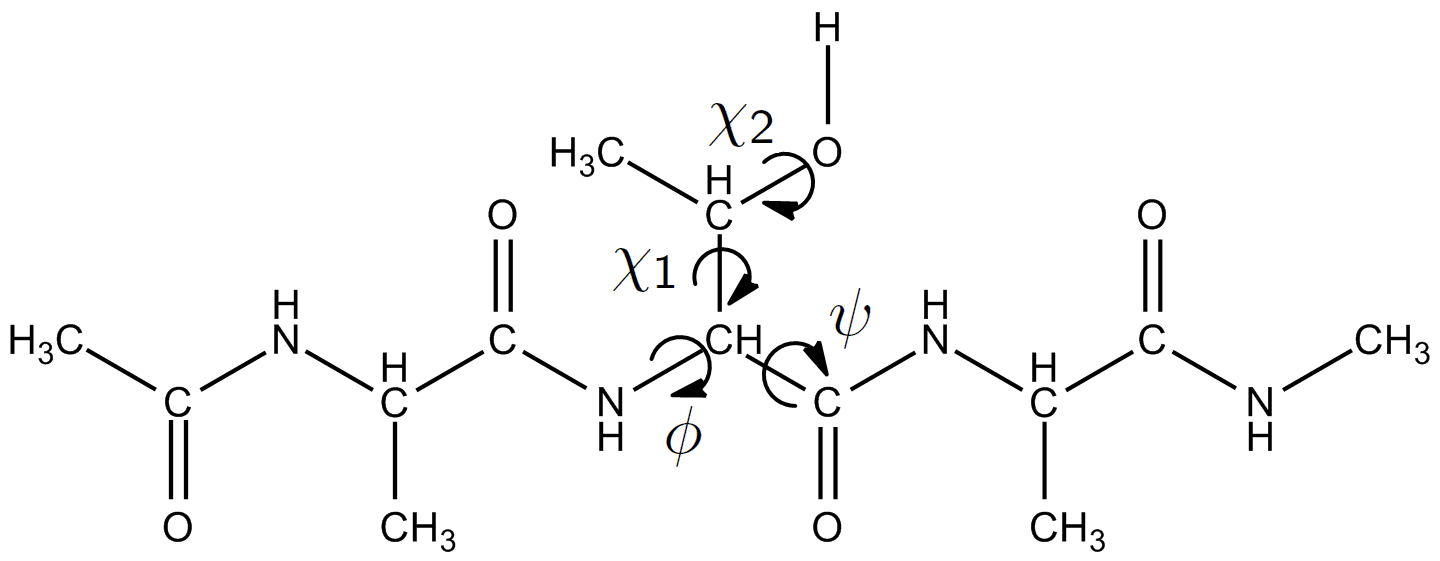}
                \caption{\small{ATA Threoine}}
                \label{fig:1d}
     \end{subfigure} \\
	 \vspace{9 mm}
	 \centering \hspace{-8 mm}\begin{subfigure}[b]{0.49\textwidth}
                \centering
                \includegraphics[width=\textwidth]{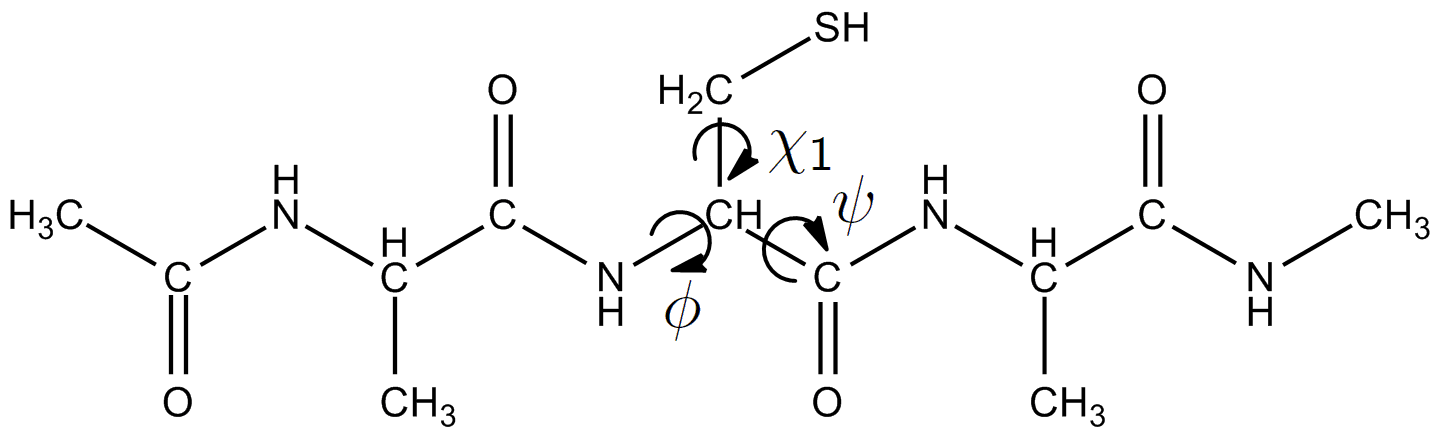}
                \caption{\small{ACA Cysteine}}
                \label{fig:1e}
     \end{subfigure} \hspace{4 mm}
	 \begin{subfigure}[b]{0.49\textwidth}
                \centering
                \includegraphics[width=\textwidth]{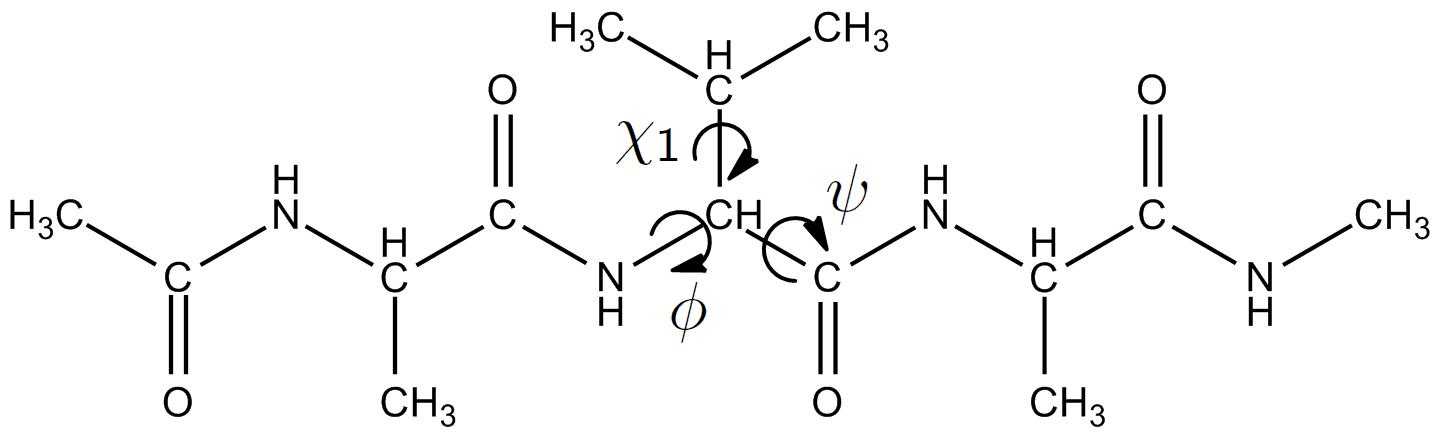}
                \caption{\small{AVA Valine}}
                \label{fig:1f}
     \end{subfigure} \\	 
	 \vspace{9 mm}
	\centering \hspace{-8 mm} \begin{subfigure}[b]{0.49\textwidth}
                \centering
                \includegraphics[width=\textwidth]{CYS.png}
                \caption{\small{ACA Cysteine}}
                \label{fig:1g}
     \end{subfigure} \hspace{4 mm}
	 \begin{subfigure}[b]{0.49\textwidth}
                \centering
                \includegraphics[width=\textwidth]{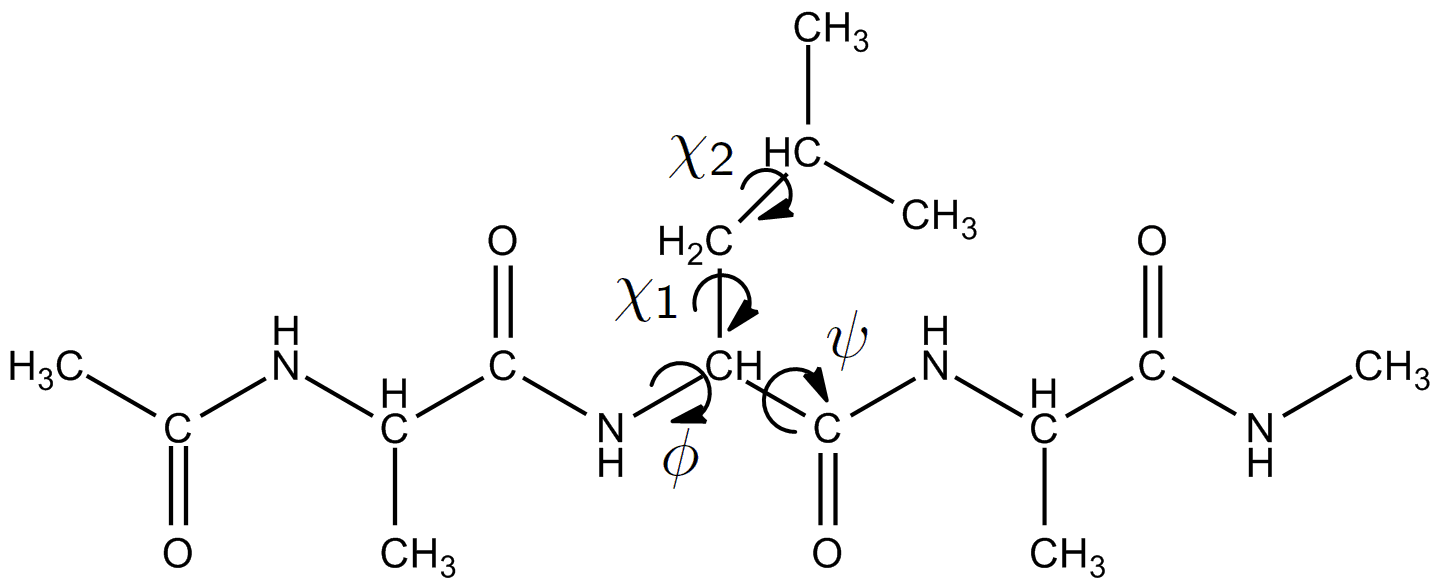}
                \caption{\small{ALA Leucine}}
                \label{fig:1h}
     \end{subfigure} \\	 
	\label{Tripeptides}
\end{figure}		 
\newpage
	 
\begin{figure}[h!]
	\centering \hspace{-8 mm}
	  \begin{subfigure}[b]{0.49\textwidth}
                \centering
                \includegraphics[width=\textwidth]{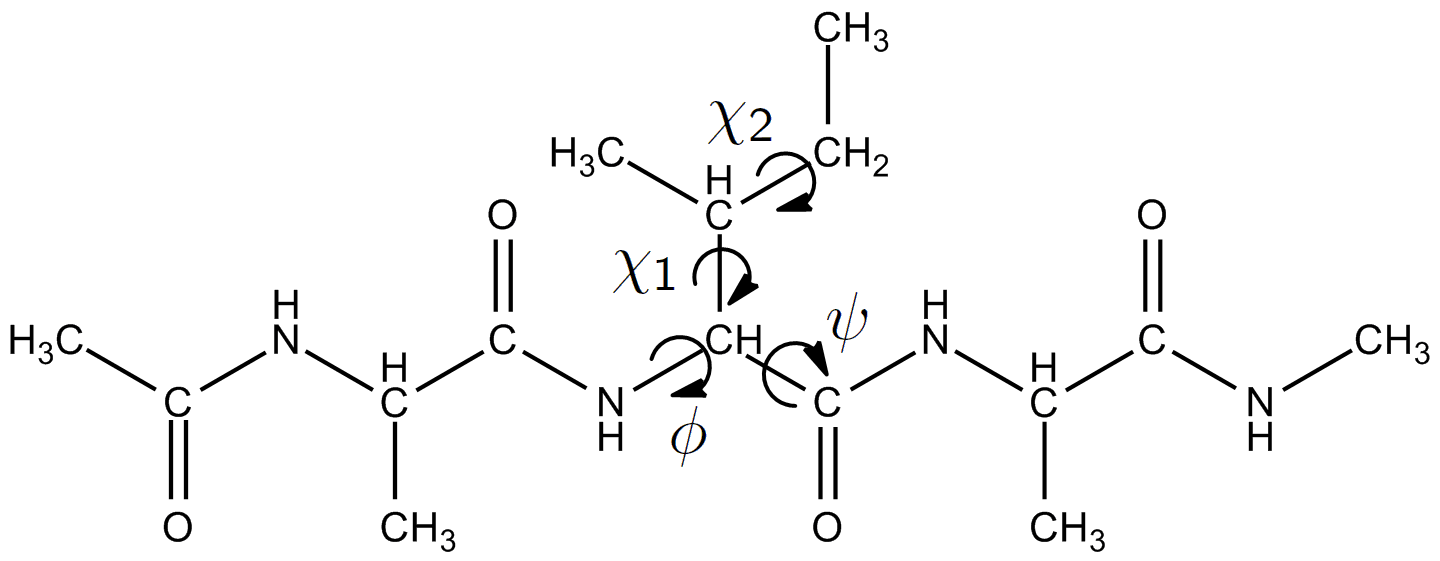}
                \caption{\small{AIA Isoleucine }}
                \label{fig:1a}
     \end{subfigure} \hspace{4 mm}
	\begin{subfigure}[b]{0.49\textwidth}
                \centering
                \includegraphics[width=\textwidth]{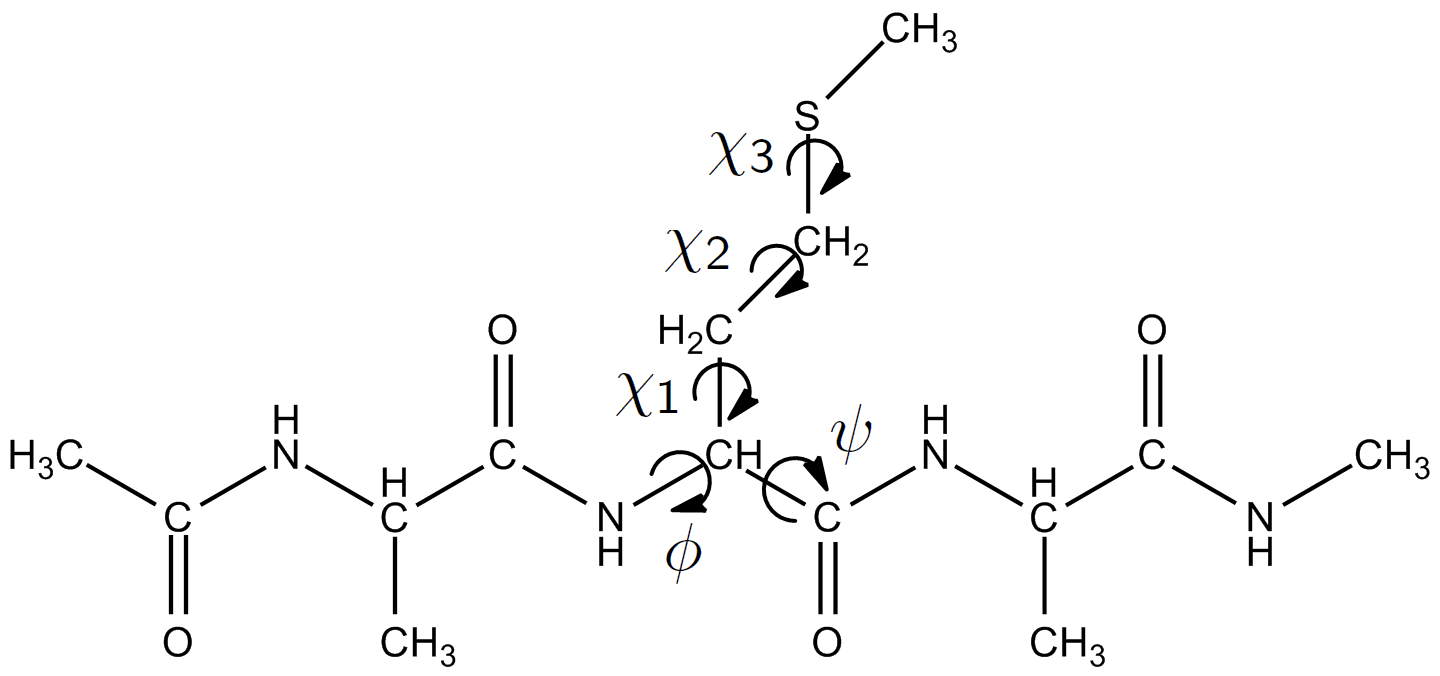}
                \caption{\small{AMA Methionine}}
                \label{fig:1b}
     \end{subfigure} \\ 
	 \vspace{9 mm}
	 \centering \hspace{-8 mm}\begin{subfigure}[b]{0.49\textwidth}
                \centering
                \includegraphics[width=\textwidth]{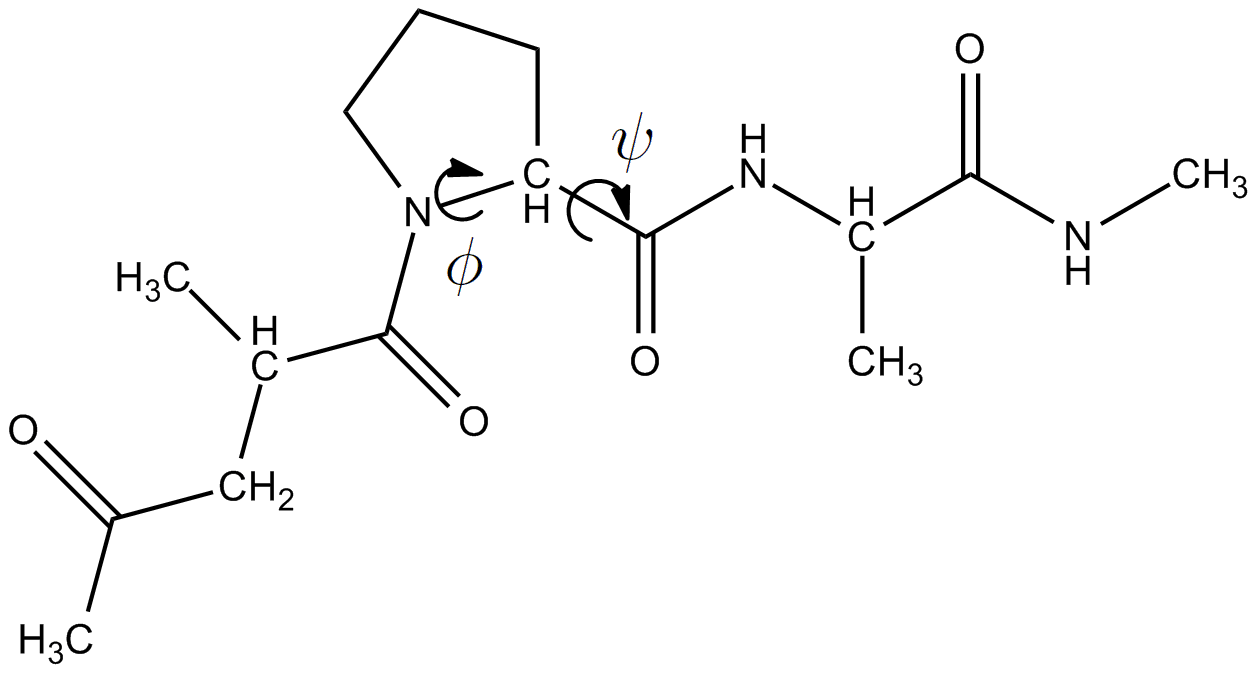}
                \caption{\small{APA Proline}}
                \label{fig:1c}
     \end{subfigure} \hspace{4 mm}
	 \begin{subfigure}[b]{0.49\textwidth}
                \centering
                \includegraphics[width=\textwidth]{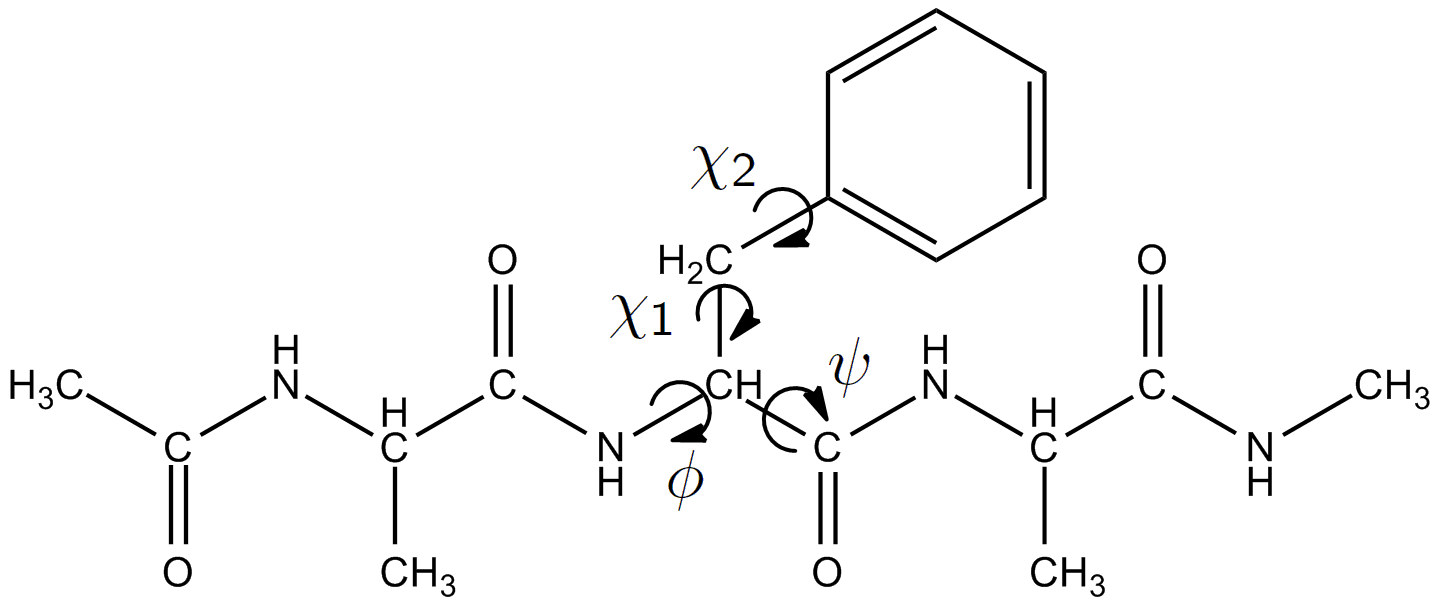}
                \caption{\small{AFA Phenylalanine}}
                \label{fig:1d}
     \end{subfigure} \\
	 \vspace{9 mm}
	 \centering \hspace{-8 mm}\begin{subfigure}[b]{0.49\textwidth}
                \centering
                \includegraphics[width=\textwidth]{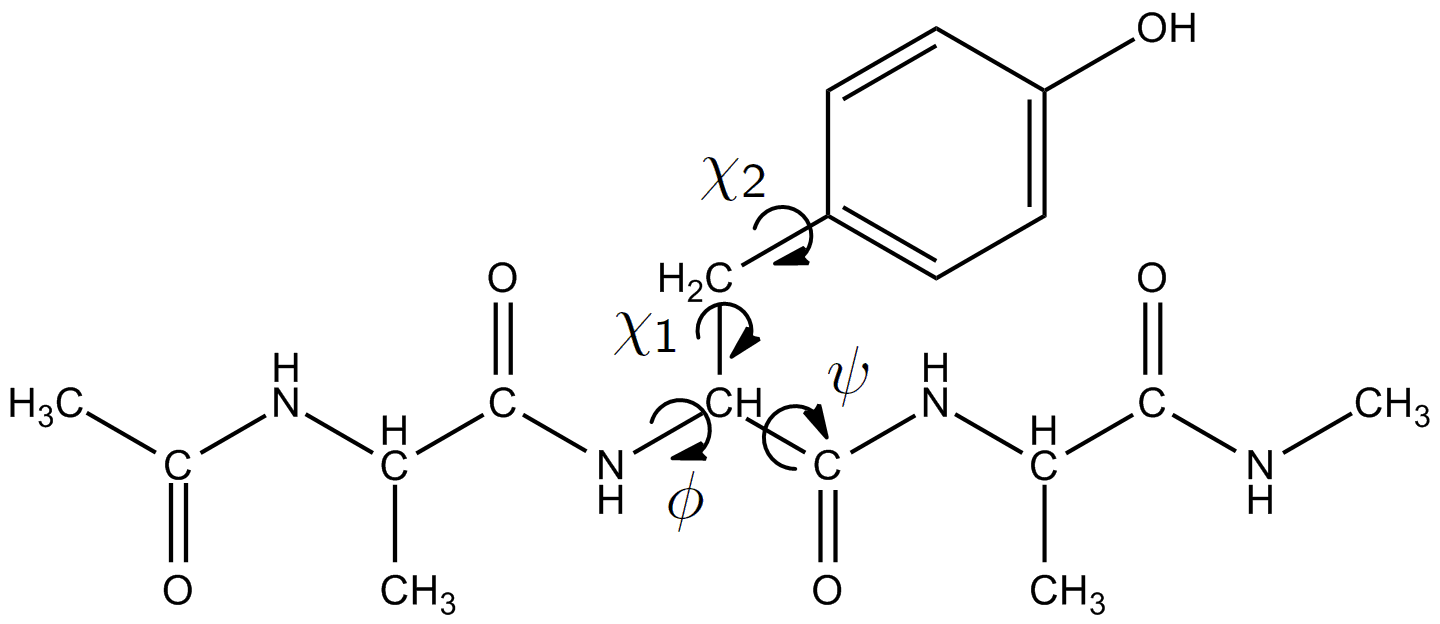}
                \caption{\small{AYA Tyrosine}}
                \label{fig:1e}
     \end{subfigure} \hspace{4 mm}
	 \begin{subfigure}[b]{0.49\textwidth}
                \centering
                \includegraphics[width=\textwidth]{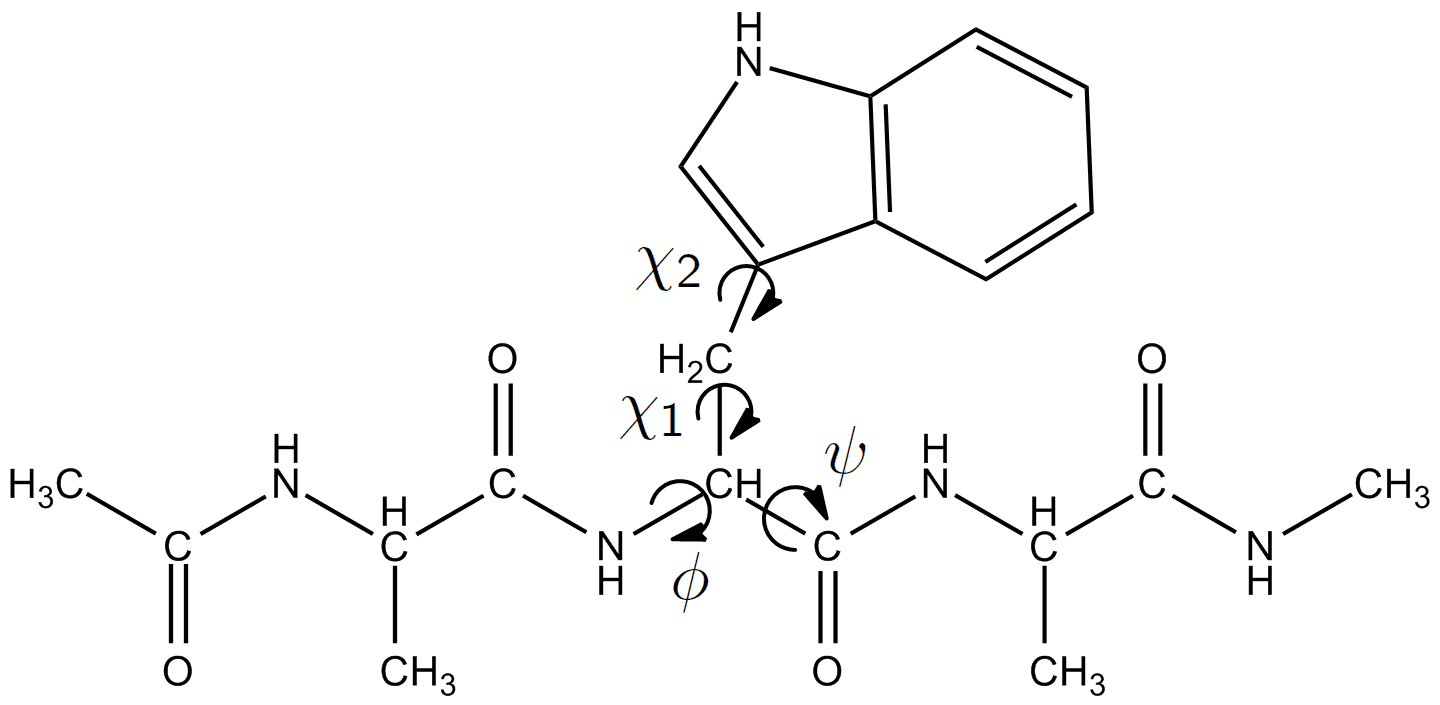}
                \caption{\small{AWA Tryptophan}}
                \label{fig:1f}
     \end{subfigure} \\	 
	 \vspace{9 mm}
	\centering \hspace{-8 mm} \begin{subfigure}[b]{0.49\textwidth}
                \centering
                \includegraphics[width=\textwidth]{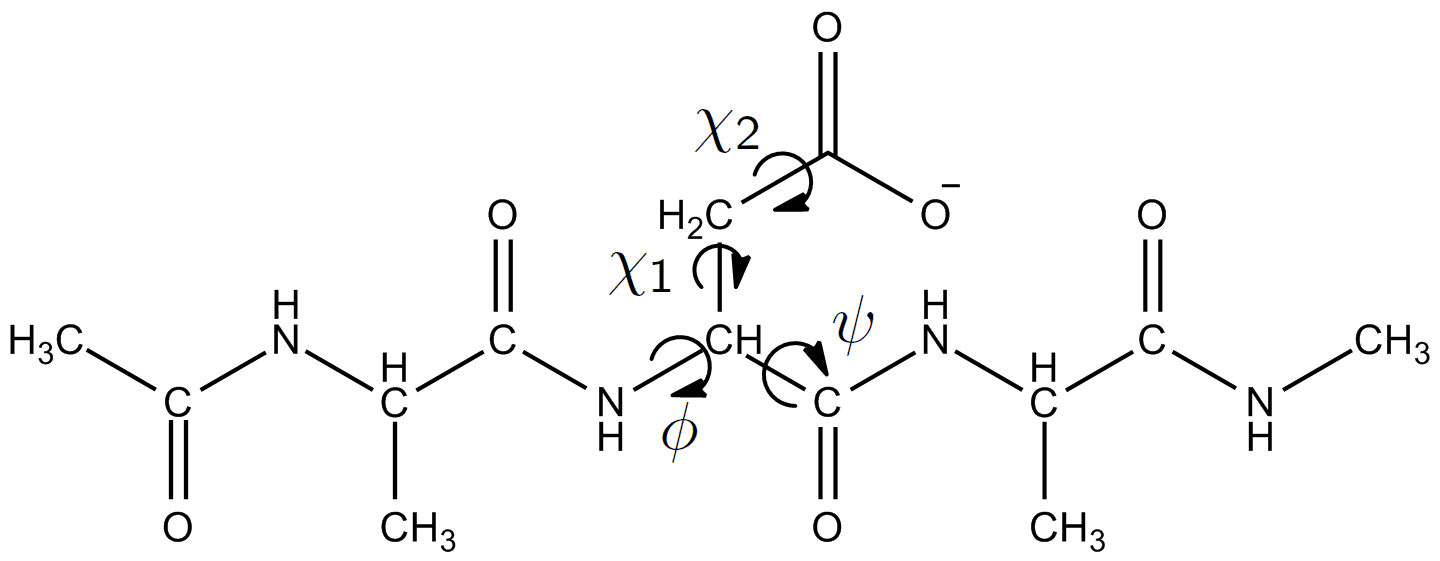}
                \caption{\small{ADA Aspartic Acid}}
                \label{fig:1g}
     \end{subfigure} \hspace{4 mm}
	 \begin{subfigure}[b]{0.49\textwidth}
                \centering
                \includegraphics[width=\textwidth]{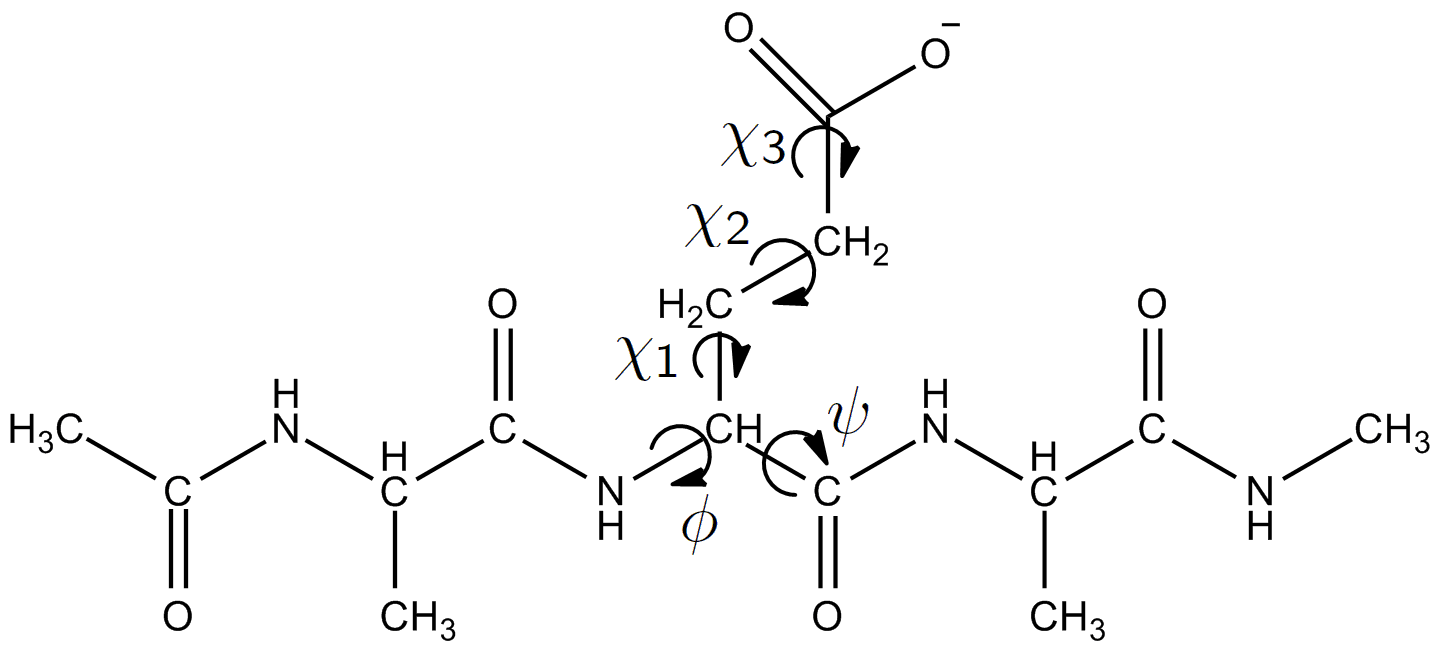}
                \caption{\small{AEA Glutamic Acid}}
                \label{fig:1h}
     \end{subfigure} \\	 
	\label{Tripeptides2}
\end{figure}

\newpage
	 
\begin{figure}[h!]
	\centering \hspace{-8 mm}
	  \begin{subfigure}[b]{0.49\textwidth}
                \centering
                \includegraphics[width=\textwidth]{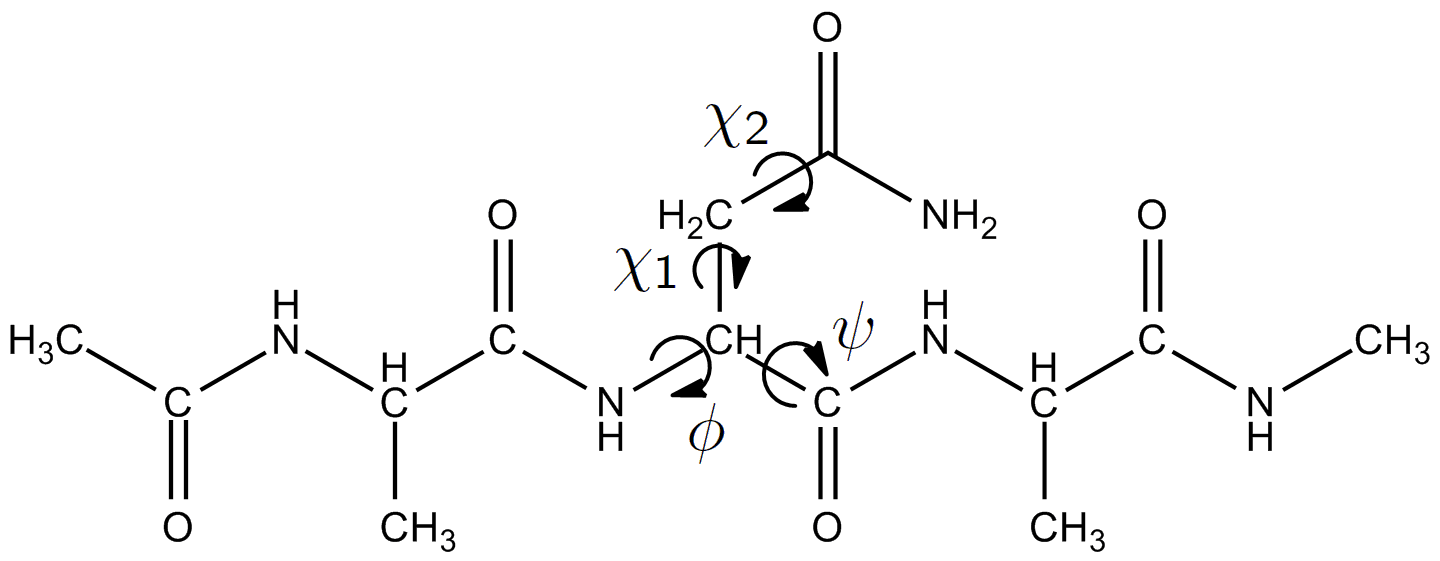}
                \caption{\small{ANA Asparagine }}
                \label{fig:1a}
     \end{subfigure} \hspace{4 mm}
	\begin{subfigure}[b]{0.49\textwidth}
                \centering
                \includegraphics[width=\textwidth]{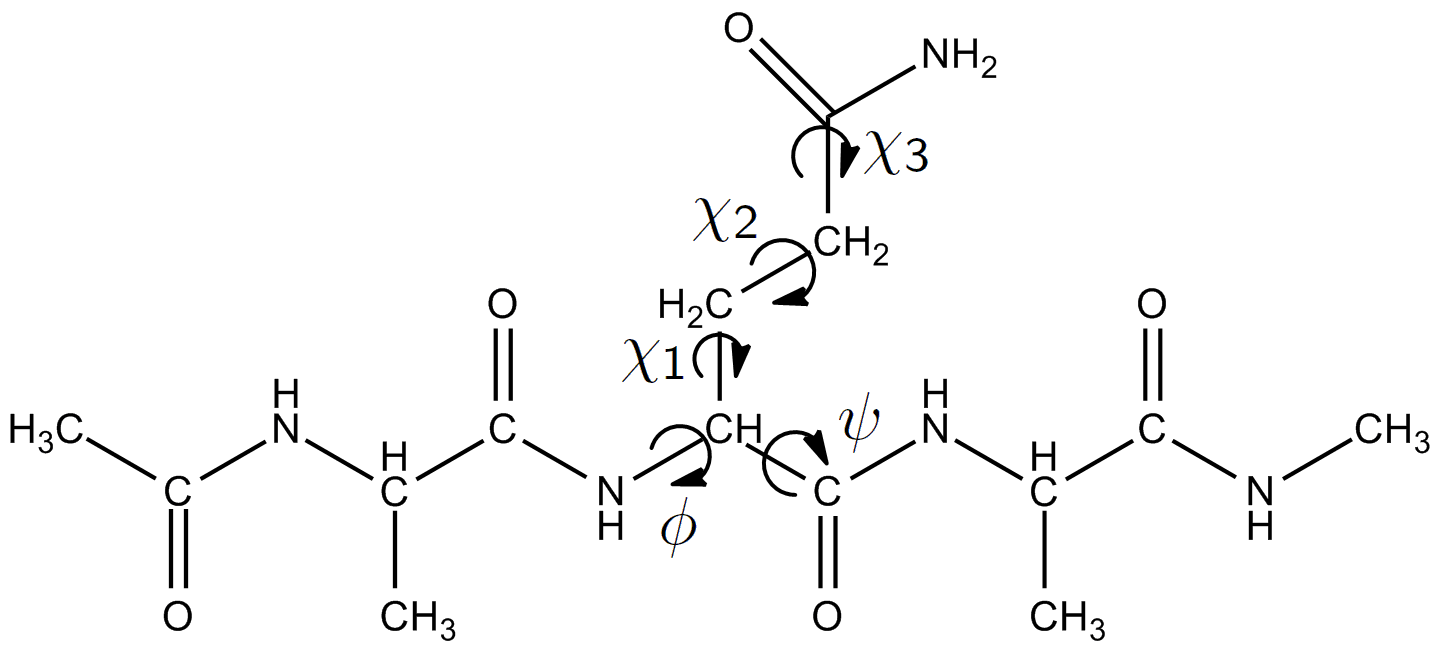}
                \caption{\small{AQA Glutamine}}
                \label{fig:1b}
     \end{subfigure} \\ 
	 \vspace{9 mm}
	 \centering \hspace{-8 mm}\begin{subfigure}[b]{0.49\textwidth}
                \centering
                \includegraphics[width=\textwidth]{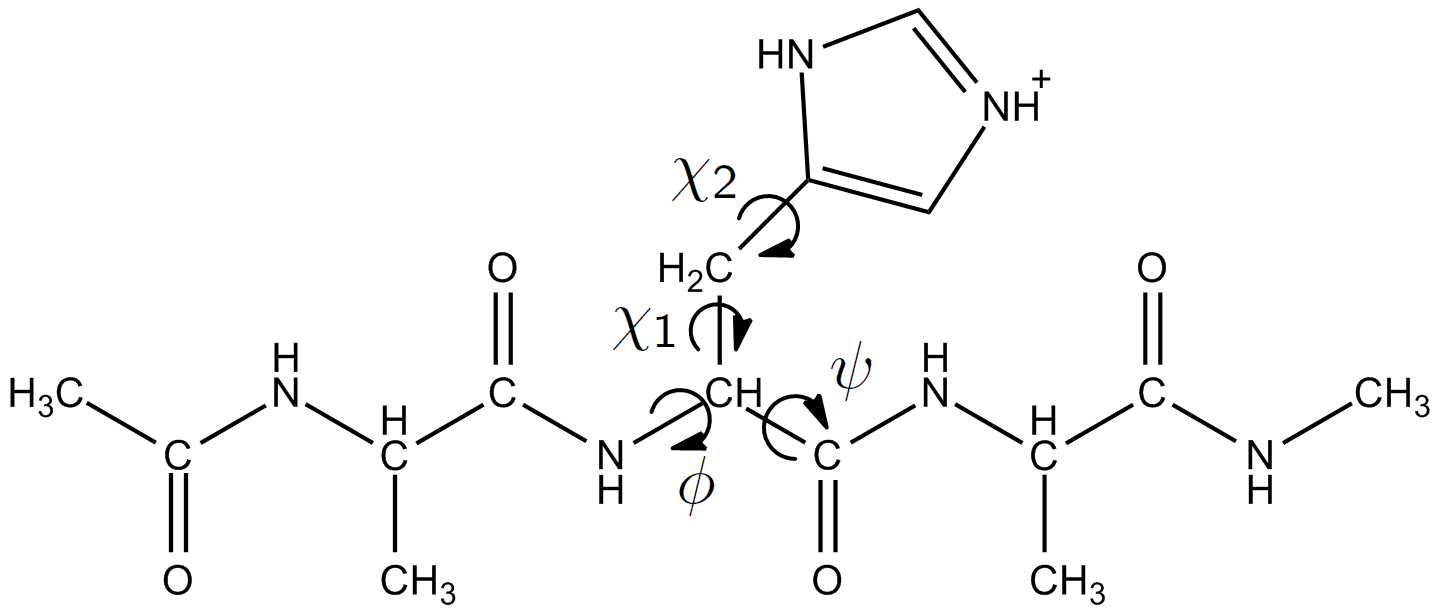}
                \caption{\small{AHA Histidine}}
                \label{fig:1c}
     \end{subfigure} \\ 
	 \begin{subfigure}[b]{0.49\textwidth}
                \centering
                \includegraphics[width=\textwidth]{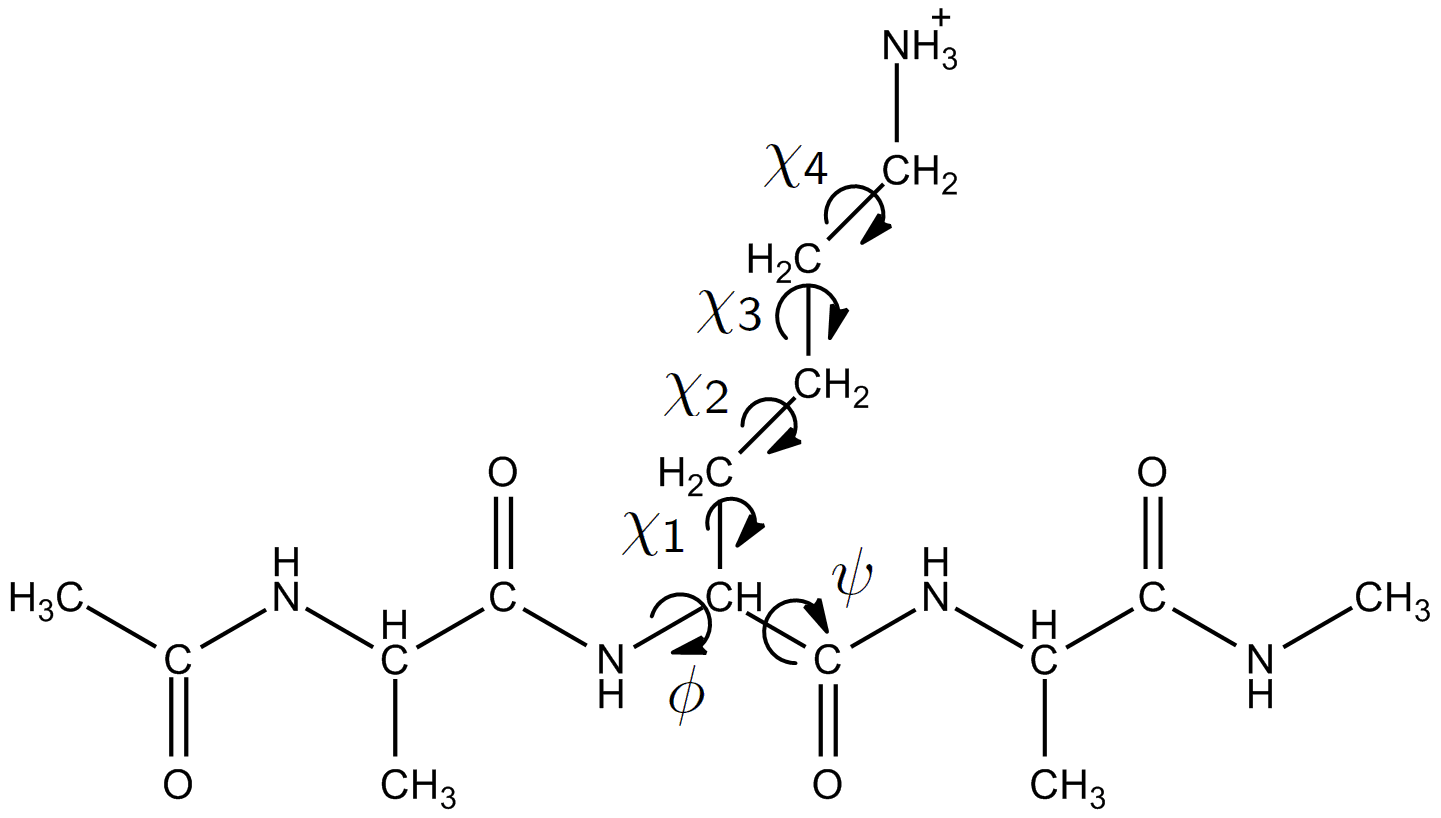}
                \caption{\small{AKA Lysine}}
                \label{fig:1d}
     \end{subfigure} \\
	 \vspace{9 mm}
	 \centering \hspace{-8 mm}\begin{subfigure}[b]{0.49\textwidth}
                \centering
                \includegraphics[width=\textwidth]{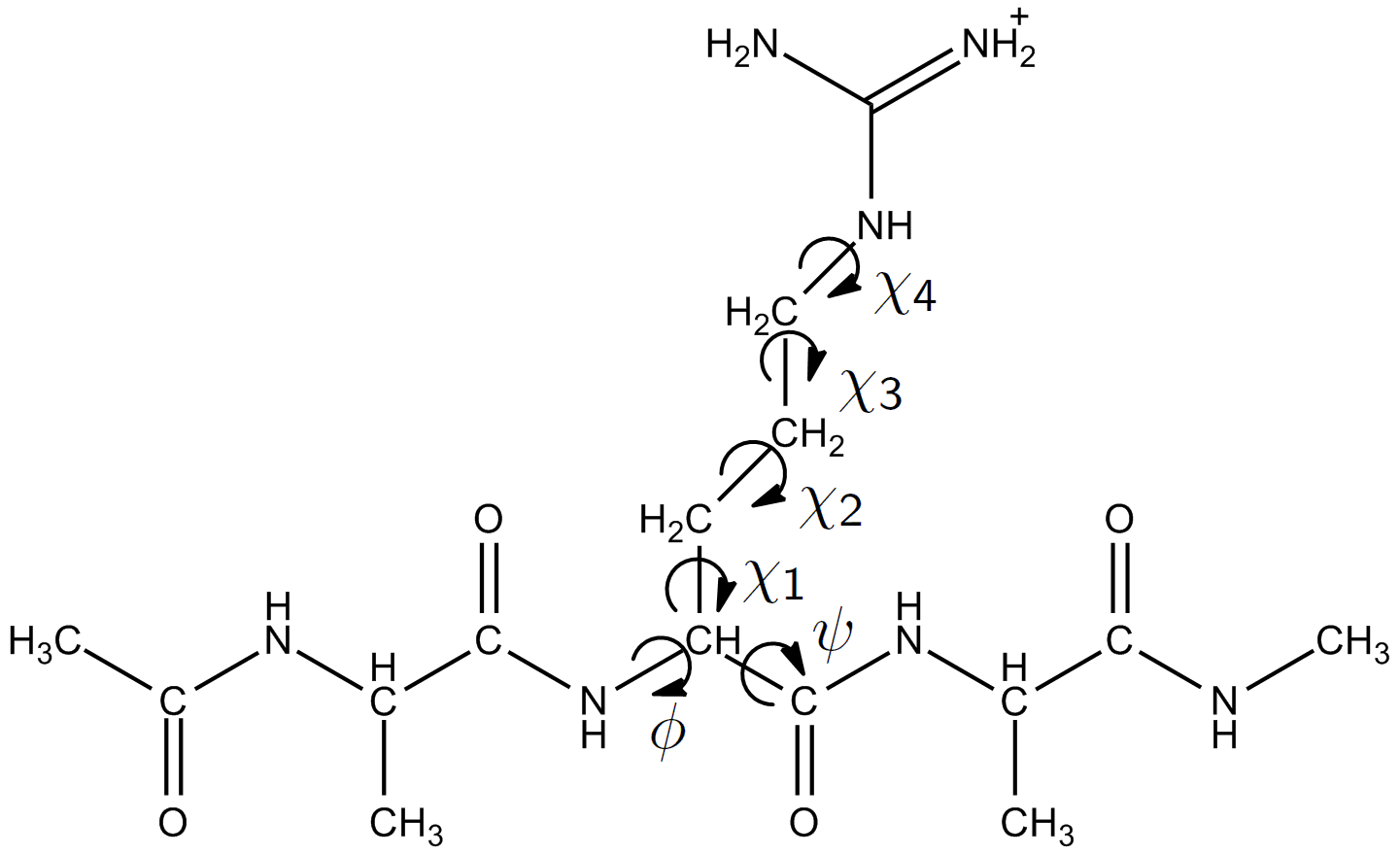}
                \caption{\small{ARA Arginie}}
                \label{fig:1e}
     \end{subfigure} \hspace{4 mm}
	\label{Tripeptides3}
\end{figure}

\end{document}